\documentclass{aa}
\usepackage{graphicx}
\usepackage{txfonts}
\usepackage{natbib}
\usepackage{xcolor}
\usepackage{xspace}
\usepackage{rotating}
\usepackage[colorlinks=true,linkcolor=blue,citecolor=blue]{hyperref}
\usepackage{etoolbox}

\makeatletter

\patchcmd{\NAT@citex}
  {\@citea\NAT@hyper@{%
     \NAT@nmfmt{\NAT@nm}%
     \hyper@natlinkbreak{\NAT@aysep\NAT@spacechar}{\@citeb\@extra@b@citeb}%
     \NAT@date}}
  {\@citea\NAT@nmfmt{\NAT@nm}%
   \NAT@aysep\NAT@spacechar\NAT@hyper@{\NAT@date}}{}{}

\patchcmd{\NAT@citex}
  {\@citea\NAT@hyper@{%
     \NAT@nmfmt{\NAT@nm}%
     \hyper@natlinkbreak{\NAT@spacechar\NAT@@open\if*#1*\else#1\NAT@spacechar\fi}%
       {\@citeb\@extra@b@citeb}%
     \NAT@date}}
  {\@citea\NAT@nmfmt{\NAT@nm}%
   \NAT@spacechar\NAT@@open\if*#1*\else#1\NAT@spacechar\fi\NAT@hyper@{\NAT@date}}
  {}{}

\makeatother
\newcommand{\hlink}[1]{\mbox{\url{http://#1}}\xspace}

\newcommand{\rfig}[1]{Fig.~\ref{#1}}
\newcommand{\rfigs}[1]{Figs.~\ref{#1}}
\newcommand{\req}[1]{Eq.~\ref{#1}}
\newcommand{\reqs}[1]{Eqs.~\ref{#1}}

\newcommand{\rtabs}[1]{Tables \ref{#1}}
\newcommand{\rapp}[1]{Appendix \ref{#1}}

\newcommand{\Rsec}[1]{Section \ref{#1}}
\newcommand{\rsec}[1]{section \ref{#1}}
\newcommand{\rsecs}[1]{sections \ref{#1}}

\newcommand{\egg}{\textsc{EGG}\xspace}
\newcommand{\herschel}{{\it Herschel}\xspace}
\newcommand{\spitzer}{{\it Spitzer}\xspace}
\newcommand{\hubble}{{\it Hubble}\xspace}
\newcommand{\hst}{{\it HST}\xspace}
\newcommand{\jwst}{{\it JWST}\xspace}

\newcommand{\um}{\mu{\rm m}}
\newcommand{\mm}{{\rm mm}}
\newcommand{\uJy}{\mu{\rm Jy}}
\newcommand{\mJy}{{\rm mJy}}

\newcommand{\sfr}{{\rm SFR}}

\newcommand{\sfrms}{{\rm SFR}_{\rm MS}}
\newcommand{\ssfr}{{\rm sSFR}}
\newcommand{\lir}{L_{\rm IR}}

\newcommand{\leight}{L_8}
\newcommand{\ireight}{{\rm IR8}}

\newcommand{\luv}{L_{\rm UV}}

\newcommand{\msun}{{\rm M}_\odot}
\newcommand{\Mpc}{{\rm Mpc}}
\newcommand{\kpc}{{\rm kpc}}

\newcommand{\dex}{{\rm dex}}
\newcommand{\mstar}{M_\ast}

\newcommand{\tdust}{T_{\rm dust}}

\newcommand{\rsb}{R_{\rm SB}}
\newcommand{\uvj}{$UVJ$\xspace}

\newcommand{\sersic}{S\'ersic\xspace}
\newcommand{\galfit}{GALFIT\xspace}

\newcommand{\bt}{B/T}

\newcommand{\irx}{{\rm IRX}}
\newcommand{\mean}[1]{\left<#1\right>}

\newcommand{\kelvin}{{\rm K}}
\newcommand{\logm}{\log_m}
\newcommand{\logz}{\log_z}

\newcommand{\Ks}{$K_{\rm s}$\xspace}

\newcommand{\stuff}{{\it Stuff}\xspace}
\newcommand{\skymaker}{{\it SkyMaker}\xspace}
\newcommand{\astrodeep}{\mbox{ASTRODEEP}\xspace}

\defcitealias{schreiber2015}{S15}
\defcitealias{chary2001}{CE01}
\defcitealias{henriques2015}{H15}
\defcitealias{somerville2012}{S12}
\defcitealias{lacey2016}{L16}

\newcommand*\dd{\ensuremath{{\rm d}}}

\begin{document}

%------------------------------
% Title page
%------------------------------
\title{\egg\thanks{\hlink{cschreib.github.io/egg/}}: hatching a mock Universe from empirical prescriptions}

\author{C.~Schreiber\inst{1,2}, D.~Elbaz\inst{2}, M.~Pannella\inst{3,2}, E.~Merlin\inst{4}, M.~Castellano\inst{4}, A.~Fontana\inst{4}, N.~Bourne\inst{5}, K.~Boutsia\inst{4}, F.~Cullen\inst{5}, J.~Dunlop\inst{5}, H.~C.~Ferguson\inst{6}, M.~J.~Micha{\l}owski\inst{5}, K.~Okumura\inst{2}, P.~Santini\inst{4}, X.~W.~Shu\inst{7,2}, T.~Wang\inst{2,8} and C.~White\inst{9}}

\authorrunning{C.~Schreiber et al.}

\institute{
    Leiden Observatory, Leiden University, NL-2300 RA Leiden, The Netherlands \\
    \email{cschreib@strw.leidenuniv.nl}
    \and
    Laboratoire AIM-Paris-Saclay, CEA/DSM/Irfu - CNRS - Universit\'e Paris Diderot, CEA-Saclay, pt courrier 131, F-91191 Gif-sur-Yvette, France
    \and
    Faculty of Physics, Ludwig-Maximilians Universit\"at, Scheinerstr.\ 1, 81679 Munich, Germany
    \and
    INAF - Osservatorio Astronomico di Roma, Via Frascati 33, I-00040 Monte Porzio Catone (RM), Italy
    \and
    SUPA, Institute for Astronomy, University of Edinburgh, Royal Observatory, Edinburgh, EH9 3HJ, UK
    \and
    Space Telescope Science Institute, 3700 San Martin Drive, Baltimore, MD 21218, USA
    \and
    Department of Physics, Anhui Normal University, Wuhu, Anhui, 241000, China
    \and
    School of Astronomy and Astrophysics, Nanjing University, Nanjing, 210093, China
    \and
    Department of Physics and Astronomy, Johns Hopkins University, 3400 North Charles Street, Baltimore, MD 21218, USA
}

\date{Received 15 june 2016 / Accepted 10 january 2017}

\abstract {
This paper introduces \egg, the Empirical Galaxy Generator, a tool designed within the \astrodeep collaboration to generate mock galaxy catalogs for deep fields with realistic fluxes and simple morphologies. The simulation procedure is based exclusively on empirical prescriptions -- rather than first principles -- to provide the most accurate match with current observations at $0 < z < 7$. We consider that galaxies can be either quiescent or star-forming, and use their stellar mass ($\mstar$) and redshift ($z$) as the fundamental properties from which all the other observables can be statistically derived. Drawing $z$ and $\mstar$ from the observed galaxy stellar mass functions, we then associate a star formation rate ($\sfr$) to each galaxy from the tight $\sfr$--$\mstar$ main sequence, while dust attenuation, optical colors and simple disk/bulge morphologies are obtained from empirical relations that we establish from the high quality \hubble and \herschel observations from the CANDELS fields. Random scatter is introduced in each step to reproduce the observed distributions of each parameter. Based on these observables, an adequate panchromatic spectral energy distribution (SED) is selected for each galaxy and synthetic photometry is produced by integrating the redshifted SED in common broad-band filters. Finally, the mock galaxies are placed on the sky at random positions with a fixed angular two-point correlation function to implement basic clustering. The resulting flux catalogs reproduce accurately the observed number counts in all broad bands from the ultraviolet up to the sub-millimeter, and can be directly fed to image simulators such as \skymaker. The images can then be used to test source extraction softwares and image-based techniques such as stacking. \egg is open-source, and is made available to the community on behalf of the \astrodeep collaboration, together with a set of pre-generated catalogs and images.
}

\keywords{galaxies: evolution -- galaxies: structure -- galaxies: statistics -- galaxies: photometry}

\maketitle

%------------------------------
%------------------------------
\section{Introduction}
%------------------------------
%------------------------------

To a large extent, most of our knowledge of astronomy and astrophysics is derived from two- (or three-) dimensional images of the sky acquired by observatories in space or on the ground. Different instruments will generally produce images of strongly varying properties, including (but not limited to) zero point calibration, point spread function (PSF), noise, or sky-to-pixel projection. Therefore, extracting observables of astronomical interest -- namely fluxes, shapes, positions and the respective uncertainties -- requires a good knowledge of the instrument and the image reduction pipeline. For this reason, these observables are typically compiled into catalogs, which can be used with minimal knowledge of the instrument or the image itself and allow a more immediate scientific analysis. Building these catalogs is not straightforward, and various techniques and tools have been introduced during the history of astronomy, ranging in complexity from aperture photometry to multi-component profile fitting \citep[e.g.,][to only mention most recent efforts]{bertin1996,labbe2006,desantis2006,laidler2007,wuyts2007,merlin2015}.

In this context, the goal of the \astrodeep collaboration\footnote{\hlink{www.astrodeep.eu/}} is to provide the astronomy community with optimally extracted flux catalogs in the most data-rich cosmological deep fields (i.e., the GOODS and CANDELS fields). To achieve this goal, new photometric codes and techniques are being developed (see, e.g., \citealt{merlin2015,cappelluti2016,shu2016,merlin2016,castellano2016}; Wang et al.~in prep.) to improve on existing practices both in efficiency and accuracy.

A key step in the conception process of such codes is to characterize their performance and accuracy before applying them to real images, checking not only the robustness of the flux measurements, but also the quality of the error estimates. In addition, the challenge of extracting photometry in far-IR and sub-millimeter images is tied to the large size of the PSF, which generates confusion noise \citep[e.g.,][and references therein]{dole2004-a}. In this situation the main issue is the choice of an optimal strategy to select the prior positions at which the sources will be extracted, since using the positions of all known galaxies would result either in over-fitting or a degenerate fit. For this reason, a secondary goal of \astrodeep is to provide the astrophysics community with realistic simulations of the sky at different wavelengths and with different angular resolutions, so that astronomers can test their procedures and tools and quantify their respective efficiency. These simulations must provide two essential components: first, simulated images including realistic noise properties, and second, the corresponding mock galaxy catalogs containing the true flux of each object. The resulting mock observations should be as close as possible to the real ones, including in particular the correct flux and color distributions.

Various approaches can be used to generate such mock galaxy catalogs. We separate them in two main classes: physically-motivated approaches on the one hand, and empirical approaches on the other hand. Procedures that belong to the first class are typically based on the output of large-scale cosmological simulations, such as Illustris \citep{genel2014}, or semi-analytic models (SAMs). These simulations provide dark matter halo populations with realistic redshift-dependent spatial and mass distribution, since the physics governing the growth of these halos is well understood. However, there are two main drawbacks linked to these approaches. First, the physical properties of the galaxies created by such simulations do not always match the observations (e.g., the systematic factor of two underestimation of the star formation activity in $z=2$ galaxies; \citealt{daddi2007-a,gruppioni2015}); and second, such simulations are computationally expensive and can render some tests impractical if a large number of random realizations is required.

The second class of approach aims at reproducing the observations by construction, sacrificing the consistency of the physics to reach higher fidelity of the mock data. The easiest way to achieve this is to construct a flux distribution from the observed number counts and use it to draw fluxes randomly for each galaxy (see, e.g., the \stuff program introduced in \citealt{bertin2009}). The process is extremely fast and can be used to generate datasets potentially larger than the observable Universe. But this first order approach also has limitations: first, linking the different photometric bands together and ensuring that colors and their scatter are properly reproduced is non-trivial; and second, it is impractical to extend this model to photometric bands with poorly constrained (or inexistent) number counts. Therefore, a more generic and successful approach would instead generate a realistic spectrum for each galaxy, from which broad band photometry can be immediately derived (e.g., \citealt{franceschini2001,bethermin2011,gruppioni2011,bethermin2012-a}). Choosing the spectrum and luminosity for a particular galaxy then becomes the central question. Galaxy spectra are known to be the sum of multiple components, including mainly stellar, nebular and dust emission. Active galactic nuclei (AGNs) in their various phases can  also become the dominant light source in some specific wavelength domains \citep[e.g.,][]{hao2005,richards2006,hatziminaoglou2010}. Constructing the spectrum of a galaxy hence requires a description of these various components.

To this end, in the present paper we propose a set of empirical relations and recipes that we calibrate on the deepest observations available to date in the \hubble CANDELS fields. Crucially, these fields are also covered by the deepest \herschel observations, which allow us to derive precise constraints on the dust-obscured properties of galaxies up to $z\sim3$. We implement these recipes in a new tool named \egg (the Empirical Galaxy Generator) to build realistic mock galaxy catalogs from the UV to the submillimeter.

Briefly, the process we employ considers that galaxies can first be segregated into two broad populations of ``star-forming'' (SFGs) and ``quiescent'' galaxies (QGs). We generate galaxies from these two populations based on their observed stellar mass functions. Then, all the other physical properties are statistically inferred from the SFG/QG classification, the stellar mass, and the redshift of these galaxies. We use these three parameters as the driving factors for a number of observables such as morphology (e.g., size, or bulge vs.~disk mass ratio), optical colors, dust attenuation and star formation rate. To fine tune the fidelity of the final data set, we introduce second order variations by adding a controlled amount of random scatter to most of the observables.

The outline of this paper is the following. In \rsec{SEC:sample} we describe the observational data set that we use to calibrate our recipes. This includes a brief description of the fields and the methods used to derive the physical parameters of each observed galaxy. In \rsec{SEC:stellar} we describe the recipes to derive the stellar emission, starting from the stellar mass functions (\rsec{SEC:mfunc}), the morphology of the stellar profile (\rsec{SEC:morpho}), and the optical colors (\rsec{SEC:opti_sed}). In \rsec{SEC:dust} we present our description of the dust emission, including the parametrization of star formation rate (\rsec{SEC:sfr_dust}) and obscuration (\rsec{SEC:dust_dust}), and the properties of dust, such as its temperature and chemical composition (\rsec{SEC:dust_prop}). \Rsec{SEC:clustering} describes the approach we use to generate realistic sky position distributions, including clustering. \Rsec{SEC:simu} gives an brief overview of how the simulation is assembled and implemented in the code. The resulting catalogs and images are compared to observations in \rsecs{SEC:counts} and \ref{SEC:pixelstat}, and to recent SAMs in \rsec{SEC:sam}. Examples of simulated images are given in \rsec{SEC:comp_img}. Finally, \rsec{SEC:future} discusses the limitations of the simulation, and directions to improve it.

In the following, we assume a $\Lambda$CDM cosmology with $H_0 = 70\ {\rm km}\,{\rm s}^{-1} {\rm Mpc}^{-1}$, $\Omega_{\rm M} = 0.3$, $\Omega_\Lambda = 0.7$ and a \cite{salpeter1955} initial mass function (IMF), to derive both star formation rates and stellar masses. All magnitudes are quoted in the AB system, such that $M_{\rm AB} = 23.9 - 2.5\log_{10}(S_{\!\nu}\ [\uJy])$.

%------------------------------
%------------------------------
\section{Sample and observations \label{SEC:sample}}
%------------------------------
%------------------------------

We base this analysis on the sample and data described in \cite{schreiber2015} \citepalias[hereafter][]{schreiber2015}. In this section, we make a brief summary of these observations.

%------------------------------
\subsection{Multi-wavelength photometry}
%------------------------------

The catalogs we use in this work are the official catalogs produced by CANDELS \citep{grogin2011,koekemoer2011} from the \emph{Hubble Space Telescope} (\hst) WFC3 $H$-band imaging of the four fields also covered by deep \herschel PACS and SPIRE observations, namely GOODS--{\it North} (Barro et al.~in prep.), GOODS--{\it South} \citep{guo2013-a}, UDS \citep{galametz2013} and COSMOS (Nayyeri et al.~in prep.). Each of these fields is about $150\,{\rm arcmin}^2$ and they are evenly distributed on the sky to mitigate cosmic variance.

The ancillary photometry varies from one field to another, being a combination of both space- and ground-based imaging from various facilities. It is described in detail in the catalog papers cited above, as well as in \citetalias{schreiber2015}. Briefly, the UV to near-IR wavelength coverage typically goes from the $U$ band up the \spitzer IRAC $8\,\um$, including at least the \hst bands F606W, F814W, and F160W and a deep $K$ (or $K_{\rm s}$) band, and all these images are among the deepest available views of the sky. These catalogs therefore cover most of the important galaxy spectral features across a wide range of redshifts, even for intrinsically faint objects.

We complement these catalogs with mid-IR photometry from \spitzer MIPS and far-IR photometry from \herschel PACS and SPIRE taken as part of the GOODS--\herschel \citep{elbaz2011} and CANDELS--\herschel programs (Inami et al.~in prep.).

%------------------------------
\subsection{Redshifts, stellar masses and rest-frame optical colors \label{SEC:zmstar}}
%------------------------------

From this observed photometry, photometric redshifts and stellar masses are computed following \cite{pannella2015}. The details of the fitting procedure can be found there and in \citetalias{schreiber2015}, and we only provide a brief overview in the following.

We use EAzY \citep{brammer2008} to derive the photometric redshifts from the CANDELS catalogs, allowing slight adjustments of the photometric zero points by iteratively comparing our photo-$z$s against the available spec-$z$s. The stellar masses are then computed using FAST \citep{kriek2009} by fixing the redshift to the best-fit photo-$z$ and fitting the observed photometry up to the IRAC $4.5\,\um$ band using the \cite{bruzual2003} stellar population synthesis model, assuming a \cite{salpeter1955} IMF and a \cite{calzetti2000} extinction law, allowing a range of attenuation with $A_V=0$ up to $4$. All galaxies are described with a delayed exponentially declining star formation history (SFH), with $\sfr(t) \sim t \exp(-t/\tau)$, and we allow both the age and the exponential time-scale $\tau$ to vary. This parametrization allows for both rising, declining and bursty SFHs, depending on the value of the age and ${\rm age}/\tau$, and therefore covers a large parameter space.

As shown in \cite{pannella2015} (their Appendix C) or \cite{santini2015} (their Fig.~2), the precise choice of the SFH does not significantly impact the stellar mass \citep[unless single bursts are used, see][]{michalowski2014}. Furthermore, the wavelength coverage in the CANDELS fields is such that the rest-frame UV-to-optical is properly sampled up to $z\sim7$, thanks in particular to the \spitzer IRAC bands, and therefore our mass estimates should not suffer from significant redshift-dependent systematics (see, e.g., \citealt{santini2015,grazian2015}).

Regardless, the present work is largely insensitive to the accuracy of these redshift and mass determinations. One way to see this is to consider that, throughout this paper, we essentially use $z$ and $\mstar$ to generate a simpler 2-dimensional space in which all galaxies are projected. Then, the recipes we derive in the next sections are designed to map a point from this ``projected'' space back into a much larger space of observables (i.e., all the fluxes and morphological parameters) to reproduce the observed distributions. In this sense, $z$ and $\mstar$ can be seen as arbitrary intermediate variables, and their physical meaning (or correctness) is not important. It should therefore be immediately apparent that any systematic error, e.g., in our stellar masses, will simply cancel out by construction and still produce the right distributions for the observables. This is only true because we derive our recipes and stellar mass functions from the same data set (except for the relation between $B/T$ and $\mstar$ which we took from \citealt{lang2014}, see \rsec{SEC:morpho}, however their data set and methodology are very similar to ours).

In this work (and as in \citetalias{schreiber2015}), we only consider galaxies with $H<26$ to ensure a high quality SED and photometric redshifts. When appropriate, we take into account the resulting selection effects of this magnitude cut (e.g., on the stellar mass functions and mass completeness limits). Sources with an uncertain photometric redshift (redshift \texttt{odds} less than $0.8$, as given by EAzY) or bad SED fitting (reduced $\chi^2$ larger than $10$) are excluded from the present analysis. These represent from $3$ to $6\%$ of our sample, depending on the stellar mass and redshift range, and their impact on our results are therefore marginal. We also explicitly remove foreground stars from the catalogs, using a combination of morphology and $BzK$ colors, as in \cite{pannella2015}.

Lastly, the rest-frame $U$, $V$ and $J$ magnitudes are computed for each galaxy using EAzY, by integrating the best-fit galaxy template from the photo-$z$ estimation. These colors are used, following \cite{williams2009}, to separate those galaxies that are ``quiescent'' (QGs) from the ``star-forming'' ones (SFGs). We use the same selection criteria as in \citetalias{schreiber2015}, i.e., a galaxy is deemed star-forming if its colors satisfy
\begin{equation}
    UVJ_{\rm SF} = \left\{\begin{array}{rcl}
        U - V &<& 1.3\,\text{, or} \\
        V - J &>& 1.6\,\text{, or} \\
        U - V &<& 0.88\times(V - J) + 0.49\,,
    \end{array}\right.\label{EQ:uvj}
\end{equation}
otherwise the galaxy is considered as quiescent.

%------------------------------
%------------------------------
\section{Stellar properties \label{SEC:stellar}}
%------------------------------
%------------------------------

%------------------------------
\subsection{Conditional stellar mass functions \label{SEC:mfunc}}
%------------------------------

% program: code/astrodeep/gencat/calibrate/plot_genmf.pro
% arguments: plot_genmf, suffix='active', /show_fit
% arguments: plot_genmf, suffix='passive', /show_fit
% file: code/astrodeep/gencat/calibrate/mf_active.eps
% file: code/astrodeep/gencat/calibrate/mf_passive.eps
\begin{figure*}[htpb]
    \centering
    \includegraphics[width=9cm]{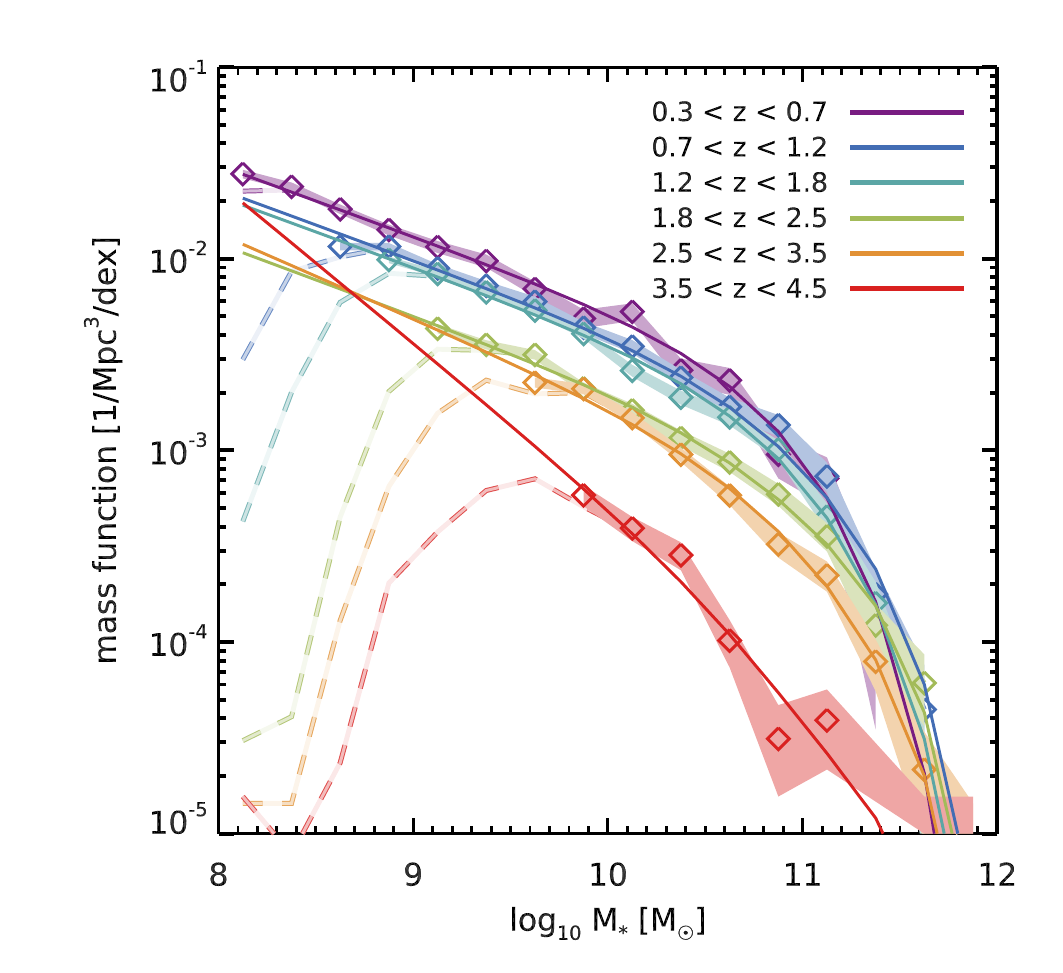}
    \includegraphics[width=9cm]{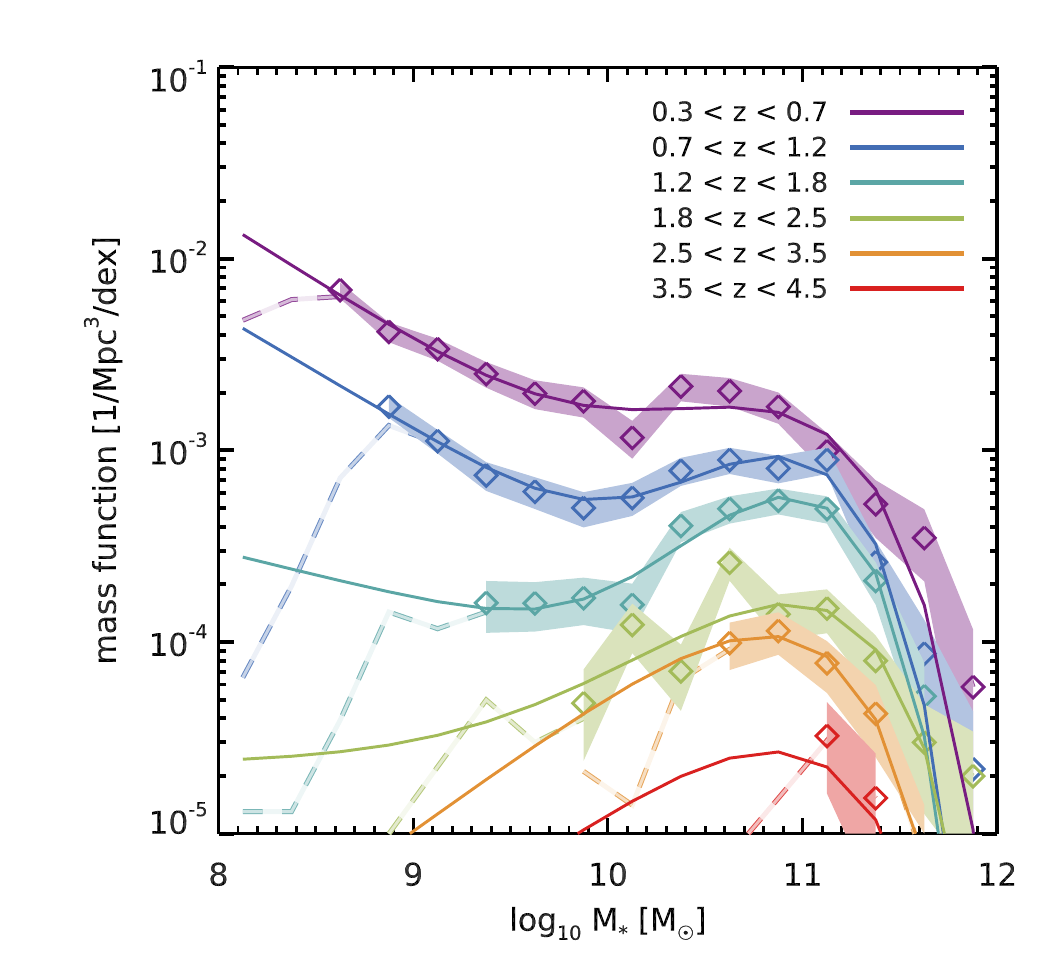}
    \caption{Conditional stellar mass function at different redshifts for SFGs (left) and QGs (right), selected with $H<26$. The dashed lines in the background indicate the raw mass functions, before completeness corrections are applied. The solid colored regions show the completeness-corrected estimate of the mass function, and the width of the region indicates the statistical uncertainty on the measurement (i.e., Poisson noise).}
    \label{FIG:mfunc}
\end{figure*}

The initial purpose of \egg is to simulate a deep field similar to the GOODS--{\it South} field. Therefore, we compute the stellar mass function in this field only, in order to most closely mimic is properties (including, in particular, cosmic variance). To do so, we use the procedure described in \citetalias{schreiber2015}, which we now briefly recall.

As stated in \rsec{SEC:zmstar}, galaxies from the GOODS--{\it South} catalog are selected with $H<26$ to ensure high quality photometry for all galaxies. This mostly eliminates the effect of the Eddington bias. As in \citetalias{schreiber2015}, we estimated the evolution of the stellar mass completeness ($90\%$) corresponding to this selection, as inferred from the observed scatter in the $\mstar$-to-$L_{H/(1+z)}$ ratio for SFGs and QGs separately. We found, for example, that at $z=1$ the completeness is as low as $5\times10^8\,\msun$ for SFGs. We use this information to correct the counts for incompleteness up to a factor of $2$, and do not attempt to measure the stellar mass function when the correction becomes larger. We then define multiple redshift bins from $z=0.3$ to $z=4.5$, and compute within each of these bins the mass distribution for both SFGs and QGs separately, according to the \uvj color-color selection (see \rsec{SEC:zmstar}). Then, we fit a double Schechter law to each distribution:
\begin{align}
    \frac{\dd^2 N(z)}{\dd \log_{10}\mstar\,\dd V} &= S(\mstar, \phi^\star_1, M^\star_1, \alpha_1) + S(\mstar, \phi^\star_2, M^\star_2, \alpha_2)\,, \nonumber \\
    S(\mstar, \phi^\star, M^\star, \alpha) &\equiv \log(10)\,\phi^\star \left(\frac{\mstar}{M^\star}\right)^{\alpha + 1} \hspace{-4pt} \exp\left(-\frac{\mstar}{M^\star}\right)\,.
\end{align}

The results are shown in \rfig{FIG:mfunc}, and the best-fit parameters are summarized in \rtabs{TAB:mfparam_active} and \ref{TAB:mfparam_passive}. Our goal here is only to find a functional form that describes well the observed data. We thus attribute no physical origin to each component of the double Schechter law, and because the fit is prone to degeneracy, we allow ourselves to arbitrarily fix some of the fit parameters. These are surrounded by brackets in the tables.

We estimate that our catalog is not complete to assess the mass function of $z=4$ QGs, and therefore do not attempt to fit it. Instead, we use the same parameters as that obtained at lower redshifts and only adjust $\phi^\star$ (the normalization) to have a fraction of quiescent galaxies equal to $15\%$ (for $\mstar > 4\times10^{10}\,\msun$), which is the extrapolation of the trend we observe at lower redshifts. This is consistent with what was previously reported by, e.g., \cite{muzzin2013}. However, it was recently suggested that this fraction could be substantially higher, e.g., up to $34\%$ of quiescent galaxies at $z=3.7$ \citep{straatman2014}. In any case, this will not change dramatically the quality of our simulated catalogs because these objects are faint and their number density very low.

To reach higher redshifts, we use the results of \cite{grazian2015} covering $4.5 < z < 7.5$. Their stellar mass functions do not distinguish SFGs from QGs, but since QGs are found in negligible numbers at these redshifts, we simply consider that the Grazian et al.~mass functions describe the SFG population. When then assume that the fraction of QGs remains $15\%$ at $z>4$ and just rescale their $z=4$ mass functions to match this constraint at all $z>4$. We extrapolate these trends up to $z=11$ to reach down to the detection limit of today's deep surveys, although we note that this extrapolation is highly uncertain and that galaxies at $z>7$ in the simulation have essentially unconstrained properties.

As for very low redshifts, we take the $z=0$ mass function from \cite{baldry2012}, assuming their SFG/QG separation from a color-magnitude diagram isolates the same populations as the \uvj selection. This will have little consequences since we are aiming for pencil-beam surveys which contain very few local galaxies.

Extrapolating these combined mass functions towards the low-mass end (assuming that the low-mass slope is not varying) we can generate a population of galaxies on a wide range of stellar masses in an arbitrary volume between $z=0$ and $z=11$.

%------------------------------
\subsection{Stellar morphology \label{SEC:morpho}}
%------------------------------

% program: code/astrodeep/gencat/calibrate/plot_morpho.pro
% file: code/astrodeep/gencat/calibrate/axis_ratio.eps
\begin{figure}
    \centering
    \includegraphics[width=9cm]{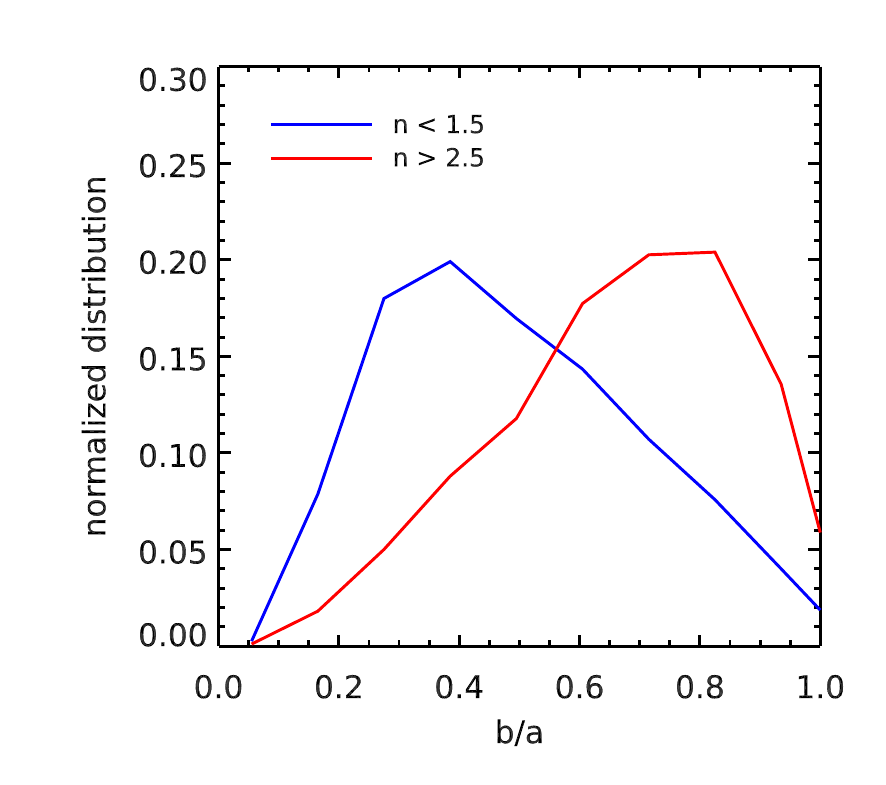}
    \caption{Observed axis-ratio distribution of disk-dominated ($n<1.5$, blue) and bulge-dominated ($n>2.5$, red) samples, combining all redshifts.}
    \label{FIG:axis_ratio}
\end{figure}

Following the approach of \stuff, we consider that galaxies are made of two components: a bulge (\sersic index $n=4$) and a disk (\sersic index $n=1$). The fraction of the stellar mass that goes into one or the other component is defined by the bulge-to-total ratio ($B/T$), and each of these components is described by several morphological parameters, including the projected axis ratio $b/a$, the half-light radius $R_{50}$, and the position angle $\theta$. In the following, we present how we calibrate the distributions of these parameters.

The bulge-to-total ratio is estimated following the results of \cite{lang2014} who conveniently measured the average $B/T$ as a function of stellar mass for both SFGs and QGs in the CANDELS fields at different redshifts. While they found the bulge fraction to increase with stellar mass for both populations, they did not observe any significant difference with redshift between $z=1$ and $z=2$, so we chose to make the $B/T$ simply depend on mass following
\begin{align}
\log_{10}(B/T)_{\rm SF} &= -0.7 + 0.27 \times \left(\logm - 10\right)\,\text{and} \nonumber \\
\log_{10}(B/T)_{\rm Q} &= -0.3 + 0.1 \times \left(\logm - 10\right)\,,
\end{align}
with $\log_m \equiv \log_{10}(\mstar/\msun)$, to which we add a log-normal scatter of $0.2\,\dex$ in order to reproduce the width of the distribution reported in \cite{lang2014}. This is a mass-weighted bulge-to-total ratio, therefore we can directly use it to compute the stellar mass inside the disk and the bulge. Estimating the contribution of each component to the luminosity of the galaxy is done in \rsec{SEC:opti_sed}.

To calibrate the morphological parameters for bulges and disks, we use the morphological catalogs of \cite{vanderwel2012} who fitted single \sersic profiles of varying index $n$ to all galaxies in the CANDELS fields using the \galfit software \citep{peng2002} on the \hst $H$-band images. In the following, we will consider two sub-samples: first, galaxies with $n<1.5$ and $\mstar > 10^9\,\msun$, and second, galaxies with $n>2.5$ and $\mstar > 3\times10^{10}\,\msun$. The cut in stellar mass is used to select galaxies bright enough that the \sersic fits are reliable. We use these sub-samples to calibrate the morphology of the disk and bulges, respectively. Indeed, for galaxies with $n<1.5$ the presence of a bulge can be neglected so that the measured properties can be attributed to the disk alone \citep[see, e.g., the Appendix of][]{lang2014}, and conversely for $n>2.5$. This latter sample of $n>2.5$ galaxies is probably less pure though, since high \sersic indices can be produced either by a dominant bulge, or by a minor bulge that has a much smaller half-light radius than the disk, as shown in the Appendix of \cite{lang2014}. However such cases are relatively rare, and the majority of $n>2.5$ galaxies can indeed be considered as bulge-dominated.

For each sub-sample, we start by measuring the projected axis ratio distribution (\rfig{FIG:axis_ratio}). We find that disk-dominated galaxies have on average lower $b/a$, which is expected from to their quasi two-dimentional intrinsic geometry: their $b/a$ distribution peaks at $0.3$, compared to $0.8$ for bulge-dominated galaxies. We consider that these distributions hold for all masses and all redshifts. \Citet{vanderwel2014} reported that the $b/a$ distribution of SFGs at $z=1.7$ shows a clear mass evolution from $10^9$ to $10^{11}\,\msun$: while their low-mass distribution is very similar to our disk-dominated distribution, they found their high-mass distribution to be bimodal. Without attempting to demonstrate it, we argue here that this trend is probably the result of the increase of the $B/T$ with stellar mass among SFGs \citep{lang2014}. On the one hand, low-mass galaxies are preferentially bulgeless, and should therefore follow the trend of pure-disks of \rfig{FIG:axis_ratio}. On the other hand, high-mass galaxies are more complex systems with a varying mixture of bulges and disks; among these, we expect to find both bulge- and disk-dominated systems, and this would explain the bimodal distribution observed by \cite{vanderwel2014}\footnote{Interestingly, by comparing their $z=1.7$ result to a similar analysis in the Sloan Digital Sky Survey (SDSS, $z=0$), \cite{vanderwel2014} showed that these distributions are also redshift-dependent, so that the bimodality extends down to lower stellar masses at lower redshifts. One possible explanation for this would be that the redshift invariance of the $B/T$--$\mstar$ relation found by \cite{lang2014} may not hold at $z<1$, indicating that SFGs in the Local Universe have more prominent bulges at fixed stellar mass.}.

% program: code/astrodeep/gencat/calibrate/plot_morpho.pro
% file: code/astrodeep/gencat/calibrate/hlrad_disk.eps
% file: code/astrodeep/gencat/calibrate/hlrad_bulge.eps
\begin{figure*}
    \centering
    \includegraphics[width=14cm]{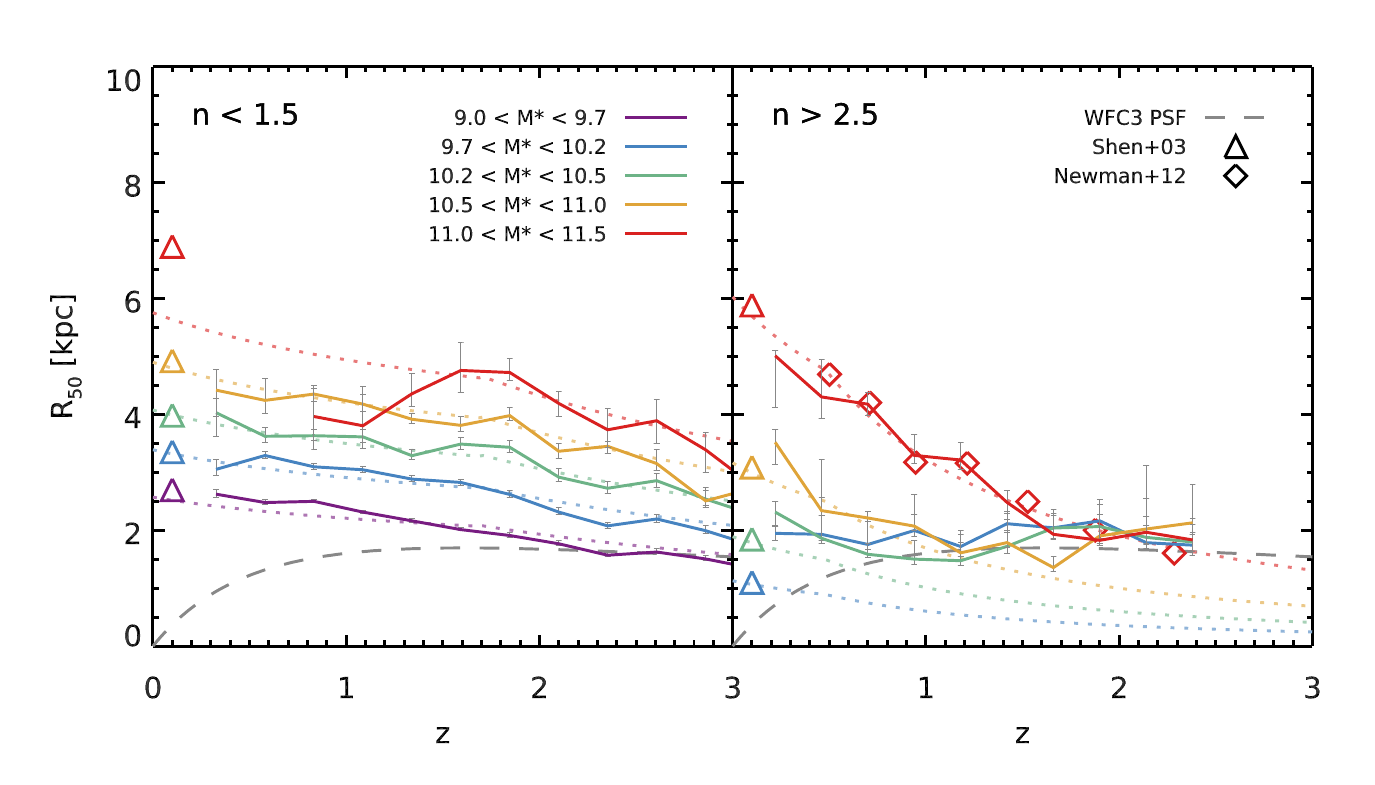}
    \caption{Observed relation between the half-light radius $R_{50}$ (along the major axis) and redshift of disk-dominated galaxies (left) and bulge-dominated galaxies (right). Different stellar mass bins are shown with different colors as indicated in the legend. The median values over all CANDELS fields are shown with solid colored lines, error bars indicating the uncertainty on the median from bootstrapping, and the prescription adopted in this work is displayed with a dotted line in the background. We also show how the size of the \hst $H$-band PSF ($0.2"$) translates into proper distance with a dashed line. Empty triangles at $z=0.1$ are the values obtained by \cite{shen2003} in the SDSS, converted from a Kroupa to a Salpeter IMF. Based on the median axis ratios we measure (\rfig{FIG:axis_ratio}) and following \cite{dutton2011}, we multiply the Shen et al.~values by a factor $1.4$ and $1.1$ for disk- and bulge-dominated galaxies, respectively, to correct for the fact that their radii were measured in circularized apertures. Finally, for bulge-dominated galaxies, we also display the size measurements of \cite{newman2012}, which were obtained by selecting quiescent galaxies based on their $\ssfr$. Their values are reported as $R_{50}/M_{11}^{0.57}$, which we renormalize to the stellar mass of our highest mass bin.}
    \label{FIG:hlrad}
\end{figure*}

The next step is the calibration of the half-light radius. The size of a galaxy correlates with its stellar mass (i.e., the mass--size relation, see, e.g., \citealt{shen2003}), and sizes at fixed mass were also globally smaller in the past \citep[e.g.,][]{ferguson2004,daddi2005}. For this reason, we bin our two sub-samples in stellar mass and observe the evolution of the median half-light radius with redshift. The observed trends are reported in \rfig{FIG:hlrad}. Defining $\log_z \equiv \log_{10}(1+z)$, we parametrize these relations with the following equations, for disks:
\begin{align}
\log_{10}(R_{50,\rm disk} [\kpc]) &= 0.2 \times (\logm - 9.35) + F_{z,{\rm disk}}\,, \\
\text{with $F_{z,{\rm disk}}$} &= \left\{\begin{array}{lcl}
    0.41 - 0.22 \times \logz  & \text{for} & z \leq 1.7 \\
    0.62 - 0.70 \times \logz  & \text{for} & z > 1.7
\end{array}\right.\,, \nonumber
\end{align}
and for bulges:
\begin{align}
\log_{10}(R_{50,\rm bulge} [\kpc]) &= 0.2 \times (\logm - 11.25) + F_{z,{\rm bulge}}\,, \\
\text{with $F_{z,{\rm bulge}}$} &= \left\{\begin{array}{lcl}
    0.78 - 0.6 \times \logz  & \text{for} & z \leq 0.5 \\
    0.90 - 1.3 \times \logz  & \text{for} & z > 0.5
\end{array}\right.\,, \nonumber
\end{align}
to which we add a log-normal scatter of $0.17$ and $0.2\,\dex$ respectively. Although these latter values are smaller than the scatter reported, e.g., by \cite{shen2003} or \cite{dutton2011}, we stress that they apply to the disk and bulge components only. When considering the size of the galaxy as a whole (i.e., the sum of the disk and the bulge, see \rapp{APP:size}), we find that the additional scatter in the bulge-to-disk ratio is sufficient to reproduce the observed width of the mass--size relation. However, to preserve the normalization of the mass--size relation in composite systems, we use the total mass $\mstar$ to derive each component's respective size.

Lastly, we attribute a position angle to each galaxy by randomly drawing from a uniform distribution, and assign the same angle to both the bulge and disk components.

%------------------------------
\subsection{Stellar spectral energy distribution \label{SEC:opti_sed}}
%------------------------------

% program: code/astrodeep/gencat/calibrate/uvj_track.pro
% file: code/astrodeep/gencat/calibrate/uvj_track.eps
% file: code/astrodeep/gencat/calibrate/uvj_proj.eps
\begin{figure*}
    \centering
    \includegraphics[width=9cm]{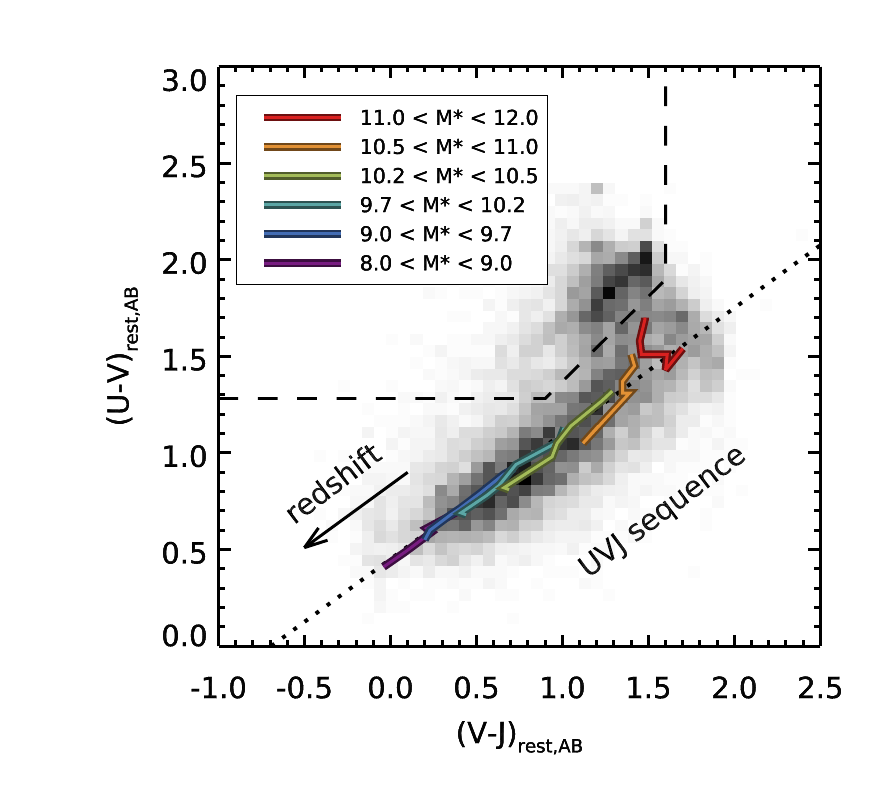}
    \includegraphics[width=9cm]{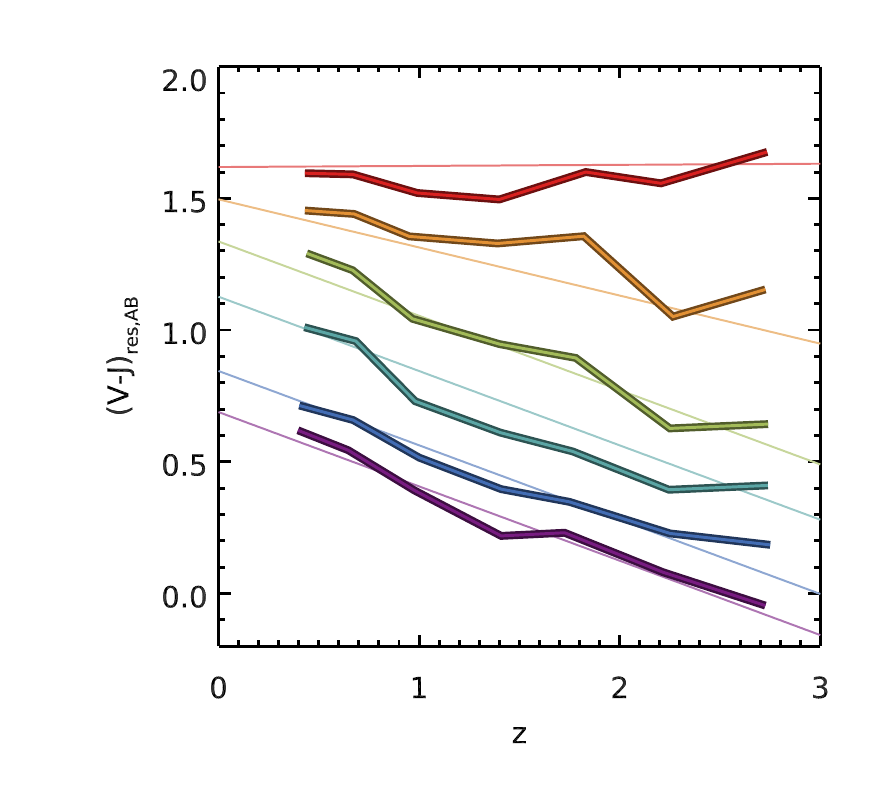}
    \caption{{\bf Left:} \uvj diagram of all galaxies in the CANDELS fields more massive than $10^{10}\,\msun$ (background gray scale). The redshift evolution of the median $U-V$ and $V-J$ colors of SFGs in different stellar mass bins is overlaid with colored lines. They all fall along a single line we dub the ``\uvj sequence'', which is illustrated by a dotted line. Finally, the adopted dividing line between SFGs and QGs is shown with a long dashed line. {\bf Right:} evolution of the $V-J$ color with redshift for each bin of mass. We show in the background the prescription adopted in this work.}
    \label{FIG:uvj_track}
\end{figure*}

Once the main physical properties are generated, we can associate a stellar SED to both the disk and bulge component of each galaxy. Instead of basing our approach solely on physical arguments, e.g., stellar age and dust content, we choose a simpler effective prescription where the SED is determined by the position of the galaxy on the \uvj diagram. This color-color diagram (already introduced in \rsec{SEC:zmstar}) provides a simple way to describe a wide range of spectral types, in particular ``blue and star-forming'', ``red and dead'' and ``red and dust-obscured'', from low to high redshifts \citep[e.g.,][]{williams2009,patel2012,straatman2014}. Our first goal is therefore to define a recipe to generate realistic colors distributions in the \uvj plane. Since SFGs and QGs are well segregated into two groups (or ``clouds''), we will treat each population separately, starting with the SFGs.

To this end, we consider all the \uvj star-forming galaxies in the CANDELS catalogs and divide this sample in multiple stellar mass bins. For each of these bins, we compute the median $U-V$ and $V-J$ colors at varying redshifts, and display the resulting tracks on the \uvj diagram in \rfig{FIG:uvj_track} (left). At fixed stellar mass, both colors go from blue to red as redshift decreases. A similar trend could already be identified in \citetalias{schreiber2015} (Fig.~1), and is expected given the decrease of specific SFR \citep{noeske2007,arnouts2013} and the increase in metallicity \citep{mannucci2010}. Interestingly, we find here that all the tracks seem to follow a straight line that we call the ``\uvj sequence'':
\begin{align}
(U-V)_{SF} &= 0.65 \times (V-J)_{SF} + 0.45\,. \label{EQ:uvj_sequence_sf}
\end{align}
See also \cite{labbe2007} where such a sequence is found among blue galaxies in a color-magnitude diagram. This sequence also happens to run parallel to the dust attenuation vector \citep{williams2009}, meaning that the position of a galaxy on this sequence could also be interpreted as a sign of varying dust content.

To better illustrate where galaxies of different mass are located on this sequence and how they evolve with time, we show in \rfig{FIG:uvj_track} (right) the evolution of the average $V-J$ color with redshift for each mass bin. Massive galaxies appear to always have the same very red colors (consistent with the fact that these are typically the most dusty, e.g., \citealt{pannella2009-a}), while less massive galaxies were substantially bluer in the past. We parametrize the resulting color tracks using the following equations:
\begin{align}
(V-J)_{\rm SF} &= a_0 + a_1 \times {\rm min}(z, 3.3)\,, \label{EQ:vj_sf} \\
\text{with $a_0$} &= 0.48 \times {\rm erf}(\logm - 10) + 1.15\,, \nonumber \\
\text{and $a_1$} &= -0.28 + 0.25 \times {\rm max}(0, \logm - 10.35)\,. \nonumber
\end{align}
We add a Gaussian scatter of $0.1$ mag to this color, and limit its value to be at most $1.7$ to prevent extreme red colors. We then use \req{EQ:uvj_sequence_sf} to obtain the $U-V$ color, and add an extra layer of Gaussian scatter of $0.12$ mag to both colors to reproduce a wider range than what is allowed by this simple prescription. This scatter is most likely caused by variations of star formation histories, specific star formation rates \citep[e.g.,][]{arnouts2013} and inclination \citep[e.g.,][]{patel2012}, but we do not attempt to study its origin here. We emphasize that the redshift evolution of SFGs on the \uvj sequence (first line in the above equation) is stopped at $z>3.3$: not only does this redshift domain go outside of the range in which the recipe was calibrated, but we also find that this step is necessary to reproduce the UV luminosity functions at $4 < z < 7$.

The prescription for QGs is relatively simpler, since these are located within a smaller region of the \uvj diagram, the so-called red cloud. We use the same approach as for SFGs, this time selecting the \uvj quiescent galaxies in the CANDELS catalog, and compute the median $U-V$ and $V-J$ colors in bins of redshift and mass. Here as well galaxies tend to reside on a sequence:
\begin{align}
(U-V)_{Q} &= 0.88 \times (V-J)_{Q} + 0.75\,, \label{EQ:uvj_sequence_q}
\end{align}
although the dynamic range is much smaller than for SFGs. In particular we find no significant redshift trend, and a moderate trend with stellar mass such that more massive galaxies are redder, probably because they are older. We parametrize this trend as
\begin{align}
(V-J)_{Q} &= 0.1 \times (\logm - 11) + 1.25\,, \label{EQ:vj_q}
\end{align}
to which we add $0.1$ mag of Gaussian scatter. The resulting color is clamped within $1.15$ and $1.45$ to prevent galaxies from exiting the red cloud. As for SFGs, the $U-V$ color is obtained by applying \req{EQ:uvj_sequence_q}, and both colors are perturbed independently with a Gaussian scatter of $0.1$ mag.

We use the above relations to derive the colors of the disk and bulge components of each galaxy. To do so, and following the observations of \cite{schreiber2016-b} at $z=1$ (see in particular the Fig.~5 from that paper), we consider that all disks are ``star-forming'' and obtain their colors from \reqs{EQ:uvj_sequence_sf} and \ref{EQ:vj_sf}. Similarly, we consider that all bulges of bulge-dominated galaxies ($\bt > 0.6$) are ``quiescent'' and described by \reqs{EQ:uvj_sequence_q} and \ref{EQ:vj_q}. Determining the colors of disks and bulges in composite systems is challenging, and can only be attempted for the brightest galaxies \citep[see, e.g.,][]{bruce2014}. In the absence of observational constraints, bulges in composite galaxies are randomly chosen to be ``star-forming'' or ``quiescent'' with uniform probability to simulate both bulges and pseudo-bulges. This prescription allows us to reproduce the observed total color distributions of galaxies as a function of $\bt$ at $z=1$ from \cite{schreiber2016-b}.

We stress here that, with the above prescriptions, ``star-forming'' colors are not necessarily ``blue''. As can be seen in \rfig{FIG:uvj_track}, massive SFGs do have red colors up to $(U-V)=1.5$ owing to dust attenuation. In addition, the random scatter we introduce in the various steps will allow a few SFGs ($\sim$$1\%$) to have colors that fall inside the quiescent region (dashed line). As a consequence, $15\%$ of our disk-dominated galaxies ($\bt < 0.3$) at $z<0.5$ and $\mstar > 3\times10^{10}\,\msun$ have red $g-r$ colors, which is comparable to the value observed by \cite{masters2010} at $z\sim0$ (see their Figure 2). Therefore, these prescriptions allow us to produce many different combinations of colors and morphologies which are representative of the real galaxy population.

The last step is to associate a complete stellar SED to each pair of colors we just generated. From the observed catalogs, we bin uniformly the \uvj plane into small buckets of $0.1$ mag and gather all the galaxies that fall inside each bucket, regardless of their redshift and stellar mass. We then compute the average of their rest-frame SED per unit mass, which is taken as the best-fit template produced by {\it FAST} and the \cite{bruzual2003} stellar library when fitting for the stellar mass (\rsec{SEC:zmstar}). We discard the buckets containing less than $10$ galaxies, and end up with an empirical library of $345$ SEDs, each corresponding to a given position in the \uvj diagram. This library does not cover the whole \uvj plane though, and therefore if a simulated galaxy has colors that fall outside of the covered region (which is rare by construction), it is attributed the SED of the closest non-empty bucket. Finally, the stellar SED of both disks and bulges is obtained by rescaling the chosen SED by their respective stellar mass.

The way we build our stellar SED library implicitly assumes that the mass-to-light ratio (M/L) is uniquely determined by the $U-V$ and $V-J$ colors. In practice though, we find that this is not a valid assumption for the lowest redshift galaxies ($z<1$), for which the procedure produces too low M/L by about $40\%$ on average, and conversely for high redshift galaxies ($z>6$) where the M/L ratio is overestimated by up to a factor $4$. To compensate for this effect, we add a correction term to the stellar mass before using it to rescale the SED ($-0.15\,\dex$ for $z<0.45$ and linearly coming back to zero at $z=1.3$, then $+0.6\,\dex$ for $z=8$ and linearly coming back to zero at $z=6$).

%------------------------------
%------------------------------
\section{Star formation and dust properties \label{SEC:dust}}
%------------------------------
%------------------------------

%------------------------------
\subsection{Star formation rate \label{SEC:sfr_dust}}
%------------------------------

Given the redshift and the stellar mass, we can attribute a star formation rate ($\sfr$) to each galaxy by following the Two Star Formation Mode model \citep[2SFM,][]{sargent2012}. This model relies on the existence of the $\sfr$--$\mstar$ main sequence and has been shown to successfully reproduce the observed flux and redshift distributions from the MIR-to-the submm and even the radio \citep{bethermin2012-a,bethermin2015-a}. Taking advantage of these results, we use here a similar prescription where the model parameters are updated with our latest \herschel measurements.

Using the $\sfr(z,\mstar)$ equation derived from \herschel stacking in \citetalias{schreiber2015}, we associate a ``main sequence'' (MS) star formation rate to each SFG with
\begin{align}
    \log_{10}(\sfrms [\msun / {\rm yr}]) &=  \logm - 9.5 + 1.5\,\logz \hspace{2.5cm} \nonumber \\
        - &0.3 \, \big[{\rm max}(0, \log - 9.36 - 2.5\,\logz)\big]^2\,,\label{EQ:sfrms}
\end{align}
where $\logm \equiv \log_{10}(\mstar [\msun])$, and $\logz \equiv \log_{10}(1+z)$.

We then obtain the $\sfr$ of each galaxy by applying a log-normal scatter of $0.3\,\dex$ to reproduce the observed width of the main sequence, which was found in \citetalias{schreiber2015} to be constant both as a function of stellar mass and redshift. In addition, $3\%$ of the galaxies are randomly chosen and placed in a ``starburst'' mode, where their $\sfr$ is enhanced by a factor of $5.24$, following \citetalias{schreiber2015}. \cite{sargent2012} showed that this last step is necessary to correctly capture the bright-end of the IR luminosity functions. In the next section we will use the offset of each galaxy with respect to the main sequence to fine-tune their dust spectrum. Following \cite{elbaz2011} we quantify this offset with the ``starburstiness''
\begin{equation}
\rsb \equiv \frac{\sfr}{\sfrms}\,,
\end{equation}
which is equal to $1$ for a purely main sequence galaxy.

For QGs, we use the IR stacks presented in the Appendix of \citetalias{schreiber2015}, where it was reported that QGs do show some IR emission, typically a factor of ten fainter than SFGs of the same mass. This light may be caused either by residual star formation, AGN torus emission, dust heated by old stars, or by incorrect classification of some SFGs. Although this is an interesting question, its answer is irrelevant for our purposes, and we choose to model this faint emission by interpreting it as residual star-formation. Therefore, QGs are attributed an $\sfr$ following
\begin{equation}
    \log_{10}(\sfr_{\rm QS} [\msun / {\rm yr}]) = 0.5\,\logm + \logz - 6.1\,,
\end{equation}
to which we add a log-normal scatter of $0.45\,\dex$. This latter value was chosen to roughly reproduce the number of $24\,\um$ detected QGs at $z=1$.

%------------------------------
\subsection{Obscuration \label{SEC:dust_dust}}
%------------------------------

To estimate the dust luminosity, we decompose the $\sfr$ into a dust-obscured component, which re-emerges in the FIR, and dust-free component, which emerges in the UV. To do so, we use the observed relation between stellar mass and dust obscuration \citep[e.g.,][]{pannella2009-a,buat2012,heinis2014}, which we calibrate here in terms of the infrared excess $\irx\equiv\log_{10}(\lir/\luv)$ \citep{meurer1999}. Using the stacked $\lir$ from \citetalias{schreiber2015}, we find that the relation between $\irx$ and $\mstar$ can be described by
\begin{equation}
    \irx = (0.45\,{\rm min}(z, 3.0) + 0.35) \times (\logm - 10.5) + 1.2\,.\label{EQ:irx}
\end{equation}
This formula is very similar to that reported by \cite{heinis2014}, except that our relation is found to be redshift dependent: at a fixed mass above $\mstar \sim 3\times10^{10}\,\msun$, attenuation becomes more important at higher redshifts, while it becomes less pronounced for lower mass galaxies. This behavior, at least at the high mass end, is consistent with the results of \cite{pannella2015} who report that the typical MS galaxy at $z=2$ is sensibly different from its analog at $z \leq 1$, which they argue is because of modifications in the geometry of the star-forming regions. We also add a scatter of $0.4\,\dex$ to this relation: although it has a negligible impact on the generated IR luminosities, \cite{bernhard2014} showed that this is a necessary ingredient to properly reproduce the bright-end of the UV luminosity function.

%------------------------------
\subsection{On the consistency between the stellar and dust emission \label{SEC:opt_fir_const}}
%------------------------------

One of the main physical inconsistencies of our model is found here: the stellar UV emission and reddening were already determined indirectly in \rsec{SEC:opti_sed}, based on the position of the galaxies in the \uvj diagram, and we do not use these values here to predict the $\sfr$ or the $\lir$. We make this choice to avoid forcing galaxies to match the ``energy balance'' between the UV and IR \citep[e.g.,][]{dacunha2008,noll2009}; we use independent empirical prescriptions to generate the UV and IR emission separately, so that the actual $L_{\rm UV}$ of our simulated galaxies differs from the value one would otherwise obtain from \reqs{EQ:sfrms} and \ref{EQ:irx}. In practice the two are still well correlated with no systematic bias over six orders of magnitude, albeit with a substantial scatter of $0.42\,\dex$. Similarly, the $\sfr$ one would derive by fitting the stellar SED from \rsec{SEC:opti_sed} is not strictly identical to the $\sfr$ of \req{EQ:sfrms}; they correlate with a scatter of $0.36\,\dex$.

We argue that, in spite of being physically inconsistent, this scatter is a feature that allows us to more accurately match the observations. Indeed, different $\sfr$ indicators tend to agree in the average sense and for large populations, but often fail to provide consistent results when applied to one single galaxy \citep[e.g.,][]{goldader2002,buat2005,elbaz2007,penner2012,oteo2013-b}. To illustrate this, we select from the observed CANDELS catalogs all the galaxies at $0.7<z<0.8$ with a detection in the \spitzer MIPS or \herschel bands, and use their $\lir$ and $\luv$ as computed in \citetalias{schreiber2015} to derive a ``UV+FIR'' $\sfr$. We then compare this value against the $\sfr$ obtained by FAST when fitting the UV-to-NIR photometry, and find a scatter of $0.42\,\dex$ (measurement uncertainties were not subtracted from this value) which is comparable to the $0.36\,\dex$ of the simulated catalog.

%------------------------------
\subsection{Dust temperature and chemical composition \label{SEC:dust_prop}}
%------------------------------

In the previous sections we have attributed an $\sfr$ to each galaxy, and estimated what fraction of the associated light comes out in the FIR. From there, the last missing ingredient to predict dust fluxes is a suitable dust SED library, with enough adjustable parameters to reproduce accurately the observed counts. A number of such SED libraries have already been published, calibrated either in $8$-to-$1000\,\um$ luminosity ($\lir$) from local galaxies \citep[][hereafer \citetalias{chary2001}]{chary2001}, far-infrared (FIR) colors \citep{dale2002}, or intensity of the interstellar radiation field ($\mean{U}$) from distant galaxies \citep{magdis2012,bethermin2015-a}.

We use here a new SED library \citep[introduced in][]{schreiber2016-b} in which both the dust temperature ($\tdust$) and the $\ireight=\lir/\leight$ (where $\leight$ is the $k$-corrected luminosity in the IRAC $8\,\um$ filter) are free parameters. There are two reasons for this choice. First, these are the two parameters that are the easiest to measure without FIR spectroscopy (which is only available for a few bright objects), and are the ones affecting most the shape of the SED. Second, $8\,\um$ is the domain where polycyclic aromatic hydrocarbon molecules (PAHs) emit the bulk of their light through strong emission lines. This is a domain that will be routinely accessed by the {\it James Webb Space Telescope} in the near future, and there will be a need for a properly calibrated library to exploit these data together with ancillary \herschel and \spitzer observations.

We calibrate the redshift and mass evolution of both $\tdust$ and $\ireight$ using the MIR-to-FIR stacks of \citetalias{schreiber2015}, to which we added stacks of the \spitzer IRS $16\,\um$ imaging \citep{teplitz2011} to better constrain the rest-frame $8\,\um$ and the PAH features (available in GOODS--{\it North} and {\it South} only). We then further refine this calibration using individual \herschel detections to constrain the scatter on these parameters, and also to calibrate how they are modified for those galaxies that are offset from the main sequence. These results will be described in more detail in another work (Schreiber et al.~in prep.).

The parametrizations for $\tdust$ and $\ireight$ for a galaxy lying exactly on the main sequence are:
% 1.5*min(0.0, out.z-2.0)*clamp(out.m - 10.7, 0.0, 1.0)
\begin{align}
\tdust^{\rm MS} [\kelvin] &= 4.65\times(z - 2) + 31 \nonumber \\
&+ 1.5 \times \min(z - 2, 0)\times{\rm clamp}(\logm - 10.7, 0, 1)\,, \label{EQ:tdust_ms} \\
\log_{10}(\ireight_{\rm MS}) &= \log_{10}\left(7.73 + 1.95 \times \min(z - 2, 0)\right)  \nonumber \\
&- 1.8 \times {\rm clamp}(\logm - 10, -1, 0) \,. \label{EQ:ir8_ms}
\end{align}
The second line in both equations is a second order dependence on the stellar mass motivated by the stacked fluxes. We make massive galaxies at $z<2$ slightly colder (owing to their reduced star formation efficiency, see \citealt{schreiber2016-b}), while low-mass galaxies are given a lower $\ireight$ essentially consistent with no PAH emission. The latter is known to happen for sub-solar metallicity objects in the local Universe \citep[e.g.,][]{madden2006,wu2006,ohalloran2006,galliano2008,ciesla2014}.

As shown in \cite{elbaz2011}, starbursting galaxies have a depleted $8\,\um$ luminosity compared to the total $\lir$, and an increased dust temperature (see also \citealt{nordon2012,magdis2012,magnelli2014,bethermin2015-a}). This is a sign that these galaxies are experiencing an episode of star formation in a more compact interstellar medium. To take this effect into account, we include a dependence of both $\tdust$ and $\ireight$ on the starburtiness (see previous section):
\begin{align}
\tdust [\kelvin] &= \tdust^{MS} + 6.6 \times \log_{10}(\rsb) \,, \label{EQ:tdust_sb} \\
\log_{10}(\ireight) &= \log_{10}(\ireight_{\rm MS}) + 0.43 \times \log_{10}(\rsb)\,, \label{EQ:ir8_sb}
\end{align}
to which we add a Gaussian scatter of $4.1\kelvin$ for $\tdust$ and a log-normal scatter of $0.1\,\dex$ for $\ireight$.

Using these two quantities, we pick a FIR SED from our library and rescale it with the $\lir$ computed in the previous section to build the dust SED of the galaxy.

%------------------------------
\section{Sky position \label{SEC:clustering}}
%------------------------------

% program: ../plots/make_clust.pro
% file: ../plots/clust.eps
\begin{figure*}
    \centering
    \includegraphics[width=6cm]{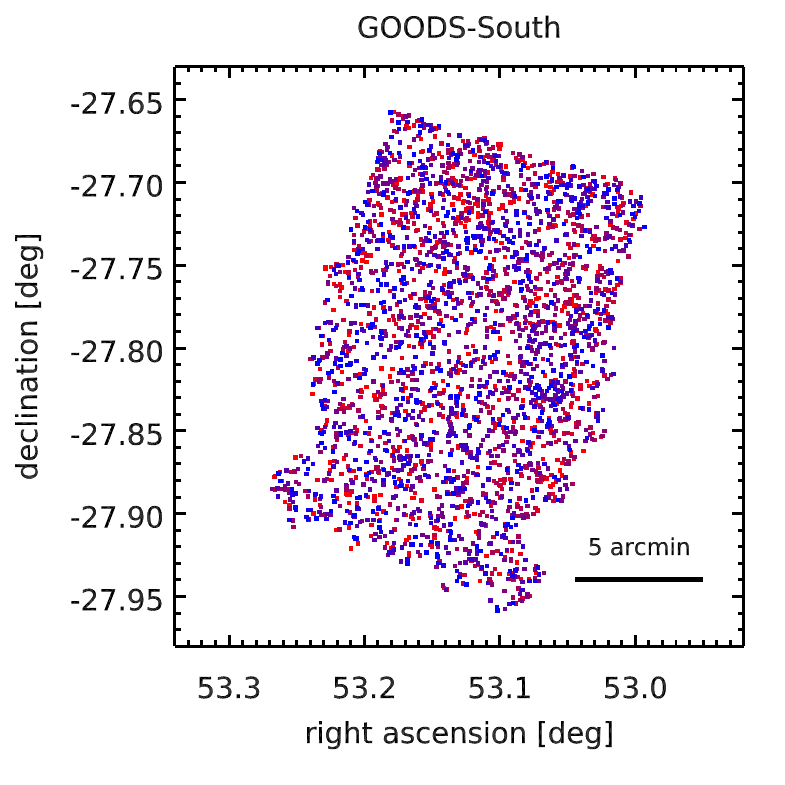}
    \includegraphics[width=6cm]{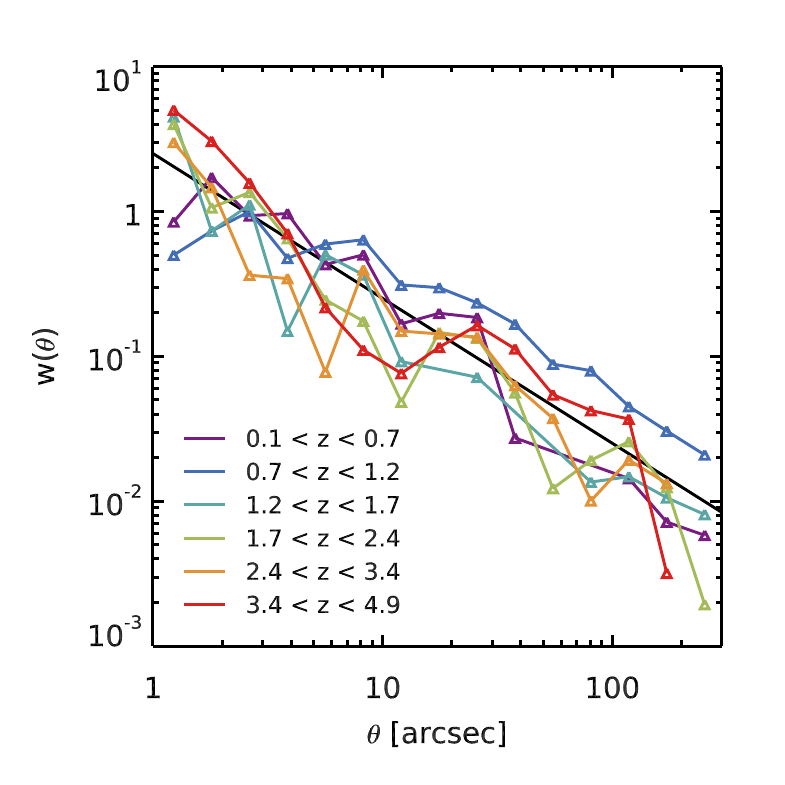}
    \includegraphics[width=6cm]{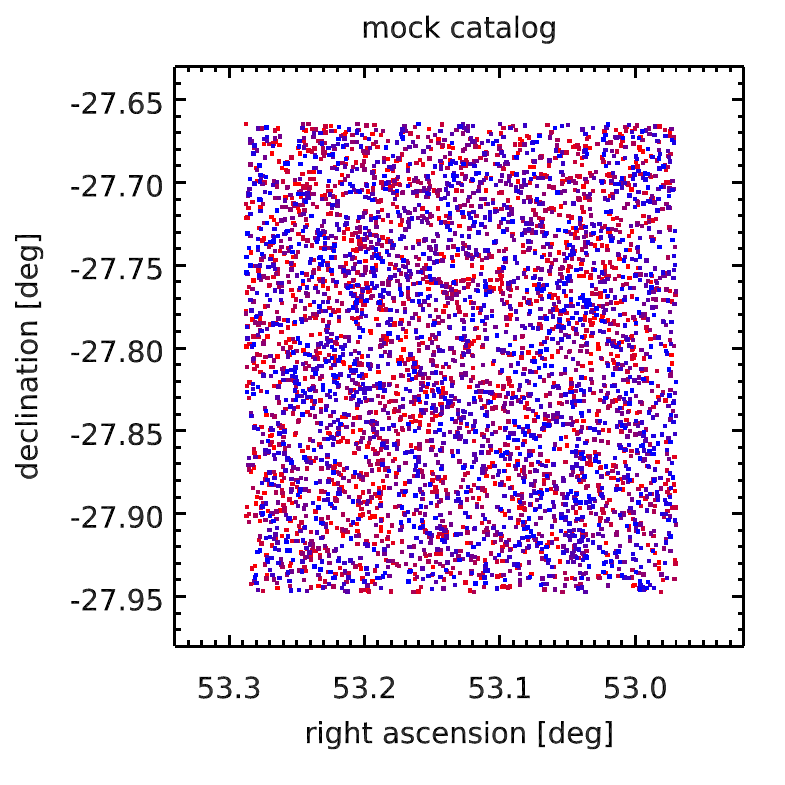}
    \caption{{\bf Left:} Sky positions of the galaxies in the GOODS--{\it South} field with $H < 23$. The colors indicates the redshift: blue points are $z<0.3$, red points are $z>1.5$, and galaxies in between are shown in shades of purple. {\bf Middle:}Angular two-point correlation function for GOODS--{\it South} galaxies with $9 < \log_{10}(\mstar/\msun) < 10.3$, in various redshift slices as indicated in the legend. The straight black line is a power law of index $-1$. Error bars are not shown for clarity. {\bf Right:} Same as left, but for the mock catalog produced by \egg.}
    \label{FIG:clustering}
\end{figure*}

The simplest approach to generate the position on the sky of each galaxy would be to draw these positions uniformly on the sphere, within the region of the sky that is covered by the simulated survey. The stellar mass functions we used in \rsec{SEC:mfunc} ensure that we get a correct sky density of object over the whole simulated area. However, within the $\Lambda$CDM cosmology, we expect galaxies to form large-scale structures by following the merging history of their dark matter halos \citep{peebles1982}. In other words, galaxies tend to cluster on the sky, and we need to simulate this effect to generate realistic sky positions. In \citetalias{schreiber2015} \citep[see also][]{bethermin2010-a}, we showed that clustering can have a significant impact on the statistical properties of confused, long-wavelength images from \spitzer and \herschel: it will tend to increase the contrast compared to a uniform position distribution, i.e., creating overdense and underdense regions within the survey area. On the other hand, we expect clustering to be no more than a cosmetic change for the high-resolution \hst images, which do not suffer from confusion.

The procedure we use here is to aim at reproducing the observed angular two-point correlation function $w(\theta)$, i.e., the excess probability of finding a galaxy at a given angular distance $\theta$ from another, as compared to a uniform position distribution. The first step is therefore to measure this two-point correlation function in the real GOODS--{\it South} field. To do so, we bin the whole catalog in redshift slices of width $\Delta = 0.25\times(1+z)$, and only two mass bins because the statistics is limited ($\mstar = 10^{9}$ to $3\times10^{10}\,\msun$, and $\mstar = 3\times10^{10}$ to $10^{12}\,\msun$), and we do not attempt to further refine the sample by separating different galaxy types. We then use the \cite{landy1993} estimator to compute the two-point correlation function of each sample, and observe a significant clustering signal between $1\arcsec$ and $5\arcmin$ at all redshifts (see \rfig{FIG:clustering}, middle panel). This signal is well described by a single power law
\begin{align}
w(\theta) \sim \theta^{\,-\gamma}\,,
\end{align}
with $\gamma = 1$, and where $\theta$ is the angular distance between two sources. As in \citetalias{schreiber2015}, we find no significant change in angular clustering amplitude with redshift between $z=0.3$ and $z=4$ (which is also consistent with the results of \citealt{bethermin2015-a}), but we do find that the amplitude of the two-point correlation function for massive galaxies is on average about three times larger. The fact that massive galaxies are more clustered should not come as a surprise knowing that we selected all massive galaxies, including red quiescent galaxies which are known to be the best tracers of the large scale structures, both locally and in the distant Universe \citep[e.g.,][]{cooper2006,cucciati2006,elbaz2007}.

However, it is important to note that these measured two-point correlation functions are affected by the uncertainties on the photometric redshifts (photo-$z$s). Indeed, within each adopted redshift bin, there is a chance that we miss some galaxies that scattered out of the bin, and another chance that we are contaminated by some galaxies scattering into the bin. The net result is that we observe a clustering amplitude that is lower than the intrinsic one. This effect can be simulated (and we do so in the following) once the uncertainty on the observed redshifts is known. Redshift uncertainties also contribute to some extent to the mass-dependent clustering that we observe, since massive bright galaxies have more robust photo-$z$s and are therefore expected to show a cleaner clustering signal. To measure this uncertainty, we cross-matched our GOODS--{\it South} catalog against the 3DHST catalog \citep[DR1,][]{skelton2014}. While the two catalogs are based on the same raw observations, the data reduction and photometry are performed independently with different tools. On the other hand, the photo-$z$s are estimated with the same code, so we will likely underestimate the real redshift uncertainty. We measure the distribution of redshift differences between the two catalogs, and take into account that what we observe is the combination of uncertainties coming from both catalogs (i.e., assuming they are independent, $\sqrt{2}$ higher than that of a single catalog). We find that the redshift uncertainty in $\Delta z/(1+z)$ is well described by the combination of two zero-mean Gaussians at high (low) masses: a first distribution of width $2.2\%$ ($2.5\%$) that describes $84\%$ ($76\%$) of galaxies, and a second distribution of width $6.6\%$ ($9.3\%$) that describes the remaining $16\%$ ($24\%$).

To produce sky positions that resemble these observations, we use the \cite{soneira1978} algorithm, which is a simple method to produce a two-point correlation function with an adjustable power-law slope. Briefly, the algorithm starts by drawing $\eta$ random positions within a circle of radius $R$ on the sky. This is the first level. At each of these positions, a new set of $\eta$ positions is randomly generated within a smaller circle of radius $R/\lambda$, and the ensemble of all these new positions make up the second level. This procedure is repeated up to a given level $L$, reducing the size the circles by a factor $\lambda$ for each step. The final level contains a set of $\eta^L$ positions on which we randomly place our simulated galaxies drawn from narrow redshift bins ($\Delta z /(1+z) \sim 0.1$) to roughly mimic the redshift-space clustering. The parameters of the Soneira \& Pebbles algorithm are tweaked to recover the right power law slope $\gamma=1$ ($\eta = 5$ and $\lambda = 6$) and to generate a large enough number of positions (i.e., by varying $L$, which does not affect the shape of the power law). The last parameter, $R$, is arbitrarily fixed\footnote{Choosing larger values for $R$ would lead to situations where most galaxies in a given redshift bin could fall out of the field of view, generating additional cosmic variance. We do not presently implement this in the code.} to $R=3\arcmin$, which truncates large-scale clustering beyond this angular scale (i.e., beyond $\sim1\,\Mpc$ at all $z > 0.5$). To fill the whole mock survey area, we randomly place several $3\arcmin$ circles within the field with a uniform probability distribution and use the Soneira \& Peebles algorithm only within each of these circles.

Using this method, we can produce a catalog of clustered positions with the right power-law slope, which we checked by measuring the clustering of the generated positions with the Landy \& Szalay estimator. However, we still have to tune the amplitude of this clustering. We choose here a simple approach where we use the Soneira \& Peebles algorithm only for a given fraction $f$ of the simulated galaxies, and use uniformly distributed positions for the remaining fraction. We choose this fraction by first generating a set of positions with $f=100\%$, i.e., maximum clustering, apply the above procedure to measure the correlation function, and compare it to the observed one. The difference of amplitude then tells us by how much we need to reduce the simulated clustering. We stress that it is important here to take into account the redshift uncertainties that affect the observed relation. To do so, we measure the two-point correlation function in the simulation using ``wrong'' redshifts, which are obtained from the ``true'' redshifts of the simulation and then perturbed within the uncertainty described above. After taking this into account, we find that $f=25\%$ for $\mstar < 3\times10^{10}\,\msun$, and $f=60\%$ for more massive galaxies\footnote{These values depend largely on the details of the clustering modeling (i.e., the choice of $R$, $\eta$ and $\lambda$) as well as the width of the redshift windows within which the angular correlation function is measured, and have therefore little meaning in the absolute sense. However the trend for massive galaxies to be relatively more clustered should reflect a real phenomenon.}. An example of the resulting sky distribution for massive galaxies is shown in \rfig{FIG:clustering} (right).

To double check, we also compute the angular correlation function of the whole catalog above $\mstar > 10^{10}\,\msun$, mixing all redshifts. Doing so, we get rid of the issue of the redshift uncertainty, and find also a good agreement with the observations.

%------------------------------
%------------------------------
\section{Generating a light cone \label{SEC:simu}}
%------------------------------
%------------------------------

% program: yEd + inkscape
\begin{figure*}
    \centering
    \includegraphics[width=18cm]{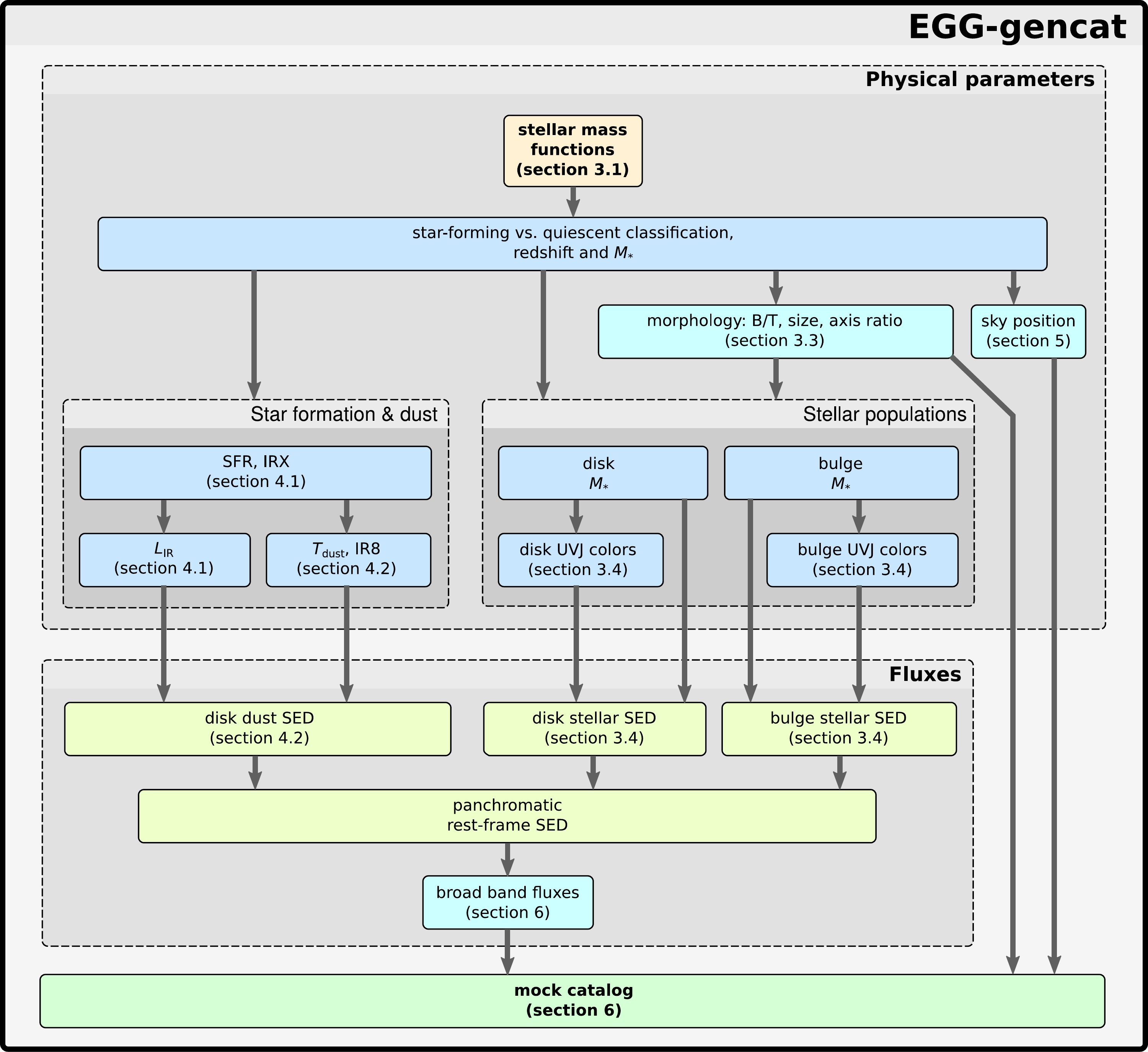}
    \caption{Flow chart of the catalog simulation procedure, as implemented in \texttt{egg-gencat}.}
    \label{FIG:procedure}
\end{figure*}

\subsection{Standard procedure}

With all these recipes, we can now generate a complete catalog of galaxies, each with its own UV-to-submm SED. In this section, we summarize the procedure that is implemented in \egg to produce a final flux catalog. For a quick overview, one can refer to the flow chart presented in \rfig{FIG:procedure}.

Given the area of the mock survey, the first step is to choose the number of galaxies that will be generated. Since we use the stellar mass function as a starting point, this amounts to choosing the lowest stellar mass that we will generate. This threshold can be chosen to be constant, e.g., down to $\mstar = 10^{8}\,\msun$, but this is in fact quite inefficient. Observations in the real GOODS--{\it South} field are flux-limited; we do detect galaxies that are less massive than $10^{8}\,\msun$ at low redshifts, but the smallest measured stellar mass at $z>2$ is closer to $10^{9}\,\msun$. Therefore, this approach can result either in a catalog that is incomplete (if the mass threshold is too high and we miss detectable galaxies at low redshift), or bloated (if the threshold is too low and we generate galaxies that will never be observed).

A more efficient approach is to use a redshift dependent threshold, so that galaxies are generated down to low stellar mass at low redshifts, and then increase this threshold to generate only the most massive galaxies at higher redshifts. To do so, we first choose a ``selection band'', e.g., the \hst F160W or the VISTA \Ks band, and a magnitude limit, e.g., $H<29$, above which the catalog will be at least $90\%$ complete. We then build a redshift grid, and for each redshift in that grid we compute the distribution of mass-to-light ratios in the selection band for all the optical SEDs in the library. We pick the $10$th percentile of this distribution, and use it to compute the minimum stellar mass at this redshift given the magnitude limit.

In addition, we impose by default a minimum redshift of $z=0.05$ to avoid having large and bright nearby galaxies in the field of view, which are by construction absent in the CANDELS fields. This will limit our ability to reproduce the bright end of the counts though, and this limit can be pushed down if one wants to simulate larger fields accurately (we do so in \rapp{APP:counts}).

\begin{figure}
    \centering
    \includegraphics[width=9cm]{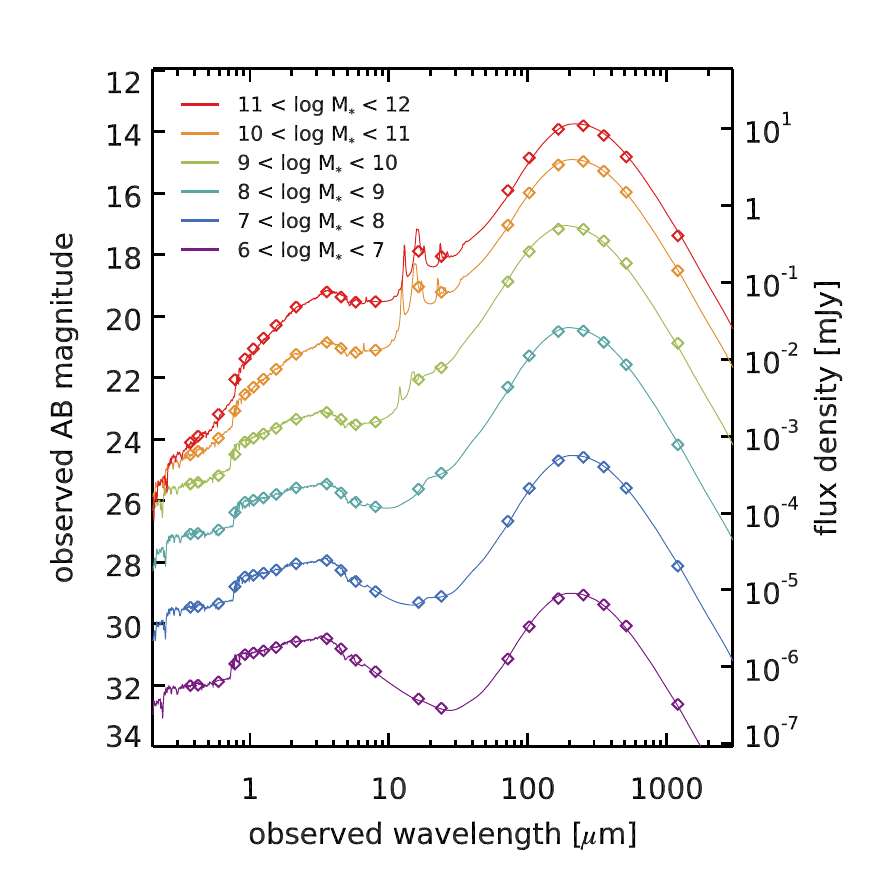}
    \caption{Average SEDs of $z=1$ SFGs in the \egg simulation. Each color corresponds to a different bin of stellar mass, as shown in the legend. The open diamonds are the corresponding broad band fluxes integrated in a set of common bands from \hubble, \spitzer, \herschel and ALMA.}
    \label{FIG:seds}
\end{figure}

Once the stellar mass and the redshift are generated from the mass functions, the program uses the method described in \rsec{SEC:clustering} to place these galaxies on the sky, and applies all the above recipes to generate the $\sfr$, the $\lir$ and other dust related parameters ($\tdust$ and $\ireight$), the \uvj colors, and the morphological parameters ($\bt$, $R_{50}$, $b/a$). Then, the optical SED is chosen based on the generated \uvj colors (\rsec{SEC:opti_sed}) and scaled by the stellar mass, while the FIR SED is chosen from the $\tdust$ and $\ireight$ (\rsec{SEC:dust_prop}) and scaled by the $\lir$. The two SEDs are redshifted and co-added to form a single, panchromatic SED that ranges from the FUV up to the submm, as shown in \rfig{FIG:seds} (the radio and X-ray domains are not yet implemented). Lastly, the SED is multiplied by the response curve of each broadband filter and integrated to generate a flux in each band. This is all done with a single call to the \texttt{egg-gencat} program, and takes $20$ seconds on a single-core of a regular desktop computer to generate a CANDELS-like field down to $H=28$.

% program: egg/paper/seds/plot_vert.pro

\subsection{Customization}

Most parameters of the simulation can be customized easily by using command line arguments, including in particular the dimensions of the mock field, the random seed, the depth, the redshift range, or the set of bands in which fluxes should be produced. New filters can be added to the default list with minimal effort. The program can also generate rest-frame magnitudes in any band if requested, or even output the full medium-resolution spectrum of each galaxy.

It is also possible to feed \egg with a pre-existing catalog of redshifts, stellar masses, star-forming/quiescent flags, and positions, e.g., coming from a real catalog: the program will then apply the same recipes and predict fluxes in any band for each input galaxy. This can be useful for proposal writing, or to test the systematics introduced when stacking FIR images for specific populations (e.g., flux boosting from clustering) as was done in \citetalias{schreiber2015}.

The program and its various options are described in full detail in the documentation, which is provided with the source code or can be browsed on line\footnote{\url{http://cschreib.github.io/egg/files/EGG.pdf}}.

%------------------------------
%------------------------------
\section{Quality of the mock catalogs \label{SEC:quality}}
%------------------------------
%------------------------------

\subsection{Number counts \label{SEC:counts}}

In this section we quantify the accuracy of the mock catalogs produced by \egg by comparing the generated number counts in various bands from the optical to the sub-mm against the observed counts in the GOODS--{\it South} or CANDELS fields and other literature data.

% program: code/astrodeep/gencat/work/plots/plot_opt.pro
% file: code/astrodeep/gencat/work/plots/mags.eps
\begin{figure*}
    \centering
    \includegraphics[width=\textwidth]{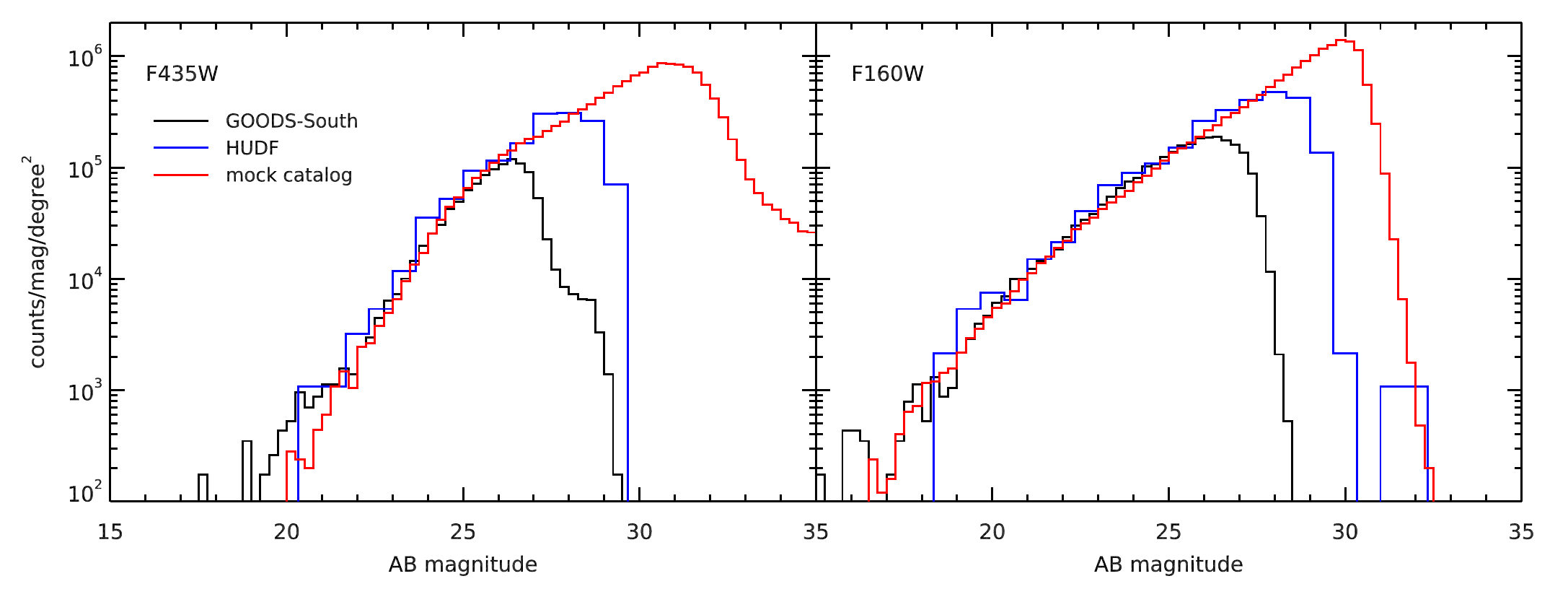}
    \caption{Observed magnitude distribution from the \hst F435W and F160W bands (other bands can be found in the Appendix). The simulated fluxes (red histogram) come from a mock field of $10\arcmin\times10\arcmin$ that is $90\%$ complete down to $H<30$. These are compared to the observed fluxes in the \hubble Ultra Deep Field (HUDF, blue) and the rest of the GOODS--{\it South} field (shallower, in black). Stars were excluded from the observed counts, since they are absent from the simulation.}
    \label{FIG:mag_hist}
\end{figure*}

% program: code/astrodeep/gencat/work/plots/plot_ir3.pro
% file: code/astrodeep/gencat/work/plots/id_flux.eps
\begin{figure*}
    \centering
    \includegraphics[width=\textwidth]{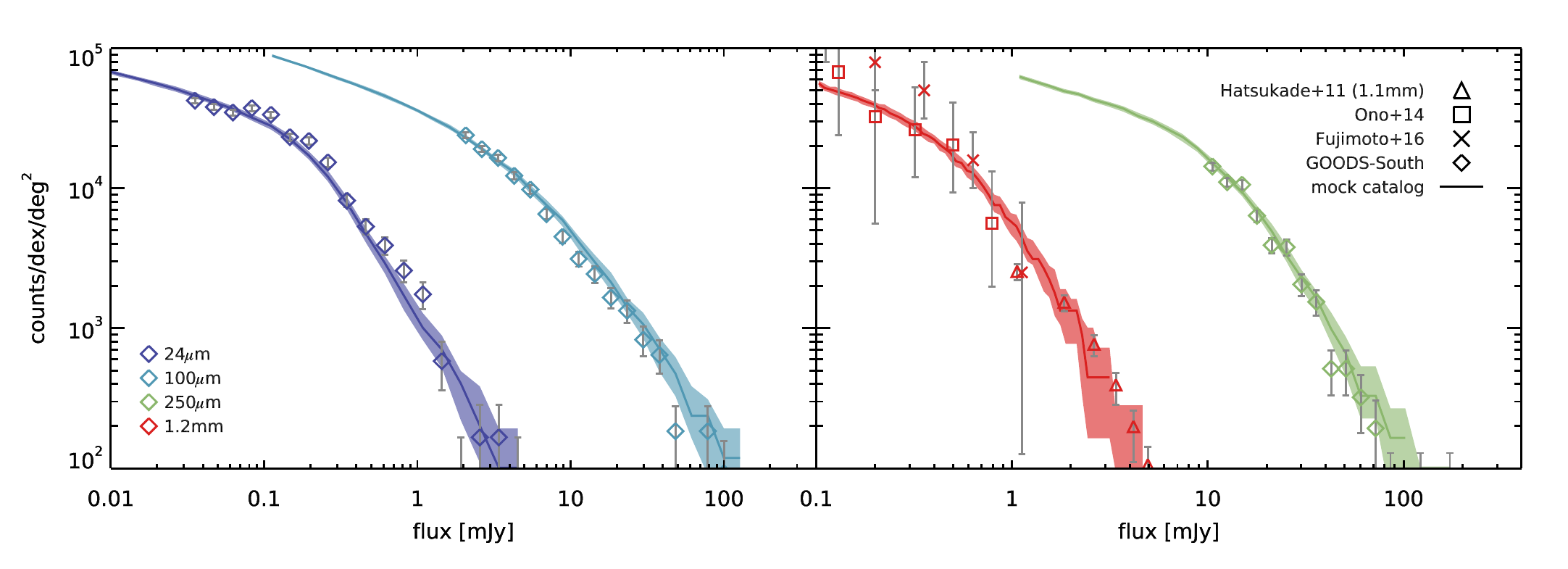}
    \caption{Number counts in the MIR ($24\,\um$), FIR ($100$ and $250\,\um$) and sub-mm ($1.2\,\mm$). Additional MIR and FIR bands can be found in the Appendix. The observed counts in the CANDELS fields are reported as open diamonds with Poisson error bars, and we compare these to the median counts of $100$ simulated catalogs, shown with a solid line of the same color. In the background, the shaded area show the range covered between the $16$th and $84$th percentile of of $100$ simulated catalogs, to illustrate how much scatter one should expect simply due to cosmic variance. Observed counts for $24\,\um$ are from GOODS only, since they are substantially deeper. For $1.2\,{\rm mm}$, observations are from \cite{ono2014} and \cite{fujimoto2016} using ALMA $1.2\,{\rm mm}$ and \cite{hatsukade2011} using AzTEC $1.1\,{\rm mm}$, reported with open squares, crosses and open triangles, respectively. The $1.1\,{\rm mm}$ fluxes were scaled down by a factor $0.77$ following \cite{fujimoto2016}. These observations were not done in GOODS--{\it South}, and are therefore affected by a different cosmic variance.}
    \label{FIG:irflux}
\end{figure*}

% program: code/astrodeep/gencat/work/plots/plot_ir_pd2.pro
% file: code/astrodeep/gencat/work/plots/id_pd.eps
\begin{figure*}
    \centering
    \includegraphics[width=\textwidth]{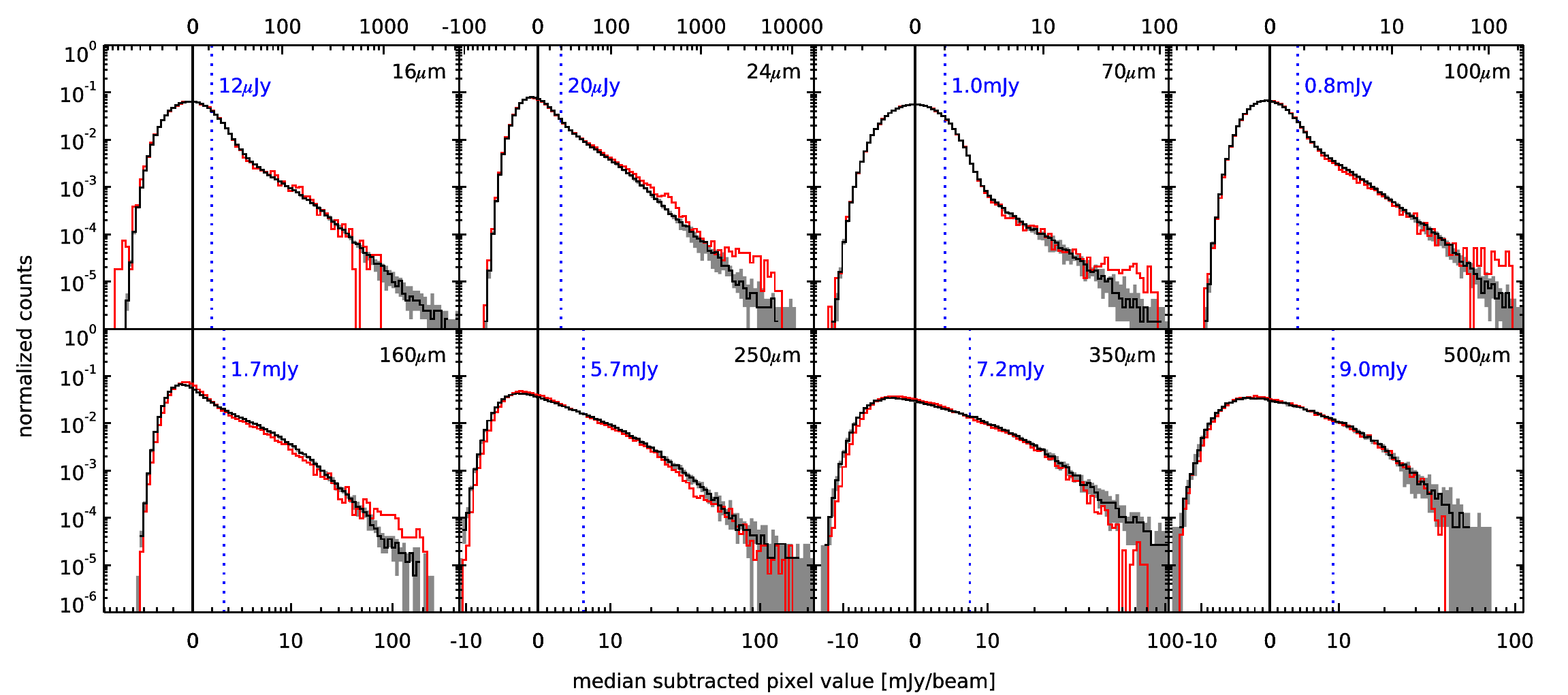}
    \caption{Pixel value distributions of the $16\,\um$ to $500\,\um$ maps, in $\uJy/{\rm beam}$ for $16$ and $24\,\um$ and $\mJy/{\rm beam}$ otherwise. We show the observed distribution in GOODS--{\it South} in red, and compare this reference to $100$ simulated catalogs generated with different random realizations. The median of these $100$ realizations is shown with a solid black line, while the range covered by the $16$th and $84$th percentiles is shaded in gray in the background. Each map is median-subtracted, and the pixel values displayed here are scaled using the hyperbolic arcsine function (see text). We show the location of the median of the map with a vertical solid black line, and the $3\sigma$ point-source detection limit with a vertical blue dotted line.}
    \label{FIG:ir_pd}
\end{figure*}

All the number counts predicted by the simulation over a large dynamic range of fluxes are provided in \rapp{APP:counts}. Here in \rfig{FIG:mag_hist}, we focus on the \hst F435W and F160W bands. We compare these to the observed counts in GOODS--{\it South}, splitting the field into two parts: the HUDF, which is deeper, and the rest of the field. We recall that stars were excluded from the observed catalogs.

The agreement is found to be excellent in the NIR. Because these wavelengths are most closely correlated to the stellar mass of the galaxies, and since the mock catalog was built to reproduce exactly the stellar mass function in GOODS--{\it South}, this should not come as a surprise. Reproducing the UV-optical (e.g., F435W) fluxes is less trivial, because these bands rather trace the emerging UV light coming from star formation, modulated by dust extinction. Nevertheless, the agreement here is also good. For all bands from F435W to \spitzer IRAC channel 4, we find a reduced $\chi^2 < 2$ between simulated and observed counts in the regimes where observations are complete.

\rfig{FIG:irflux} shows instead the MIR, FIR and sub-mm counts in a few selected bands (other bands can be found in \rapp{APP:counts}). Since the number of detected galaxies in this wavelength domain is low, both observed and simulated counts are more strongly affected by statistical fluctuations than in the optical. To mitigate this effect, we use here the four CANDELS fields covered with deep \herschel data, rather than just GOODS--{\it South}. In addition, to estimate the amplitude of these fluctuations, we produce $100$ realizations of the mock catalog, each time using a different seed to initialize the random number generator. The spread in counts among all realizations gives a first order estimate of the cosmic variance. We then compute the average counts among all realizations, and compare this against the observations. This comparison shows that our simulation is able to capture the right shape and normalization of the counts, including in particular the turnover of the MIPS $24\,\um$ counts around $200\,\uJy$. The observations in the other bands are too shallow to probe this regime, but we confirm at least that the power law slope at the bright end is correctly reproduced.

In the same figure we show a prediction of the number counts at $1.2\,\mm$, a wavelength domain in which our FIR SEDs and recipes are not calibrated. The agreement with published number counts from recent ALMA and single dish AzTEC observations is also satisfactory, reinforcing the validity of our approach. \egg also predicts $25^{+8}_{-7}$ detections above $S_{1.3\,\mm}>150\,\uJy$ in the $4.5\,{\rm arcmin}^2$ of the HUDF, which is consistent with the $16$ detections found by \cite{dunlop2016}. The error bars at the faint end are very permissive though, owing to the small area covered by ALMA to date. Single dish data covers much larger fields, but are limited both in depth and angular resolution. Poor angular resolution causes blending issues, and affect the measured flux catalogs by merging multiple moderately bright sources -- too close to be reliably separated -- into a single brighter one (see, e.g., \citealt{hodge2013}). This can artificially boost the counts at the bright end, and make the comparison with models difficult. To some extent, a similar issue must be affecting the \herschel SPIRE fluxes.

\subsection{Pixel statistics \label{SEC:pixelstat}}

To get rid of the uncertainty caused by blending, we directly analyze in \rfig{FIG:ir_pd} the pixel value distributions of the simulated maps, avoiding the problems of having to find the right (number of) counterparts for the FIR sources. This procedure also takes into account the effect of clustering, which will tend to increase the contrast of the map without actually changing the number counts. To build the simulated \spitzer and \herschel images, we use the \texttt{egg-gennoise} tool to produce empty maps only containing Gaussian noise. The amplitude of this noise is adjusted to match the RMS of empty regions in the observed maps. We then rescale the fluxes of our simulated galaxies to match the zero point of the image, and paint them on the map using \texttt{egg-genmap}. Since the angular resolution of these images is low, we consider our galaxies to be unresolved and model them as point sources, convolved with the observed PSF from GOODS--{\it South}. The resulting maps are finally median subtracted, as are the observed maps.

To display and analyze the full dynamic ranges of the images, we build histograms of the pixel values divided by the noise RMS, then rescaled by the hyperbolic arcsine function (asinh). This is similar to a logarithmic scale, except that it behaves linearly close to zero, which allows a proper representation of both negative fluctuations from the background and extreme values from bright objects.

We find the agreement with the observed maps to be good over most of the dynamic range, except possibly at the very bright end. In fact, the downside of this approach is that the bright pixel counts are very sensitive to statistical fluctuations, and a single bright (but usually rare) object can drastically impact the measured distributions. This effect can be seen in \rfig{FIG:ir_pd} for the $24$ to $160\,\um$ bands. In practice, the brightest pixels in the observed GOODS-{\it South} $24\,\um$ map belong to: a) a $z=0.3$ strong starburst (factor of ten above the MS) with an AGN; b) a pair of interacting $z=0.07$ galaxies; and c) another $z=0.03$ galaxy. Except for the AGN, these objects could be found in the simulation, although they will be rare because the volume probed at these redshifts is small. Furthermore, the minimum redshift we imposed in the simulation ($z=0.05$) prevents us from properly sampling this low redshift population.

\subsection{Comparison to semi-analytic models \label{SEC:sam}}

% program: code/egg/calibrate/mags/plot_opt.pro
% file: code/egg/calibrate/mags/mags_all.eps
\begin{figure*}
    \centering
    \includegraphics[width=\textwidth]{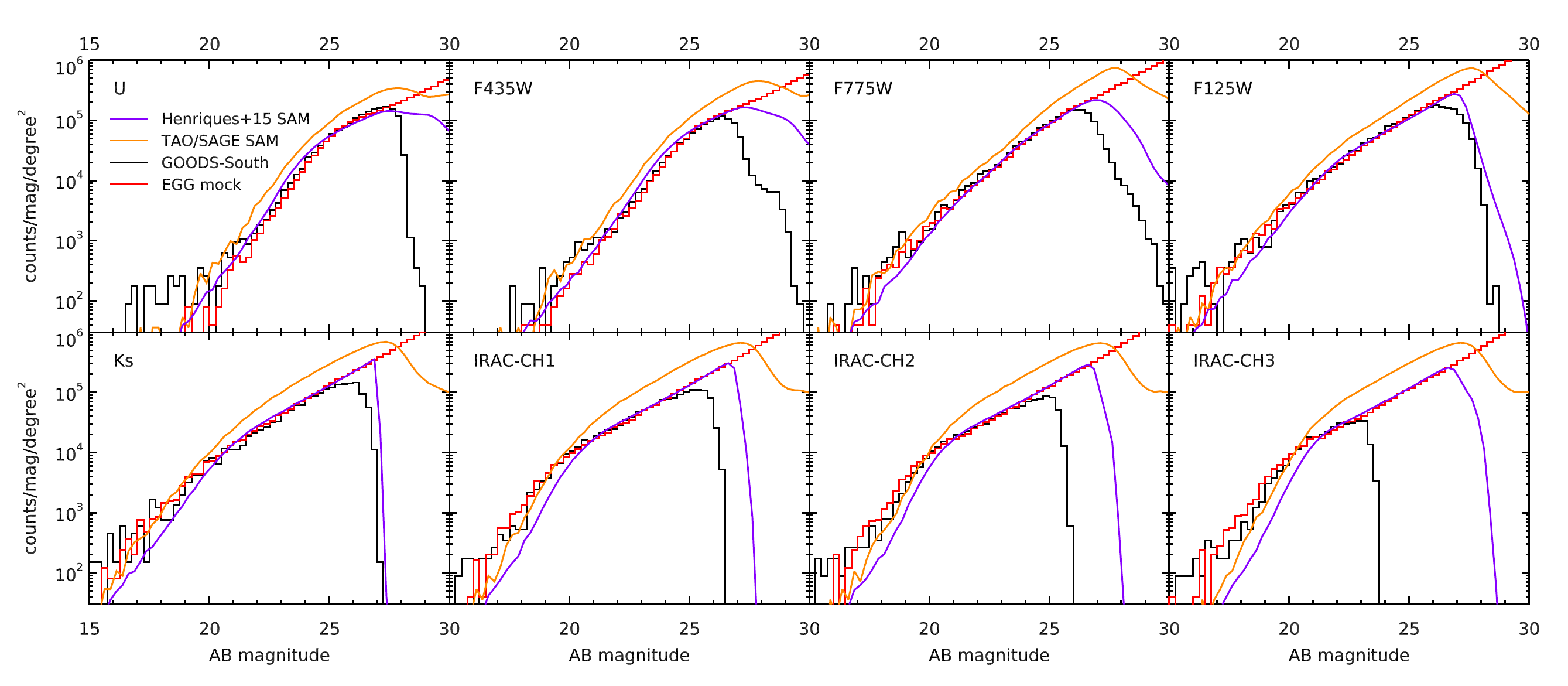}
    \caption{Same as \rfig{FIG:mag_hist}, but showing more bands and adding the counts predicted by the TAO/SAGE SAM (orange curve) and the \citetalias{henriques2015} SAM (purple curve).}
    \label{FIG:mags_sam}
\end{figure*}

\begin{figure*}
    \centering
    \includegraphics[width=0.6\textwidth]{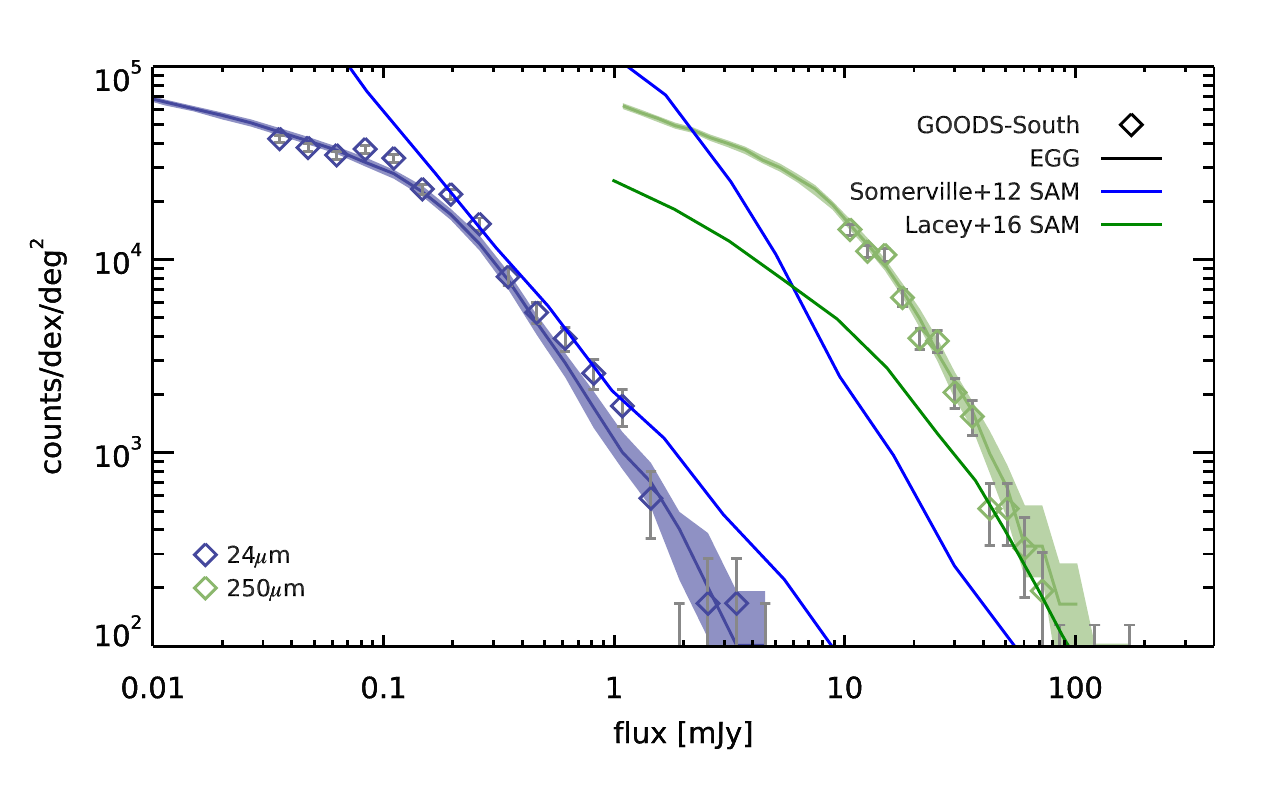}
    \caption{Same as \rfig{FIG:irflux}, but showing only the $24$ and $250\,\um$ bands and adding the counts predicted by the \citetalias{somerville2012} SAM (blue curves) and the \citetalias{lacey2016} SAM (green curve).}
    \label{FIG:ir_sam}
\end{figure*}

We now compare EGG to mock catalogs generated from state-of-the-art semi-analytic models. In particular, the predicted number counts in the UV-NIR are compared in \rfig{FIG:mags_sam}.

We investigated first the Theoretical Astronomy Observatory (TAO\footnote{\url{https://tao.asvo.org.au/tao/}}; \citealt{bernyk2016}), which uses the SAGE SAM \citep{croton2016} and an additional module to produce galaxy SEDs and broadband fluxes. We use the TAO v3.0 web interface to generate a mock catalogs of about $0.2\,\deg^2$, covering $0.01<z<10$, and keeping all galaxies with a \Ks-band magnitude brighter than $31$. We generated apparent AB magnitudes in all \hubble, \spitzer and \herschel passbands, as well as the $U$ and \Ks bands. Producing the catalog took about $27$ hours (against $7$ minutes for EGG with a similar setup).

While we requested a generous limiting magnitude of $31$ in the \Ks band, we find that the TAO mock catalog is incomplete below $K\sim27$, or equivalently below a stellar mass of $\sim10^8\,\msun$ at nearly all redshifts, because of the finite mass resolution of the underlying dark matter simulation. This is impractical if one is interested in simulating faint galaxies, however this limit is still below the depth of CANDELS and allows a fair comparison with observations, which we provide in \rfig{FIG:mags_sam}.

We find that the shape of the counts is well reproduced by TAO, however these predicted counts are systematically brighter than the observed counts by about $1$ magnitude in all passbands from the F435W to the F160W. In the \spitzer IRAC passbands, the situation gets worse and the shape of the TAO counts deviates significantly from the observations, overproducing galaxies at magnitudes fainter than $\sim20$. In contrast, the counts predicted by EGG follow very closely the observations with no significant deviation.

We also compared EGG against the \cite{henriques2015} SAM (hereafter \citetalias{henriques2015}), which can produce fluxes of galaxies from the $U$ up to the \spitzer IRAC bands. Among the available pre-built catalogs\footnote{\url{http://galformod.mpa-garching.mpg.de/public/LGalaxies/downloads.php}}, we consider a single $3\,\deg^2$ light cone selected with $K<27$. We find that the counts predicted by the \citetalias{henriques2015} model are in excellent agreement with the observations in all bands, except at the bright end in the NIR where the model is underpredicting the counts. As discussed in \cite{henriques2012}, this is most likely caused by the lack of dust emission in their model, which would increases the NIR-MIR fluxes of the bright local galaxies in the field of view.

Reproducing the far-IR counts is generally more challenging for SAMs. As pointed out by \cite{gruppioni2015}, they typically fail at producing the most actively star-forming galaxies, which are necessarily dusty. This inevitably impedes their ability to reproduce far-IR counts. While TAO is currently advertised to generate fluxes from the UV to the FIR, we found that the current implementation fails at generating any sensible dust emission in the mock galaxies; fluxes are too low by several orders of magnitude and SEDs are clearly unphysical. This is most likely an implementation issue, and we therefore do not attempt to compare these fluxes here. Instead, we compare our results to the IR counts predicted by the \cite{somerville2012} (hereafter \citetalias{somerville2012}) and \cite{lacey2016} (hereafter \citetalias{lacey2016}) SAMs in \rfig{FIG:ir_sam}.

The \citetalias{somerville2012} model reproduces correctly the bright counts at $24\,\um$, but clearly overpredicts the faint end, below $0.2\,\mJy$. At longer wavelengths, the SPIRE $250\,\um$ counts are about an order of magnitude lower than observations for $10\,\mJy < S_{250} < 100\,\mJy$. The \citetalias{lacey2016} model provides a better fit at the bright end, but still globally underpredicts the counts.

In summary, different SAMs have different strengths. Some may give an accurate description of the counts in a given wavelength regime, but will fail (or not provide any prediction) in another. Ultimately, while they have an important predictive power that is lacking to EGG, no single SAM is currently able to reproduce the emission of galaxies from the UV to the FIR at the redshifts considered here.

\subsection{Example of simulated images \label{SEC:comp_img}}

\begin{figure*}
    \centering
    \includegraphics[width=0.85\textwidth]{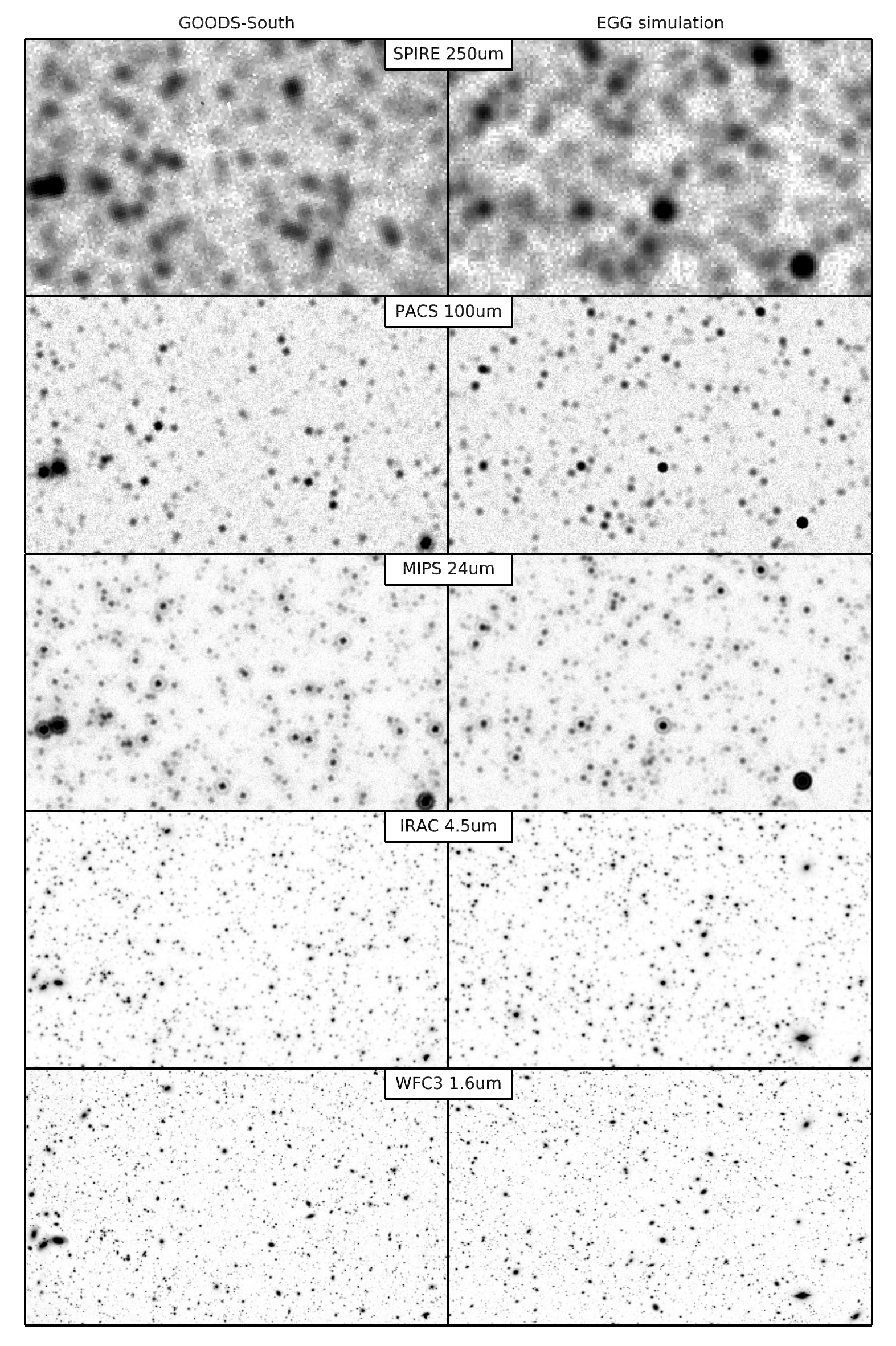}
    \caption{Comparison of real images from GOODS--{\it South} (left) and a random excerpt from a simulated field produced by \egg (right). Each image is $9\arcmin\times5.4\arcmin$. From top to bottom (FWHM of the PSF): \herschel SPIRE $500\,\um$ ($36.6\arcsec$) and PACS $100\,\um$ ($6.7\arcsec$); \spitzer MIPS $24\,\um$ ($5.7\arcsec$) and IRAC-ch2 $4.5\,\um$ ($1.6\arcsec$); and \hubble WFC3-F160W $1.6\,\um$ ($0.19\arcsec$). Images of a given band are shown with the same color bar.}
    \label{FIG:comp_sim}
\end{figure*}

\begin{sidewaysfigure*}
    \centering
    \includegraphics[width=\textheight]{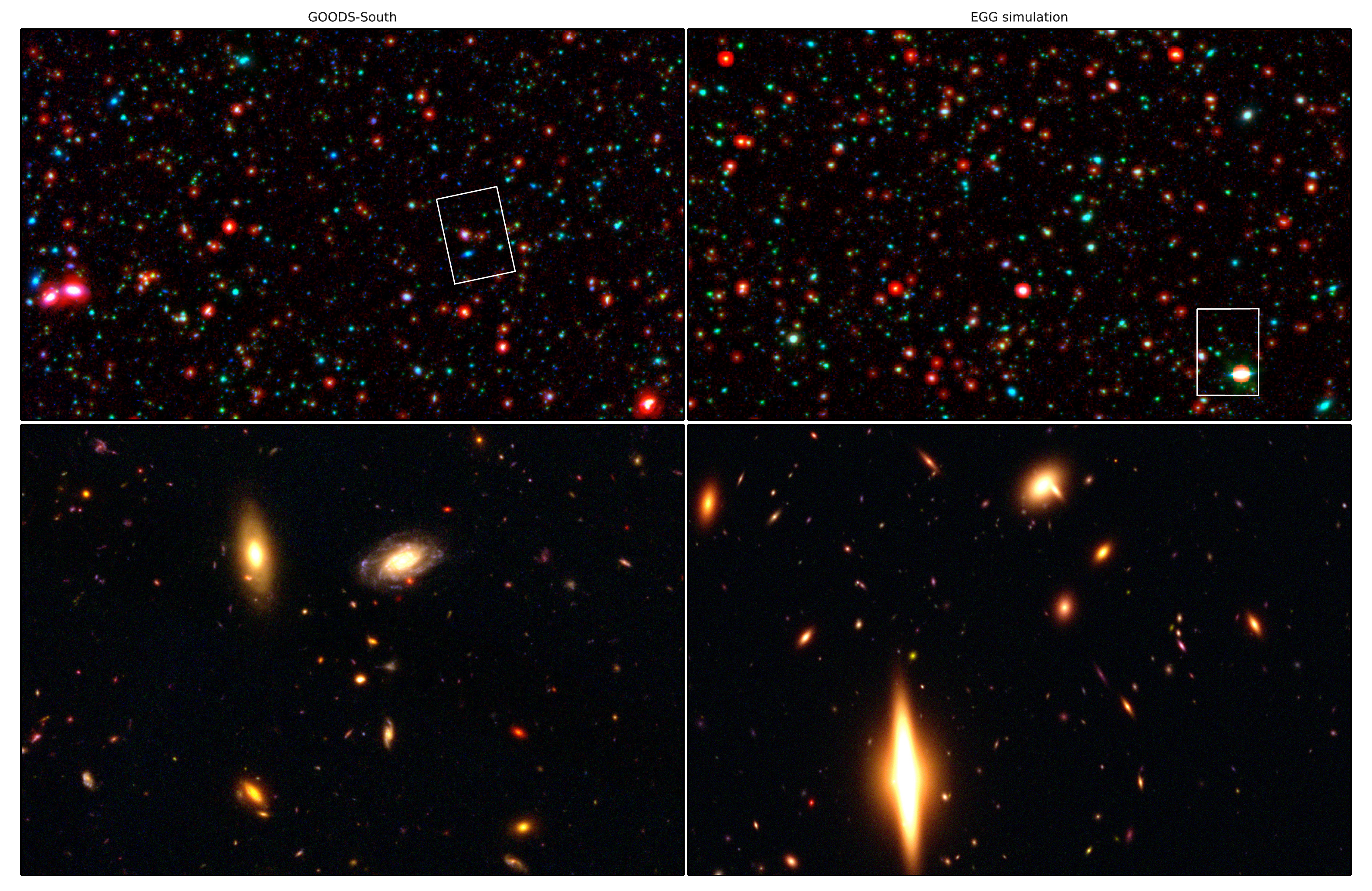}
    \caption{{\bf Top:} Three-color version of \rfig{FIG:comp_sim}, with \hubble WFC3-F160W (blue), \spitzer IRAC-ch2 (green) and \herschel PACS $100\,\um$ (red). The real images from GOODS--{\it South} are shown on the left, and the simulated images are shown on the right with the same color bar. {\bf Bottom:} Zoom-in on a small region of the above image, showing the \hubble ACS-F435W (blue), ACS-F850LP (green) and WFC3-F160W (red) bands. These regions were chosen to contain large disk galaxies.}
    \label{FIG:comp_sim_rgb}
\end{sidewaysfigure*}

A more qualitative way to assess the realism of the mock catalogs is to produce mock images of the sky and compare them by eye to observed ones. For this test, we use the FIR images introduced in the previous section, and build synthetic \hubble and \spitzer-IRAC images by feeding the flux catalogs produced by \texttt{egg-gencat} to \skymaker \citep{bertin2009}. Contrary to \texttt{egg-genmap}, \skymaker takes into account the detailed morphology we generated for each galaxy in \rsec{SEC:morpho}. We tune the background surface brightness and exposure time in \skymaker to match the noise amplitude of the observed images and use the observed PSFs from the GOODS--{\it South} field; the \texttt{egg-2skymaker} script is provided with \egg to make the transition to \skymaker straightforward and to automatically prepare the configuration files.

We display individual bands separately in \rfig{FIG:comp_sim}, and a three-color image in \rfig{FIG:comp_sim_rgb} (top). From afar, no obvious difference can be spotted. Only when zooming in on the high resolution \hst image, as in \rfig{FIG:comp_sim_rgb} (bottom), can one identify galaxies with complex morphologies (e.g., clumps or spiral arms) that are by construction absent from the simulation.

These images and the associated catalogs will be made publicly available on the \astrodeep website.

%------------------------------
%------------------------------
\section{Limitations and missing ingredients \label{SEC:future}}
%------------------------------
%------------------------------

The main objective of \egg is to simulate deep cosmological fields similar to the CANDELS fields. Because our recipes are generic and easily extensible, \egg can also be used to simulate wider areas up to a few square degrees, like the COSMOS field \citep{scoville2007}, or fields deeper than what \hubble has currently observed. However, this requires some caution:
\begin{itemize}
\item As shown in \rapp{APP:counts}, our simulated number counts are consistent with the observations in $48\,\deg^2$ from GAMA \citep{hill2011} up to bright magnitudes from the $g$ to $z$ bands, but tend to under-predict the counts in the $u$ band (at magnitudes $<20$ by a factor $2$) and over-predict the counts in the $K$ band (at magnitudes $<18$ by a similar factor). Improving the agreement here would require a finer tuning of our recipes for low redshifts ($z<0.3$), which we have mostly neglected in the present work.
\item The counts in the MIR-to-submm at the extreme bright end can be significantly affected by strongly lensed galaxies \citep[e.g.,][]{bethermin2011}, which are not modeled here.
\item As noted in \rsec{SEC:clustering}, our method to implement clustering will not generate any correlation on scales larger than $3\arcmin$, which is about an order of magnitude too small for galaxies in the local Universe \citep[e.g.,][]{connolly2002}.
\item When going deeper than $H\sim29$, the catalogs rely on extrapolations of our recipes in domains that still remain to be observationally constrained (very low mass galaxies at low redshift, or very high redshift galaxies). For example, when constructing the stellar mass functions, we assume for simplicity that the fraction of quiescent galaxies remains constant at all $z>4$ at the level of $15\%$. This assumption is totally unconstrained at present. Similarly, we had to explicitly adjust the $M/L$ ratios of high redshift galaxies to be able to reproduce the $z>5$ UV luminosity functions, and similar correction might be required for $z>8$ galaxies that will be detected by future facilities.
\end{itemize}
Therefore, while \egg can technically produce a catalog of any area and depth, it is important to consider the aforementioned limitations before attempting to do so.

Another limitation already mentioned in the previous section, and illustrated in \rfig{FIG:comp_sim_rgb} (bottom), is that the morphologies of our galaxies are idealistic. In particular, disks are only described as smooth profiles, based only on their projected axis ratio (i.e., inclination) and half-light radius; no detail is simulated on the spiral arm structure, bars, or the presence of clumps and dust lanes. Given a statistical description of their occurrence \citep[e.g.,][]{simmons2014,guo2015,willett2015}, spiral arms, bars and clumps can be straightforwardly added as a post-processing step outside of \egg by redistributing the flux of the disk, and mock images of these galaxies can be created using \galfit. Dust lanes on the other hand are more complex to simulate, since they should modify both morphology and colors as function of the disk's inclination \citep[e.g.,][]{tuffs2004}. Reddening is presently independent on inclination in \egg, so the recipe to generate axis ratios would need to be modified to take into account the chosen SED of the galaxy (or conversely). It is also questionable whether high redshift galaxies can be described by a simple disk+bulge model. High redshift galaxies at $z>3$ tend to have more irregular morphologies (see, e.g., \citealt{kartaltepe2015,huertas-company2016}), although part of this is related to the fact that \hubble only observes their rest-frame UV emission, which is affected by variable dust extinction and bright clumps; it is unknown whether disks already existed at these epochs, and only upcoming rest-optical imaging from \jwst will answer this question.

Future improvements to the tool will aim toward the inclusion of emission lines in the generated spectra, including lines in both the UV-optical (e.g., Ly$\alpha$, H$\alpha$, H$\beta$, [\ion{O}{ii}], [\ion{O}{iii}]) and the FIR (i.e., mainly [\ion{C}{ii}] and the CO ladder). While being a convenient mean to estimate line fluxes and integration time when designing observing proposals, this will most importantly allow the simulation of grism data, which will be particularly useful for Euclid. Adding emission lines to the simulation will also improve the accuracy of the rest-optical SEDs of high-redshift galaxies ($z>5$) which are know to have emission lines of particularly large equivalent widths \citep[e.g.,][]{stark2013}. At this stage, the inclusion of AGNs will become an important step to properly reproduce the line ratios and luminosity functions \citep[e.g.,][]{kauffmann2003-a,hao2005}. This would also allow us to refine the counts at the bright end in the mid-IR (see \rsec{SEC:counts}). Once AGNs are added to the simulation, the spectrum of each galaxy can be safely extended to cover more extreme wavelength regimes, including in particular the X-ray and the radio (the simulation of the X-ray domain has already been successfully attempted in a companion paper; \citealt{cappelluti2016}).

Lastly, another improvement that will come in a future version is the inclusion of foreground stars from the Milky Way, which can be used to evaluate the efficiency of standard algorithms for stars-galaxies separation in deep fields. Stars also dominate the bright end of the counts, and the halos of the brightest ones can severely impact one's ability to recover accurate fluxes in their surroundings; it can therefore be an important component especially in the simulation of wider fields where such stars are common.

%------------------------------
%------------------------------
\section{Conclusions}
%------------------------------
%------------------------------

We presented \egg, a simple tool to generate realistic galaxy mock catalogs. Produced and released to the astrophysics community as part of the \astrodeep collaboration, this program is designed to run on personal computers and can be used to create {\it ab initio} complete flux catalogs across the whole UV-to-submm spectrum for hundreds of thousands of galaxies, taking into account the emission of both stars and dust. The whole process takes only a few seconds of computation time on a single CPU.

The approach chosen to generate each galaxy is purely empirical, in that it is based on observed distributions and scaling relations, the stellar mass and the redshift of the galaxy being the two main drivers of the generated properties and spectra. The strength of this approach is that, almost by construction, we are able to accurately reproduce not only the flux distributions within the whole mock catalog, but also the relations between the individual fluxes of each galaxies. The catalogs created by this tool can therefore be used to optimize and/or validate the usual flux extraction techniques on real images, as well as to provide hints regarding the proper choice of priors when extracting highly confused images.

Lastly, since our simulated counts accurately reproduce the observations, they can also be extrapolated to fainter limits to predict the outcome of future surveys, e.g., with \jwst or ALMA. To this end, we also provide tabulated number counts (both differential and cumulative) on the EGG website (see \rapp{APP:counts}).

%------------------------------
\begin{acknowledgements}

The authors want to thank the anonymous referee for his/her comments that clearly improved the consistency and overall quality of this paper.

\egg is written using {\tt phy++}, a free and open source C++ library for fast and robust numerical astrophysics (\hlink{cschreib.github.io/phypp/}).

This work is based on observations taken by the CANDELS Multi-Cycle Treasury Program with the NASA/ESA \hst, which is operated by the Association of Universities for Research in Astronomy, Inc., under NASA contract NAS5-26555.

This research was supported by the French Agence Nationale de la Recherche (ANR) project ANR-09-BLAN-0224 as well as from the European Union Seventh Framework Programme (FP7/2007-2013) under grant agreement number 312725.
\end{acknowledgements}
%------------------------------

%------------------------------
\bibliographystyle{aa}
\bibliography{../bbib/full}

\begin{thebibliography}{116}
\expandafter\ifx\csname natexlab\endcsname\relax\def\natexlab#1{#1}\fi

\bibitem[{Arnouts {et~al.}(2013)Arnouts, Le~Floc'h, Chevallard, Johnson,
  Ilbert, Treyer, Aussel, Capak, Sanders, Scoville, {McCracken}, Milliard,
  Pozzetti, \& Salvato}]{arnouts2013}
Arnouts, S., Le~Floc'h, E., Chevallard, J., {et~al.} 2013, \aap, 558, 67

\bibitem[{Baldry {et~al.}(2012)Baldry, Driver, Loveday, Taylor, Kelvin, Liske,
  Norberg, Robotham, Brough, Hopkins, Bamford, Peacock, {Bland-Hawthorn},
  Conselice, Croom, Jones, Parkinson, Popescu, Prescott, Sharp, \&
  Tuffs}]{baldry2012}
Baldry, I.~K., Driver, S.~P., Loveday, J., {et~al.} 2012, \mnras, 421, 621

\bibitem[{Bernhard {et~al.}(2014)Bernhard, B\'{e}thermin, Sargent, Buat,
  Mullaney, Pannella, Heinis, \& Daddi}]{bernhard2014}
Bernhard, E., B\'{e}thermin, M., Sargent, M., {et~al.} 2014, \mnras, 442, 509

\bibitem[{Bernyk {et~al.}(2016)Bernyk, Croton, Tonini, Hodkinson, Hassan,
  Garel, Duffy, Mutch, Poole, \& Hegarty}]{bernyk2016}
Bernyk, M., Croton, D.~J., Tonini, C., {et~al.} 2016, \apjs, 223, 9

\bibitem[{Bertin(2009)}]{bertin2009}
Bertin, E. 2009, \memsai, 80, 422

\bibitem[{Bertin \& Arnouts(1996)}]{bertin1996}
Bertin, E. \& Arnouts, S. 1996, \aaps, 117, 393

\bibitem[{B\'{e}thermin {et~al.}(2015)B\'{e}thermin, Daddi, Magdis, Lagos,
  Sargent, Albrecht, Aussel, Bertoldi, Buat, Galametz, Heinis, Ilbert, Karim,
  Koekemoer, Lacey, Le~Floc'h, Navarrete, Pannella, Schreiber, Smol\v{c}i\'{c},
  Symeonidis, \& Viero}]{bethermin2015-a}
B\'{e}thermin, M., Daddi, E., Magdis, G., {et~al.} 2015, \aap, 573, A113

\bibitem[{B\'{e}thermin {et~al.}(2012)B\'{e}thermin, Daddi, Magdis, Sargent,
  Hezaveh, Elbaz, Le~Borgne, Mullaney, Pannella, Buat, Charmandaris, Lagache,
  \& Scott}]{bethermin2012-a}
B\'{e}thermin, M., Daddi, E., Magdis, G., {et~al.} 2012, \apjl, 757, L23

\bibitem[{B\'{e}thermin {et~al.}(2010)B\'{e}thermin, Dole, Cousin, \&
  Bavouzet}]{bethermin2010-a}
B\'{e}thermin, M., Dole, H., Cousin, M., \& Bavouzet, N. 2010, \aap, 516, 43

\bibitem[{B\'{e}thermin {et~al.}(2011)B\'{e}thermin, Dole, Lagache, Le~Borgne,
  \& Penin}]{bethermin2011}
B\'{e}thermin, M., Dole, H., Lagache, G., Le~Borgne, D., \& Penin, A. 2011,
  \aap, 529, A4

\bibitem[{Bouwens {et~al.}(2015)Bouwens, Illingworth, Oesch, Trenti, Labb\'{e},
  Bradley, Carollo, van Dokkum, Gonzalez, Holwerda, Franx, Spitler, Smit, \&
  Magee}]{bouwens2015}
Bouwens, R.~J., Illingworth, G.~D., Oesch, P.~A., {et~al.} 2015, \apj, 803, 34

\bibitem[{Bowler {et~al.}(2014)Bowler, Dunlop, {McLure}, Rogers, {McCracken},
  {Milvang-Jensen}, Furusawa, Fynbo, Taniguchi, Afonso, Bremer, \&
  Le~F\`{e}vre}]{bowler2014}
Bowler, R. A.~A., Dunlop, J.~S., {McLure}, R.~J., {et~al.} 2014, \mnras, 440,
  2810

\bibitem[{Brammer {et~al.}(2008)Brammer, van Dokkum, \& Coppi}]{brammer2008}
Brammer, G.~B., van Dokkum, P.~G., \& Coppi, P. 2008, \apj, 686, 1503

\bibitem[{Bruce {et~al.}(2014)Bruce, Dunlop, {McLure}, Cirasuolo, Buitrago,
  Bowler, Targett, Bell, {McIntosh}, Dekel, Faber, Ferguson, Grogin, Hartley,
  Kocevski, Koekemoer, Koo, \& {McGrath}}]{bruce2014}
Bruce, V.~A., Dunlop, J.~S., {McLure}, R.~J., {et~al.} 2014, \mnras, 444, 1001

\bibitem[{Bruzual \& Charlot(2003)}]{bruzual2003}
Bruzual, G. \& Charlot, S. 2003, \mnras, 344, 1000

\bibitem[{Buat {et~al.}(2005)Buat, {Iglesias-P\'{a}ramo}, Seibert, Burgarella,
  Charlot, Martin, Xu, Heckman, Boissier, Boselli, Barlow, Bianchi, Byun,
  Donas, Forster, Friedman, Jelinski, Lee, Madore, Malina, Milliard, Morissey,
  Neff, Rich, Schiminovitch, Siegmund, Small, Szalay, Welsh, \&
  Wyder}]{buat2005}
Buat, V., {Iglesias-P\'{a}ramo}, J., Seibert, M., {et~al.} 2005, \apjl, 619,
  L51

\bibitem[{Buat {et~al.}(2012)Buat, Noll, Burgarella, Giovannoli, Charmandaris,
  Pannella, Hwang, Elbaz, Dickinson, Magdis, Reddy, \& Murphy}]{buat2012}
Buat, V., Noll, S., Burgarella, D., {et~al.} 2012, \aap, 545, A141

\bibitem[{Calzetti {et~al.}(2000)Calzetti, Armus, Bohlin, Kinney, Koornneef, \&
  {Storchi-Bergmann}}]{calzetti2000}
Calzetti, D., Armus, L., Bohlin, R.~C., {et~al.} 2000, \apj, 533, 682

\bibitem[{Cappelluti {et~al.}(2016)Cappelluti, Comastri, Fontana, Zamorani,
  Amorin, Castellano, Merlin, Santini, Elbaz, Schreiber, Shu, Wang, Dunlop,
  Bourne, Bruce, Buitrago, Micha\l{}owski, Derriere, Ferguson, Faber, \&
  Vito}]{cappelluti2016}
Cappelluti, N., Comastri, A., Fontana, A., {et~al.} 2016, \apj, 823, 95

\bibitem[{Castellano {et~al.}(2016)Castellano, Amor\'{i}n, Merlin, Fontana,
  {McLure}, {M\'{a}rmol-Queralt\'{o}}, Mortlock, Parsa, Dunlop, Elbaz,
  Balestra, Boucaud, Bourne, Boutsia, Brammer, Bruce, Buitrago, Capak,
  Cappelluti, Ciesla, Comastri, Cullen, Derriere, Faber, Giallongo, Grazian,
  Grillo, Mercurio, Micha\l{}owski, Nonino, Paris, Pentericci, Pilo, Rosati,
  Santini, Schreiber, Shu, \& Wang}]{castellano2016}
Castellano, M., Amor\'{i}n, R., Merlin, E., {et~al.} 2016, \aap, 590, A31

\bibitem[{Chary \& Elbaz(2001)}]{chary2001}
Chary, R. \& Elbaz, D. 2001, \apj, 556, 562

\bibitem[{Ciesla {et~al.}(2014)Ciesla, Boquien, Boselli, Buat, Cortese, Bendo,
  Heinis, Galametz, Eales, Smith, Baes, Bianchi, De~Looze, di~Serego~Alighieri,
  Galliano, Hughes, Madden, Pierini, {R\'{e}my-Ruyer}, Spinoglio, Vaccari,
  Viaene, \& Vlahakis}]{ciesla2014}
Ciesla, L., Boquien, M., Boselli, A., {et~al.} 2014, \aap, 565, A128

\bibitem[{Connolly {et~al.}(2002)Connolly, Scranton, Johnston, Dodelson,
  Eisenstein, Frieman, Gunn, Hui, Jain, Kent, Loveday, Nichol, {O'Connell},
  Postman, Scoccimarro, Sheth, Stebbins, Strauss, Szalay, Szapudi, Tegmark,
  Vogeley, Zehavi, Annis, Bahcall, Brinkmann, Csabai, Doi, Fukugita, Hennessy,
  Hindsley, Ichikawa, Ivezi\'{c}, Kim, Knapp, Kunszt, Lamb, Lee, Lupton,
  {McKay}, Munn, Peoples, Pier, Rockosi, Schlegel, Stoughton, Tucker, Yanny, \&
  York}]{connolly2002}
Connolly, A.~J., Scranton, R., Johnston, D., {et~al.} 2002, \apj, 579, 42

\bibitem[{Cooper {et~al.}(2006)Cooper, Newman, Croton, Weiner, Willmer, Gerke,
  Madgwick, Faber, Davis, Coil, Finkbeiner, Guhathakurta, \& Koo}]{cooper2006}
Cooper, M.~C., Newman, J.~A., Croton, D.~J., {et~al.} 2006, \mnras, 370, 198

\bibitem[{Croton {et~al.}(2016)Croton, Stevens, Tonini, Garel, Bernyk, Bibiano,
  Hodkinson, Mutch, Poole, \& Shattow}]{croton2016}
Croton, D.~J., Stevens, A. R.~H., Tonini, C., {et~al.} 2016, \apjs, 222, 22

\bibitem[{Cucciati {et~al.}(2006)Cucciati, Iovino, Marinoni, Ilbert, Bardelli,
  Franzetti, Le~F\`{e}vre, Pollo, Zamorani, Cappi, Guzzo, {McCracken}, Meneux,
  Scaramella, Scodeggio, Tresse, Zucca, Bottini, Garilli, Le~Brun, Maccagni,
  Picat, Vettolani, Zanichelli, Adami, Arnaboldi, Arnouts, Bolzonella, Charlot,
  Ciliegi, Contini, Foucaud, Gavignaud, Marano, Mazure, Merighi, Paltani,
  Pell\`{o}, Pozzetti, Radovich, Bondi, Bongiorno, Busarello, de~la Torre,
  Gregorini, Lamareille, Mathez, Mellier, Merluzzi, Ripepi, Rizzo, Temporin, \&
  Vergani}]{cucciati2006}
Cucciati, O., Iovino, A., Marinoni, C., {et~al.} 2006, \aap, 458, 39

\bibitem[{da~Cunha {et~al.}(2008)da~Cunha, Charlot, \& Elbaz}]{dacunha2008}
da~Cunha, E., Charlot, S., \& Elbaz, D. 2008, \mnras, 388, 1595

\bibitem[{Daddi {et~al.}(2007)Daddi, Dickinson, Morrison, Chary, Cimatti,
  Elbaz, Frayer, Renzini, Pope, Alexander, Bauer, Giavalisco, Huynh, Kurk, \&
  Mignoli}]{daddi2007-a}
Daddi, E., Dickinson, M., Morrison, G., {et~al.} 2007, \apj, 670, 156

\bibitem[{Daddi {et~al.}(2005)Daddi, Renzini, Pirzkal, Cimatti, Malhotra,
  Stiavelli, Xu, Pasquali, Rhoads, Brusa, di~Serego~Alighieri, Ferguson,
  Koekemoer, Moustakas, Panagia, \& Windhorst}]{daddi2005}
Daddi, E., Renzini, A., Pirzkal, N., {et~al.} 2005, \apj, 626, 680

\bibitem[{Dale \& Helou(2002)}]{dale2002}
Dale, D.~A. \& Helou, G. 2002, \apj, 576, 159

\bibitem[{De~Santis {et~al.}(2006)De~Santis, Grazian, \&
  Fontana}]{desantis2006}
De~Santis, C., Grazian, A., \& Fontana, A. 2006,
  Mem.~Soc.~Astron.~Italiana~Supp., 9, 454

\bibitem[{Dole {et~al.}(2004)Dole, Rieke, Lagache, Puget, {Alonso-Herrero},
  Bai, Blaylock, Egami, Engelbracht, Gordon, Hines, Kelly, Le~Floc'h, Misselt,
  Morrison, Muzerolle, Papovich, {P\'{e}rez-Gonz\'{a}lez}, Rieke, Rigby,
  Neugebauer, Stansberry, Su, Young, Beichman, \& Richards}]{dole2004-a}
Dole, H., Rieke, G.~H., Lagache, G., {et~al.} 2004, \apjs, 154, 93

\bibitem[{Dunlop {et~al.}(2016)Dunlop, {McLure}, Biggs, Geach, Michalowski,
  Ivison, Rujopakarn, van Kampen, Kirkpatrick, Pope, Scott, Swinbank, Targett,
  Aretxaga, Austermann, Best, Bruce, Chapin, Charlot, Cirasuolo, Coppin, Ellis,
  Finkelstein, Hayward, Hughes, Ibar, Khochfar, Koprowski, Narayanan, Papovich,
  Peacock, Robertson, Vernstrom, van~der Werf, Wilson, \& Yun}]{dunlop2016}
Dunlop, J.~S., {McLure}, R.~J., Biggs, A.~D., {et~al.} 2016, {ArXiv} e-prints,
  1606.arXiv:1606.00227

\bibitem[{Dutton {et~al.}(2011)Dutton, van~den Bosch, Faber, Simard, Kassin,
  Koo, Bundy, Huang, Weiner, Cooper, Newman, Mozena, \& Koekemoer}]{dutton2011}
Dutton, A.~A., van~den Bosch, F.~C., Faber, S.~M., {et~al.} 2011, \mnras, 410,
  1660

\bibitem[{Elbaz {et~al.}(2007)Elbaz, Daddi, Le~Borgne, Dickinson, Alexander,
  Chary, Starck, Brandt, Kitzbichler, {MacDonald}, Nonino, Popesso, Stern, \&
  Vanzella}]{elbaz2007}
Elbaz, D., Daddi, E., Le~Borgne, D., {et~al.} 2007, \aap, 468, 33

\bibitem[{Elbaz {et~al.}(2011)Elbaz, Dickinson, Hwang, {D\'{i}az-Santos},
  Magdis, Magnelli, Le~Borgne, Galliano, Pannella, Chanial, Armus,
  Charmandaris, Daddi, Aussel, Popesso, Kartaltepe, Altieri, Valtchanov, Coia,
  Dannerbauer, Dasyra, Leiton, Mazzarella, Alexander, Buat, Burgarella, Chary,
  Gilli, Ivison, Juneau, Le~Floc'h, Lutz, Morrison, Mullaney, Murphy, Pope,
  Scott, Brodwin, Calzetti, Cesarsky, Charlot, Dole, Eisenhardt, Ferguson,
  F\"{o}rster~Schreiber, Frayer, Giavalisco, Huynh, Koekemoer, Papovich, Reddy,
  Surace, Teplitz, Yun, \& Wilson}]{elbaz2011}
Elbaz, D., Dickinson, M., Hwang, H.~S., {et~al.} 2011, \aap, 533, 119

\bibitem[{Ferguson {et~al.}(2004)Ferguson, Dickinson, Giavalisco, Kretchmer,
  Ravindranath, Idzi, Taylor, Conselice, Fall, Gardner, Livio, Madau,
  Moustakas, Papovich, Somerville, Spinrad, \& Stern}]{ferguson2004}
Ferguson, H.~C., Dickinson, M., Giavalisco, M., {et~al.} 2004, \apjl, 600, L107

\bibitem[{Finkelstein {et~al.}(2015)Finkelstein, Ryan, Papovich, Dickinson,
  Song, Somerville, Ferguson, Salmon, Giavalisco, Koekemoer, Ashby, Behroozi,
  Castellano, Dunlop, Faber, Fazio, Fontana, Grogin, Hathi, Jaacks, Kocevski,
  Livermore, {McLure}, Merlin, Mobasher, Newman, Rafelski, Tilvi, \&
  Willner}]{finkelstein2015}
Finkelstein, S.~L., Ryan, Russell~E., J., Papovich, C., {et~al.} 2015, \apj,
  810, 71

\bibitem[{Franceschini {et~al.}(2001)Franceschini, Aussel, Cesarsky, Elbaz, \&
  Fadda}]{franceschini2001}
Franceschini, A., Aussel, H., Cesarsky, C.~J., Elbaz, D., \& Fadda, D. 2001,
  \aap, 378, 1

\bibitem[{Fujimoto {et~al.}(2016)Fujimoto, Ouchi, Ono, Shibuya, Ishigaki,
  Nagai, \& Momose}]{fujimoto2016}
Fujimoto, S., Ouchi, M., Ono, Y., {et~al.} 2016, \apjs, 222, 1

\bibitem[{Galametz {et~al.}(2013)Galametz, Grazian, Fontana, Ferguson, Ashby,
  Barro, Castellano, Dahlen, Donley, Faber, Grogin, Guo, Huang, Kocevski,
  Koekemoer, Lee, {McGrath}, Peth, Willner, Almaini, Cooper, Cooray, Conselice,
  Dickinson, Dunlop, Fazio, Foucaud, Gardner, Giavalisco, Hathi, Hartley, Koo,
  Lai, de~Mello, {McLure}, Lucas, Paris, Pentericci, Santini, Simpson,
  Sommariva, Targett, Weiner, Wuyts, \& {the {CANDELS} Team}}]{galametz2013}
Galametz, A., Grazian, A., Fontana, A., {et~al.} 2013, \apjs, 206, 10

\bibitem[{Galliano {et~al.}(2008)Galliano, Dwek, \& Chanial}]{galliano2008}
Galliano, F., Dwek, E., \& Chanial, P. 2008, \apj, 672, 214

\bibitem[{Genel {et~al.}(2014)Genel, Vogelsberger, Springel, Sijacki, Nelson,
  Snyder, {Rodriguez-Gomez}, Torrey, \& Hernquist}]{genel2014}
Genel, S., Vogelsberger, M., Springel, V., {et~al.} 2014, \mnras, 445, 175

\bibitem[{Goldader {et~al.}(2002)Goldader, Meurer, Heckman, Seibert, Sanders,
  Calzetti, \& Steidel}]{goldader2002}
Goldader, J.~D., Meurer, G., Heckman, T.~M., {et~al.} 2002, \apj, 568, 651

\bibitem[{Grazian {et~al.}(2015)Grazian, Fontana, Santini, Dunlop, Ferguson,
  Castellano, Amorin, Ashby, Barro, Behroozi, Boutsia, Caputi, Chary, Dekel,
  Dickinson, Faber, Fazio, Finkelstein, Galametz, Giallongo, Giavalisco,
  Grogin, Guo, Kocevski, Koekemoer, Koo, Lee, Lu, Merlin, Mobasher, Nonino,
  Papovich, Paris, Pentericci, Reddy, Renzini, Salmon, Salvato, Sommariva,
  Song, \& Vanzella}]{grazian2015}
Grazian, A., Fontana, A., Santini, P., {et~al.} 2015, \aap, 575, A96

\bibitem[{Grogin {et~al.}(2011)Grogin, Kocevski, Faber, Ferguson, Koekemoer,
  Riess, Acquaviva, Alexander, Almaini, Ashby, Barden, Bell, Bournaud, Brown,
  Caputi, Casertano, Cassata, Castellano, Challis, Chary, Cheung, Cirasuolo,
  Conselice, Roshan~Cooray, Croton, Daddi, Dahlen, Dav\'{e}, de~Mello, Dekel,
  Dickinson, Dolch, Donley, Dunlop, Dutton, Elbaz, Fazio, Filippenko,
  Finkelstein, Fontana, Gardner, Garnavich, Gawiser, Giavalisco, Grazian, Guo,
  Hathi, H\"{a}ussler, Hopkins, Huang, Huang, Jha, Kartaltepe, Kirshner, Koo,
  Lai, Lee, Li, Lotz, Lucas, Madau, {McCarthy}, {McGrath}, {McIntosh},
  {McLure}, Mobasher, Moustakas, Mozena, Nandra, Newman, Niemi, Noeske,
  Papovich, Pentericci, Pope, Primack, Rajan, Ravindranath, Reddy, Renzini,
  Rix, Robaina, Rodney, Rosario, Rosati, Salimbeni, Scarlata, Siana, Simard,
  Smidt, Somerville, Spinrad, Straughn, Strolger, Telford, Teplitz, Trump,
  van~der Wel, Villforth, Wechsler, Weiner, Wiklind, Wild, Wilson, Wuyts, Yan,
  \& Yun}]{grogin2011}
Grogin, N.~A., Kocevski, D.~D., Faber, S.~M., {et~al.} 2011, \apjs, 197, 35

\bibitem[{Gruppioni {et~al.}(2015)Gruppioni, Calura, Pozzi, Delvecchio, Berta,
  De~Lucia, Fontanot, Franceschini, Marchetti, Menci, Monaco, \&
  Vaccari}]{gruppioni2015}
Gruppioni, C., Calura, F., Pozzi, F., {et~al.} 2015, \mnras, 451, 3419

\bibitem[{Gruppioni {et~al.}(2013)Gruppioni, Pozzi, Rodighiero, Delvecchio,
  Berta, Pozzetti, Zamorani, Andreani, Cimatti, Ilbert, Le~Floc'h, Lutz,
  Magnelli, Marchetti, Monaco, Nordon, Oliver, Popesso, Riguccini, Roseboom,
  Rosario, Sargent, Vaccari, Altieri, Aussel, Bongiovanni, Cepa, Daddi,
  {Dom\'{i}nguez-S\'{a}nchez}, Elbaz, F\"{o}rster~Schreiber, Genzel, Iribarrem,
  Magliocchetti, Maiolino, Poglitsch, P\'{e}rez~Garc\'{i}a, {Sanchez-Portal},
  Sturm, Tacconi, Valtchanov, Amblard, Arumugam, Bethermin, Bock, Boselli,
  Buat, Burgarella, {Castro-Rodr\'{i}guez}, Cava, Chanial, Clements, Conley,
  Cooray, Dowell, Dwek, Eales, Franceschini, Glenn, Griffin, Hatziminaoglou,
  Ibar, Isaak, Ivison, Lagache, Levenson, Lu, Madden, Maffei, Mainetti, Nguyen,
  {O'Halloran}, Page, Panuzzo, Papageorgiou, Pearson, {P\'{e}rez-Fournon},
  Pohlen, Rigopoulou, {Rowan-Robinson}, Schulz, Scott, Seymour, Shupe, Smith,
  Stevens, Symeonidis, Trichas, Tugwell, Vigroux, Wang, Wright, Xu, Zemcov,
  Bardelli, Carollo, Contini, Le~F\'{e}vre, Lilly, Mainieri, Renzini,
  Scodeggio, \& Zucca}]{gruppioni2013}
Gruppioni, C., Pozzi, F., Rodighiero, G., {et~al.} 2013, \mnras, 432, 23

\bibitem[{Gruppioni {et~al.}(2011)Gruppioni, Pozzi, Zamorani, \&
  Vignali}]{gruppioni2011}
Gruppioni, C., Pozzi, F., Zamorani, G., \& Vignali, C. 2011, \mnras, 416, 70

\bibitem[{Guo {et~al.}(2015)Guo, Ferguson, Bell, Koo, Conselice, Giavalisco,
  Kassin, Lu, Lucas, Mandelker, {McIntosh}, Primack, Ravindranath, Barro,
  Ceverino, Dekel, Faber, Fang, Koekemoer, Noeske, Rafelski, \&
  Straughn}]{guo2015}
Guo, Y., Ferguson, H.~C., Bell, E.~F., {et~al.} 2015, \apj, 800, 39

\bibitem[{Guo {et~al.}(2013)Guo, Ferguson, Giavalisco, Barro, Willner, Ashby,
  Dahlen, Donley, Faber, Fontana, Galametz, Grazian, Huang, Kocevski,
  Koekemoer, Koo, {McGrath}, Peth, Salvato, Wuyts, Castellano, Cooray,
  Dickinson, Dunlop, Fazio, Gardner, Gawiser, Grogin, Hathi, Hsu, Lee, Lucas,
  Mobasher, Nandra, Newman, \& van~der Wel}]{guo2013-a}
Guo, Y., Ferguson, H.~C., Giavalisco, M., {et~al.} 2013, \apjs, 207, 24

\bibitem[{Hao {et~al.}(2005)Hao, Strauss, Fan, Tremonti, Schlegel, Heckman,
  Kauffmann, Blanton, Gunn, Hall, Ivezi\'{c}, Knapp, Krolik, Lupton, Richards,
  Schneider, Strateva, Zakamska, Brinkmann, \& Szokoly}]{hao2005}
Hao, L., Strauss, M.~A., Fan, X., {et~al.} 2005, \aj, 129, 1795

\bibitem[{Hatsukade {et~al.}(2011)Hatsukade, Kohno, Aretxaga, Austermann,
  Ezawa, Hughes, Ikarashi, Iono, Kawabe, Khan, Matsuo, Matsuura, Nakanishi,
  Oshima, Perera, Scott, Shirahata, Takeuchi, Tamura, Tanaka, Tosaki, Wilson,
  \& Yun}]{hatsukade2011}
Hatsukade, B., Kohno, K., Aretxaga, I., {et~al.} 2011, \mnras, 411, 102

\bibitem[{Hatziminaoglou {et~al.}(2010)Hatziminaoglou, Omont, Stevens, Amblard,
  Arumugam, Auld, Aussel, Babbedge, Blain, Bock, Boselli, Buat, Burgarella,
  {Castro-Rodr\'{i}guez}, Cava, Chanial, Clements, Conley, Conversi, Cooray,
  Dowell, Dwek, Dye, Eales, Elbaz, Farrah, Fox, Franceschini, Gear, Glenn,
  Gonz\'{a}lez~Solares, Griffin, Halpern, Ibar, Isaak, Ivison, Lagache,
  Levenson, Lu, Madden, Maffei, Mainetti, Marchetti, Mortier, Nguyen,
  {O'Halloran}, Oliver, Page, Panuzzo, Papageorgiou, Pearson,
  {P\'{e}rez-Fournon}, Pohlen, Rawlings, Rigopoulou, Rizzo, Roseboom,
  {Rowan-Robinson}, Sanchez~Portal, Schulz, Scott, Seymour, Shupe, Smith,
  Symeonidis, Trichas, Tugwell, Vaccari, Valtchanov, Vigroux, Wang, Ward,
  Wright, Xu, \& Zemcov}]{hatziminaoglou2010}
Hatziminaoglou, E., Omont, A., Stevens, J.~A., {et~al.} 2010, \aap, 518, L33

\bibitem[{Heinis {et~al.}(2014)Heinis, Buat, B\'{e}thermin, Bock, Burgarella,
  Conley, Cooray, Farrah, Ilbert, Magdis, Marsden, Oliver, Rigopoulou, Roehlly,
  Schulz, Symeonidis, Viero, Xu, \& Zemcov}]{heinis2014}
Heinis, S., Buat, V., B\'{e}thermin, M., {et~al.} 2014, \mnras, 437, 1268

\bibitem[{Henriques {et~al.}(2012)Henriques, White, Lemson, Thomas, Guo,
  Marleau, \& Overzier}]{henriques2012}
Henriques, B. M.~B., White, S. D.~M., Lemson, G., {et~al.} 2012, \mnras, 421,
  2904

\bibitem[{Henriques {et~al.}(2015)Henriques, White, Thomas, Angulo, Guo,
  Lemson, Springel, \& Overzier}]{henriques2015}
Henriques, B. M.~B., White, S. D.~M., Thomas, P.~A., {et~al.} 2015, \mnras,
  451, 2663

\bibitem[{Hill {et~al.}(2011)Hill, Kelvin, Driver, Robotham, Cameron, Cross,
  Andrae, Baldry, Bamford, {Bland-Hawthorn}, Brough, Conselice, Dye, Hopkins,
  Liske, Loveday, Norberg, Peacock, Croom, Frenk, Graham, Jones, Kuijken,
  Madore, Nichol, Parkinson, Phillipps, Pimbblet, Popescu, Prescott, Seibert,
  Sharp, Sutherland, Thomas, Tuffs, \& van Kampen}]{hill2011}
Hill, D.~T., Kelvin, L.~S., Driver, S.~P., {et~al.} 2011, \mnras, 412, 765

\bibitem[{Hodge {et~al.}(2013)Hodge, Karim, Smail, Swinbank, Walter, Biggs,
  Ivison, Weiss, Alexander, Bertoldi, Brandt, Chapman, Coppin, Cox, Danielson,
  Dannerbauer, De~Breuck, Decarli, Edge, Greve, Knudsen, Menten, Rix,
  Schinnerer, Simpson, Wardlow, \& van~der Werf}]{hodge2013}
Hodge, J.~A., Karim, A., Smail, I., {et~al.} 2013, \apj, 768, 91

\bibitem[{{Huertas-Company} {et~al.}(2016){Huertas-Company}, Bernardi,
  {P\'{e}rez-Gonz\'{a}lez}, Ashby, Barro, Conselice, Daddi, Dekel, Dimauro,
  Faber, Grogin, Kartaltepe, Kocevski, Koekemoer, Koo, Mei, \&
  Shankar}]{huertas-company2016}
{Huertas-Company}, M., Bernardi, M., {P\'{e}rez-Gonz\'{a}lez}, P.~G., {et~al.}
  2016, \mnras, 462, 4495

\bibitem[{Kartaltepe {et~al.}(2015)Kartaltepe, Mozena, Kocevski, {McIntosh},
  Lotz, Bell, Faber, Ferguson, Koo, Bassett, Bernyk, Blancato, Bournaud,
  Cassata, Castellano, Cheung, Conselice, Croton, Dahlen, de~Mello, {DeGroot},
  Donley, Guedes, Grogin, Hathi, Hilton, Hollon, Koekemoer, Liu, Lucas, Martig,
  {McGrath}, {McPartland}, Mobasher, Morlock, {O'Leary}, Peth, Pforr,
  Pillepich, Rosario, Soto, Straughn, Telford, Sunnquist, Trump, Weiner, Wuyts,
  Inami, Kassin, Lani, Poole, \& Rizer}]{kartaltepe2015}
Kartaltepe, J.~S., Mozena, M., Kocevski, D., {et~al.} 2015, \apjs, 221, 11

\bibitem[{Kauffmann {et~al.}(2003)Kauffmann, Heckman, White, Charlot, Tremonti,
  Brinchmann, Bruzual, Peng, Seibert, Bernardi, Blanton, Brinkmann, Castander,
  Cs\'{a}bai, Fukugita, Ivezic, Munn, Nichol, Padmanabhan, Thakar, Weinberg, \&
  York}]{kauffmann2003-a}
Kauffmann, G., Heckman, T.~M., White, S. D.~M., {et~al.} 2003, \mnras, 341, 33

\bibitem[{Koekemoer {et~al.}(2011)Koekemoer, Faber, Ferguson, Grogin, Kocevski,
  Koo, Lai, Lotz, Lucas, {McGrath}, Ogaz, Rajan, Riess, Rodney, Strolger,
  Casertano, Castellano, Dahlen, Dickinson, Dolch, Fontana, Giavalisco,
  Grazian, Guo, Hathi, Huang, van~der Wel, Yan, Acquaviva, Alexander, Almaini,
  Ashby, Barden, Bell, Bournaud, Brown, Caputi, Cassata, Challis, Chary,
  Cheung, Cirasuolo, Conselice, Roshan~Cooray, Croton, Daddi, Dav\'{e},
  de~Mello, de~Ravel, Dekel, Donley, Dunlop, Dutton, Elbaz, Fazio, Filippenko,
  Finkelstein, Frazer, Gardner, Garnavich, Gawiser, Gruetzbauch, Hartley,
  H\"{a}ussler, Herrington, Hopkins, Huang, Jha, Johnson, Kartaltepe,
  Khostovan, Kirshner, Lani, Lee, Li, Madau, {McCarthy}, {McIntosh}, {McLure},
  {McPartland}, Mobasher, Moreira, Mortlock, Moustakas, Mozena, Nandra, Newman,
  Nielsen, Niemi, Noeske, Papovich, Pentericci, Pope, Primack, Ravindranath,
  Reddy, Renzini, Rix, Robaina, Rosario, Rosati, Salimbeni, Scarlata, Siana,
  Simard, Smidt, Snyder, Somerville, Spinrad, Straughn, Telford, Teplitz,
  Trump, Vargas, Villforth, Wagner, Wandro, Wechsler, Weiner, Wiklind, Wild,
  Wilson, Wuyts, \& Yun}]{koekemoer2011}
Koekemoer, A.~M., Faber, S.~M., Ferguson, H.~C., {et~al.} 2011, \apjs, 197, 36

\bibitem[{Kriek {et~al.}(2009)Kriek, van Dokkum, Labb\'{e}, Franx, Illingworth,
  Marchesini, \& Quadri}]{kriek2009}
Kriek, M., van Dokkum, P.~G., Labb\'{e}, I., {et~al.} 2009, \apj, 700, 221

\bibitem[{Labb\'{e} {et~al.}(2006)Labb\'{e}, Bouwens, Illingworth, \&
  Franx}]{labbe2006}
Labb\'{e}, I., Bouwens, R., Illingworth, G.~D., \& Franx, M. 2006, \apjl, 649,
  L67

\bibitem[{Labb\'{e} {et~al.}(2007)Labb\'{e}, Franx, Rudnick, Schreiber, van
  Dokkum, Moorwood, Rix, R\"{o}ttgering, Trujillo, \& van~der Werf}]{labbe2007}
Labb\'{e}, I., Franx, M., Rudnick, G., {et~al.} 2007, \apj, 665, 944

\bibitem[{Lacey {et~al.}(2016)Lacey, Baugh, Frenk, Benson, Bower, Cole,
  {Gonzalez-Perez}, Helly, Lagos, \& Mitchell}]{lacey2016}
Lacey, C.~G., Baugh, C.~M., Frenk, C.~S., {et~al.} 2016, \mnras, 462, 3854

\bibitem[{Laidler {et~al.}(2007)Laidler, Papovich, Grogin, Idzi, Dickinson,
  Ferguson, Hilbert, Clubb, \& Ravindranath}]{laidler2007}
Laidler, V.~G., Papovich, C., Grogin, N.~A., {et~al.} 2007, \pasp, 119, 1325

\bibitem[{Landy \& Szalay(1993)}]{landy1993}
Landy, S.~D. \& Szalay, A.~S. 1993, \apj, 412, 64

\bibitem[{Lang {et~al.}(2014)Lang, Wuyts, Somerville, F\"{o}rster~Schreiber,
  Genzel, Bell, Brammer, Dekel, Faber, Ferguson, Grogin, Kocevski, Koekemoer,
  Lutz, {McGrath}, Momcheva, Nelson, Primack, Rosario, Skelton, Tacconi, van
  Dokkum, \& Whitaker}]{lang2014}
Lang, P., Wuyts, S., Somerville, R.~S., {et~al.} 2014, \apj, 788, 11

\bibitem[{Madden {et~al.}(2006)Madden, Galliano, Jones, \&
  Sauvage}]{madden2006}
Madden, S.~C., Galliano, F., Jones, A.~P., \& Sauvage, M. 2006, \aap, 446, 877

\bibitem[{Magdis {et~al.}(2012)Magdis, Daddi, B\'{e}thermin, Sargent, Elbaz,
  Pannella, Dickinson, Dannerbauer, da~Cunha, Walter, Rigopoulou, Charmandaris,
  Hwang, \& Kartaltepe}]{magdis2012}
Magdis, G.~E., Daddi, E., B\'{e}thermin, M., {et~al.} 2012, \apj, 760, 6

\bibitem[{Magnelli {et~al.}(2014)Magnelli, Lutz, Saintonge, Berta, Santini,
  Symeonidis, Altieri, Andreani, Aussel, B\'{e}thermin, Bock, Bongiovanni,
  Cepa, Cimatti, Conley, Daddi, Elbaz, F\"{o}rster~Schreiber, Genzel, Ivison,
  Le~Floc'h, Magdis, Maiolino, Nordon, Oliver, Page, P\'{e}rez~Garc\'{i}a,
  Poglitsch, Popesso, Pozzi, Riguccini, Rodighiero, Rosario, Roseboom,
  {Sanchez-Portal}, Scott, Sturm, Tacconi, Valtchanov, Wang, \&
  Wuyts}]{magnelli2014}
Magnelli, B., Lutz, D., Saintonge, A., {et~al.} 2014, \aap, 561, 86

\bibitem[{Mannucci {et~al.}(2010)Mannucci, Cresci, Maiolino, Marconi, \&
  Gnerucci}]{mannucci2010}
Mannucci, F., Cresci, G., Maiolino, R., Marconi, A., \& Gnerucci, A. 2010,
  \mnras, 408, 2115

\bibitem[{Masters {et~al.}(2010)Masters, Mosleh, Romer, Nichol, Bamford,
  Schawinski, Lintott, Andreescu, Campbell, Crowcroft, Doyle, Edmondson,
  Murray, Raddick, Slosar, Szalay, \& Vandenberg}]{masters2010}
Masters, K.~L., Mosleh, M., Romer, A.~K., {et~al.} 2010, \mnras, 405, 783

\bibitem[{Merlin {et~al.}(2016)Merlin, Amor\'{i}n, Castellano, Fontana,
  Buitrago, Dunlop, Elbaz, Boucaud, Bourne, Boutsia, Brammer, Bruce, Capak,
  Cappelluti, Ciesla, Comastri, Cullen, Derriere, Faber, Ferguson, Giallongo,
  Grazian, Lotz, Micha\l{}owski, Paris, Pentericci, Pilo, Santini, Schreiber,
  Shu, \& Wang}]{merlin2016}
Merlin, E., Amor\'{i}n, R., Castellano, M., {et~al.} 2016, \aap, 590, A30

\bibitem[{Merlin {et~al.}(2015)Merlin, Fontana, Ferguson, Dunlop, Elbaz,
  Bourne, Bruce, Buitrago, Castellano, Schreiber, Grazian, {McLure}, Okumura,
  Shu, Wang, Amor\'{i}n, Boutsia, Cappelluti, Comastri, Derriere, Faber, \&
  Santini}]{merlin2015}
Merlin, E., Fontana, A., Ferguson, H.~C., {et~al.} 2015, \aap, 582, A15

\bibitem[{Meurer {et~al.}(1999)Meurer, Heckman, \& Calzetti}]{meurer1999}
Meurer, G.~R., Heckman, T.~M., \& Calzetti, D. 1999, \apj, 521, 64

\bibitem[{Micha\l{}owski {et~al.}(2014)Micha\l{}owski, Hayward, Dunlop, Bruce,
  Cirasuolo, Cullen, \& Hernquist}]{michalowski2014}
Micha\l{}owski, M.~J., Hayward, C.~C., Dunlop, J.~S., {et~al.} 2014, \aap, 571,
  A75

\bibitem[{Muzzin {et~al.}(2013)Muzzin, Marchesini, Stefanon, Franx,
  {McCracken}, {Milvang-Jensen}, Dunlop, Fynbo, Brammer, Labb\'{e}, \& van
  Dokkum}]{muzzin2013}
Muzzin, A., Marchesini, D., Stefanon, M., {et~al.} 2013, \apj, 777, 18

\bibitem[{Newman {et~al.}(2012)Newman, Ellis, Bundy, \& Treu}]{newman2012}
Newman, A.~B., Ellis, R.~S., Bundy, K., \& Treu, T. 2012, \apj, 746, 162

\bibitem[{Noeske {et~al.}(2007)Noeske, Weiner, Faber, Papovich, Koo,
  Somerville, Bundy, Conselice, Newman, Schiminovich, Le~Floc'h, Coil, Rieke,
  Lotz, Primack, Barmby, Cooper, Davis, Ellis, Fazio, Guhathakurta, Huang,
  Kassin, Martin, Phillips, Rich, Small, Willmer, \& Wilson}]{noeske2007}
Noeske, K.~G., Weiner, B.~J., Faber, S.~M., {et~al.} 2007, \apjl, 660, L43

\bibitem[{Noll {et~al.}(2009)Noll, Burgarella, Giovannoli, Buat, Marcillac, \&
  {Mu\~{n}oz-Mateos}}]{noll2009}
Noll, S., Burgarella, D., Giovannoli, E., {et~al.} 2009, \aap, 507, 1793

\bibitem[{Nordon {et~al.}(2012)Nordon, Lutz, Genzel, Berta, Wuyts, Magnelli,
  Altieri, Andreani, Aussel, Bongiovanni, Cepa, Cimatti, Daddi, Fadda,
  F\"{o}rster~Schreiber, Lagache, Maiolino, P\'{e}rez~Garc\'{i}a, Poglitsch,
  Popesso, Pozzi, Rodighiero, Rosario, Saintonge, {Sanchez-Portal}, Santini,
  Sturm, Tacconi, Valtchanov, \& Yan}]{nordon2012}
Nordon, R., Lutz, D., Genzel, R., {et~al.} 2012, \apj, 745, 182

\bibitem[{{O'Halloran} {et~al.}(2006){O'Halloran}, Satyapal, \&
  Dudik}]{ohalloran2006}
{O'Halloran}, B., Satyapal, S., \& Dudik, R.~P. 2006, \apj, 641, 795

\bibitem[{Ono {et~al.}(2014)Ono, Ouchi, Kurono, \& Momose}]{ono2014}
Ono, Y., Ouchi, M., Kurono, Y., \& Momose, R. 2014, \apj, 795, 5

\bibitem[{Oteo {et~al.}(2013)Oteo, Magdis, Bongiovanni, {P\'{e}rez-Garc\'{i}a},
  Cepa, Cedr\'{e}s, Ederoclite, {S\'{a}nchez-Portal}, Aguerri, Alfaro, Altieri,
  Andreani, {Aparicio-Villegas}, Aussel, Ben\'{i}tez, Berta, Broadhurst,
  {Cabrera-Ca\~{n}o}, Castander, Cervi\~{n}o, Cimatti, {Cristobal-Hornillos},
  Daddi, Elbaz, {Fernandez-Soto}, Schreiber, Genzel, {Gonzalez-Delgado},
  Husillos, Infante, Le~Floc'h, Lutz, Magnelli, Maiolino, M\'{a}rquez,
  Mart\'{i}nez, Masegosa, Matute, Moles, Molino, Olmo, Perea,
  {P\'{e}rez-Mart\'{i}nez}, {Pintos-Castro}, Poglitsch, Polednikova, Popesso,
  Povi\'{c}, Pozzi, Prada, Quintana, Riguccini, Sturm, Tacconi, Valtchanov, \&
  Viironen}]{oteo2013-b}
Oteo, I., Magdis, G., Bongiovanni, A., {et~al.} 2013, \mnras, 435, 158

\bibitem[{Pannella {et~al.}(2009)Pannella, Carilli, Daddi, {McCracken}, Owen,
  Renzini, Strazzullo, Civano, Koekemoer, Schinnerer, Scoville,
  Smol\v{c}i\'{c}, Taniguchi, Aussel, Kneib, Ilbert, Mellier, Salvato,
  Thompson, \& Willott}]{pannella2009-a}
Pannella, M., Carilli, C.~L., Daddi, E., {et~al.} 2009, \apjl, 698, L116

\bibitem[{Pannella {et~al.}(2015)Pannella, Elbaz, Daddi, Dickinson, Hwang,
  Schreiber, Strazzullo, Aussel, Bethermin, Buat, Charmandaris, Cibinel,
  Juneau, Ivison, Le~Borgne, Le~Floc{\textquoteright}h, Leiton, Lin, Magdis,
  Morrison, Mullaney, Onodera, Renzini, Salim, Sargent, Scott, Shu, \&
  Wang}]{pannella2015}
Pannella, M., Elbaz, D., Daddi, E., {et~al.} 2015, \apj, 807, 141

\bibitem[{Patel {et~al.}(2012)Patel, Holden, Kelson, Franx, van~der Wel, \&
  Illingworth}]{patel2012}
Patel, S.~G., Holden, B.~P., Kelson, D.~D., {et~al.} 2012, \apjl, 748, L27

\bibitem[{Peebles(1982)}]{peebles1982}
Peebles, P. J.~E. 1982, \apjl, 263, L1

\bibitem[{Peng {et~al.}(2002)Peng, Ho, Impey, \& Rix}]{peng2002}
Peng, C.~Y., Ho, L.~C., Impey, C.~D., \& Rix, H. 2002, \aj, 124, 266

\bibitem[{Penner {et~al.}(2012)Penner, Dickinson, Pope, Dey, Magnelli,
  Pannella, Altieri, Aussel, Buat, Bussmann, Charmandaris, Coia, Daddi,
  Dannerbauer, Elbaz, Hwang, Kartaltepe, Lin, Magdis, Morrison, Popesso, Scott,
  \& Valtchanov}]{penner2012}
Penner, K., Dickinson, M., Pope, A., {et~al.} 2012, \apj, 759, 28

\bibitem[{Richards {et~al.}(2006)Richards, Lacy, {Storrie-Lombardi}, Hall,
  Gallagher, Hines, Fan, Papovich, Vanden~Berk, Trammell, Schneider,
  Vestergaard, York, Jester, Anderson, Budav\'{a}ri, \& Szalay}]{richards2006}
Richards, G.~T., Lacy, M., {Storrie-Lombardi}, L.~J., {et~al.} 2006, \apjs,
  166, 470

\bibitem[{Salpeter(1955)}]{salpeter1955}
Salpeter, E.~E. 1955, \apj, 121, 161

\bibitem[{Santini {et~al.}(2015)Santini, Ferguson, Fontana, Mobasher, Barro,
  Castellano, Finkelstein, Grazian, Hsu, Lee, Lee, Pforr, Salvato, Wiklind,
  Wuyts, Almaini, Cooper, Galametz, Weiner, Amorin, Boutsia, Conselice, Dahlen,
  Dickinson, Giavalisco, Grogin, Guo, Hathi, Kocevski, Koekemoer, Kurczynski,
  Merlin, Mortlock, Newman, Paris, Pentericci, Simons, \&
  Willner}]{santini2015}
Santini, P., Ferguson, H.~C., Fontana, A., {et~al.} 2015, \apj, 801, 97

\bibitem[{Sargent {et~al.}(2012)Sargent, B\'{e}thermin, Daddi, \&
  Elbaz}]{sargent2012}
Sargent, M.~T., B\'{e}thermin, M., Daddi, E., \& Elbaz, D. 2012, \apjl, 747,
  L31

\bibitem[{Schreiber {et~al.}(2016)Schreiber, Elbaz, Pannella, Ciesla, Wang,
  Koekemoer, Rafelski, \& Daddi}]{schreiber2016-b}
Schreiber, C., Elbaz, D., Pannella, M., {et~al.} 2016, \aap, 589, A35

\bibitem[{Schreiber {et~al.}(2015)Schreiber, Pannella, Elbaz, B\'{e}thermin,
  Inami, Dickinson, Magnelli, Wang, Aussel, Daddi, Juneau, Shu, Sargent, Buat,
  Faber, Ferguson, Giavalisco, Koekemoer, Magdis, Morrison, Papovich, Santini,
  \& Scott}]{schreiber2015}
Schreiber, C., Pannella, M., Elbaz, D., {et~al.} 2015, \aap, 575, A74

\bibitem[{Scoville {et~al.}(2007)Scoville, Aussel, Brusa, Capak, Carollo,
  Elvis, Giavalisco, Guzzo, Hasinger, Impey, Kneib, {LeFevre}, Lilly, Mobasher,
  Renzini, Rich, Sanders, Schinnerer, Schminovich, Shopbell, Taniguchi, \&
  Tyson}]{scoville2007}
Scoville, N., Aussel, H., Brusa, M., {et~al.} 2007, \apjs, 172, 1

\bibitem[{Shen {et~al.}(2003)Shen, Mo, White, Blanton, Kauffmann, Voges,
  Brinkmann, \& Csabai}]{shen2003}
Shen, S., Mo, H.~J., White, S. D.~M., {et~al.} 2003, \mnras, 343, 978

\bibitem[{Shu {et~al.}(2016)Shu, Elbaz, Bourne, Schreiber, Wang, Dunlop,
  Fontana, Leiton, Pannella, Okumura, Micha\l{}owski, Santini, Merlin,
  Buitrago, Bruce, Amorin, Castellano, Derriere, Comastri, Cappelluti, Wang, \&
  Ferguson}]{shu2016}
Shu, X.~W., Elbaz, D., Bourne, N., {et~al.} 2016, \apjs, 222, 4

\bibitem[{Simmons {et~al.}(2014)Simmons, Melvin, Lintott, Masters, Willett,
  Keel, Smethurst, Cheung, Nichol, Schawinski, Rutkowski, Kartaltepe, Bell,
  Casteels, Conselice, Almaini, Ferguson, Fortson, Hartley, Kocevski,
  Koekemoer, {McIntosh}, Mortlock, Newman, Ownsworth, Bamford, Dahlen, Faber,
  Finkelstein, Fontana, Galametz, Grogin, Gr\"{u}tzbauch, Guo, H\"{a}u{\ss}ler,
  Jek, Kaviraj, Lucas, Peth, Salvato, Wiklind, \& Wuyts}]{simmons2014}
Simmons, B.~D., Melvin, T., Lintott, C., {et~al.} 2014, \mnras, 445, 3466

\bibitem[{Skelton {et~al.}(2014)Skelton, Whitaker, Momcheva, Brammer, van
  Dokkum, Labb\'{e}, Franx, van~der Wel, Bezanson, Da~Cunha, Fumagalli,
  F\"{o}rster~Schreiber, Kriek, Leja, Lundgren, Magee, Marchesini, Maseda,
  Nelson, Oesch, Pacifici, Patel, Price, Rix, Tal, Wake, \&
  Wuyts}]{skelton2014}
Skelton, R.~E., Whitaker, K.~E., Momcheva, I.~G., {et~al.} 2014, \apjs, 214, 24

\bibitem[{Somerville {et~al.}(2012)Somerville, Gilmore, Primack, \&
  Dom\'{i}nguez}]{somerville2012}
Somerville, R.~S., Gilmore, R.~C., Primack, J.~R., \& Dom\'{i}nguez, A. 2012,
  \mnras, 423, 1992

\bibitem[{Soneira \& Peebles(1978)}]{soneira1978}
Soneira, R.~M. \& Peebles, P. J.~E. 1978, \aj, 83, 845

\bibitem[{Stark {et~al.}(2013)Stark, Schenker, Ellis, Robertson, {McLure}, \&
  Dunlop}]{stark2013}
Stark, D.~P., Schenker, M.~A., Ellis, R., {et~al.} 2013, \apj, 763, 129

\bibitem[{Straatman {et~al.}(2014)Straatman, Labb\'{e}, Spitler, Allen,
  Altieri, Brammer, Dickinson, van Dokkum, Inami, Glazebrook, Kacprzak,
  Kawinwanichakij, Kelson, {McCarthy}, Mehrtens, Monson, Murphy, Papovich,
  Persson, Quadri, Rees, Tomczak, Tran, \& Tilvi}]{straatman2014}
Straatman, C. M.~S., Labb\'{e}, I., Spitler, L.~R., {et~al.} 2014, \apjl, 783,
  L14

\bibitem[{Teplitz {et~al.}(2011)Teplitz, Chary, Elbaz, Dickinson, Bridge,
  Colbert, Le~Floc'h, Frayer, Howell, Koo, Papovich, Phillips, Scarlata, Siana,
  Spinrad, \& Stern}]{teplitz2011}
Teplitz, H.~I., Chary, R., Elbaz, D., {et~al.} 2011, \aj, 141, 1

\bibitem[{Tuffs {et~al.}(2004)Tuffs, Popescu, V\"{o}lk, Kylafis, \&
  Dopita}]{tuffs2004}
Tuffs, R.~J., Popescu, C.~C., V\"{o}lk, H.~J., Kylafis, N.~D., \& Dopita, M.~A.
  2004, \aap, 419, 821

\bibitem[{van~der Wel {et~al.}(2012)van~der Wel, Bell, H\"{a}ussler, {McGrath},
  Chang, Guo, {McIntosh}, Rix, Barden, Cheung, Faber, Ferguson, Galametz,
  Grogin, Hartley, Kartaltepe, Kocevski, Koekemoer, Lotz, Mozena, Peth, \&
  Peng}]{vanderwel2012}
van~der Wel, A., Bell, E.~F., H\"{a}ussler, B., {et~al.} 2012, \apjs, 203, 24

\bibitem[{van~der Wel {et~al.}(2014)van~der Wel, Chang, Bell, Holden, Ferguson,
  Giavalisco, Rix, Skelton, Whitaker, Momcheva, Brammer, Kassin, Martig, Dekel,
  Ceverino, Koo, Mozena, van Dokkum, Franx, Faber, \& Primack}]{vanderwel2014}
van~der Wel, A., Chang, Y., Bell, E.~F., {et~al.} 2014, \apjl, 792, L6

\bibitem[{Willett {et~al.}(2015)Willett, Schawinski, Simmons, Masters, Skibba,
  Kaviraj, Melvin, Wong, Nichol, Cheung, Lintott, \& Fortson}]{willett2015}
Willett, K.~W., Schawinski, K., Simmons, B.~D., {et~al.} 2015, \mnras, 449, 820

\bibitem[{Williams {et~al.}(2009)Williams, Quadri, Franx, van Dokkum, \&
  Labb\'{e}}]{williams2009}
Williams, R.~J., Quadri, R.~F., Franx, M., van Dokkum, P., \& Labb\'{e}, I.
  2009, \apj, 691, 1879

\bibitem[{Wu {et~al.}(2006)Wu, Charmandaris, Hao, Brandl, {Bernard-Salas},
  Spoon, \& Houck}]{wu2006}
Wu, Y., Charmandaris, V., Hao, L., {et~al.} 2006, \apj, 639, 157

\bibitem[{Wuyts {et~al.}(2007)Wuyts, Labb\'{e}, Franx, Rudnick, van Dokkum,
  Fazio, F\"{o}rster~Schreiber, Huang, Moorwood, Rix, R\"{o}ttgering, \&
  van~der Werf}]{wuyts2007}
Wuyts, S., Labb\'{e}, I., Franx, M., {et~al.} 2007, \apj, 655, 51

\end{thebibliography}
%------------------------------

\appendix

\section{Fit parameters for the conditional stellar mass functions}

We provide in \rtabs{TAB:mfparam_active} and \ref{TAB:mfparam_passive} the values of the  parameters of the two Schechter functions we fit to the observed stellar mass functions, as described in \rsec{SEC:mfunc}.

\begin{table*}
\begin{center}
\begin{tabular}{ccccccc}
\hline\hline \\[-2.5mm]
$z$      & $\phi^\star_1$         & $\log_{10}(M^\star_1)$ & $\alpha_1$ & $\phi^\star_2$         & $\log_{10}(M^\star_2)$ & $\alpha_2$ \\
         & $\dex^{-1}\,\Mpc^{-3}$ & $\log_{10}(\msun)$     &            & $\dex^{-1}\,\Mpc^{-3}$ & $\log_{10}(\msun)$     &            \\ \hline \\[-2.5mm]
0.3--0.7 & $8.90 \times 10^{-4}$  & [11]                   & [-1.4]     & $8.31 \times 10^{-5}$  & 10.64                  & [0.5] \\
0.7--1.2 & $7.18 \times 10^{-4}$  & [11]                   & [-1.4]     & $4.04 \times 10^{-4}$  & 10.73                  & [0.5] \\
1.2--1.8 & $4.66 \times 10^{-4}$  & [11]                   & [-1.5]     & $4.18 \times 10^{-4}$  & 10.67                  & [0.5] \\
1.8--2.5 & $2.14 \times 10^{-4}$  & [11]                   & [-1.57]    & $4.06 \times 10^{-4}$  & 10.84                  & [0.5] \\
2.5--3.5 & $2.12 \times 10^{-4}$  & [11]                   & [-1.6]     & $9.07 \times 10^{-5}$  & 10.94                  & [0.5] \\
3.5--4.5 & $4.45 \times 10^{-5}$  & [11]                   & [-1.7]     & $8.60 \times 10^{-6}$  & 11.69                  & [0.5] \\ \hline
\end{tabular}
\end{center}
\caption{Double Schechter function parameters for the SFG population. Parameters that were chosen manually are enclosed in brackets.\label{TAB:mfparam_active}}
\end{table*}

\begin{table*}
\begin{center}
\begin{tabular}{ccccccc}
\hline\hline \\[-2.5mm]
$z$ & $\phi^\star_1$             & $\log_{10}(M^\star_1)$ & $\alpha_1$ & $\phi^\star_2$          & $\log_{10}(M^\star_2)$ & $\alpha_2$ \\
    & $\dex^{-1}\,\Mpc^{-3}$     & $\log_{10}(\msun)$     &            & $\dex^{-1}\,\Mpc^{-3}$  & $\log_{10}(\msun)$     &            \\ \hline \\[-2.5mm]
0.3--0.7 & $7.77 \times 10^{-5}$ & [11]                   & [-1.65]    & $1.54 \times 10^{-3}$   & 11.04                  & -0.48 \\
0.7--1.2 & $3.54 \times 10^{-5}$ & [11]                   & [-1.60]    & $1.04 \times 10^{-3}$   & 10.86                  & 0.06 \\
1.2--1.9 & $2.30 \times 10^{-5}$ & [11]                   & [-1.25]    & $6.25 \times 10^{-4}$   & 10.83                  & 0.30 \\
1.9--2.5 & [$10^{-5}$]           & [11]                   & [-1]       & $1.73 \times 10^{-4}$   & 11.05                  & -0.17 \\
2.5--3.5 & [$0$]                 & [11]                   & [-1]       & $1.22 \times 10^{-4}$   & 10.94                  & -0.26 \\
3.5--4.5 & [$0$]                 & [11]                   & [-1.35]    & [$3 \times 10^{-5}$]    & [11]                   & [-0.30] \\ \hline
\end{tabular}
\end{center}
\caption{Double Schechter function parameters for the QG population. Parameters that were chosen manually are enclosed in brackets.\label{TAB:mfparam_passive}}
\end{table*}

\section{Estimating the effective half-light radius of a double \sersic profile \label{APP:size}}

Using \galfit, we create a suite of ideal double \sersic profiles ($n=1$ disk + $n=4$ bulge) of varying relative fluxes and sizes. We then use \galfit again to fit the generated image with a single \sersic profile of variable index $n$ and size $R_{50,\rm total}$. Using a simpler growth curve analysis, we confirm that the resulting size is an unbiased measure of the effective half-light radius of the galaxy. Assuming the two components have the same projected axis ratio, we find that this value can be approximated as
\begin{equation}
R_{50,\rm total} = R_{50,\rm disk} \times \big(1 - (B/T)^{\alpha}\big) + R_{50,\rm bulge} \times (B/T)^{\alpha}\,,
\end{equation}
where $\alpha=1-0.8\times\log_{10}(R_{50,\rm disk}/R_{50,\rm bulge})$.

\section{Number counts and map statistics for additional bands \label{APP:counts}}

\begin{figure*}
    \centering
    \includegraphics[width=0.95\textwidth]{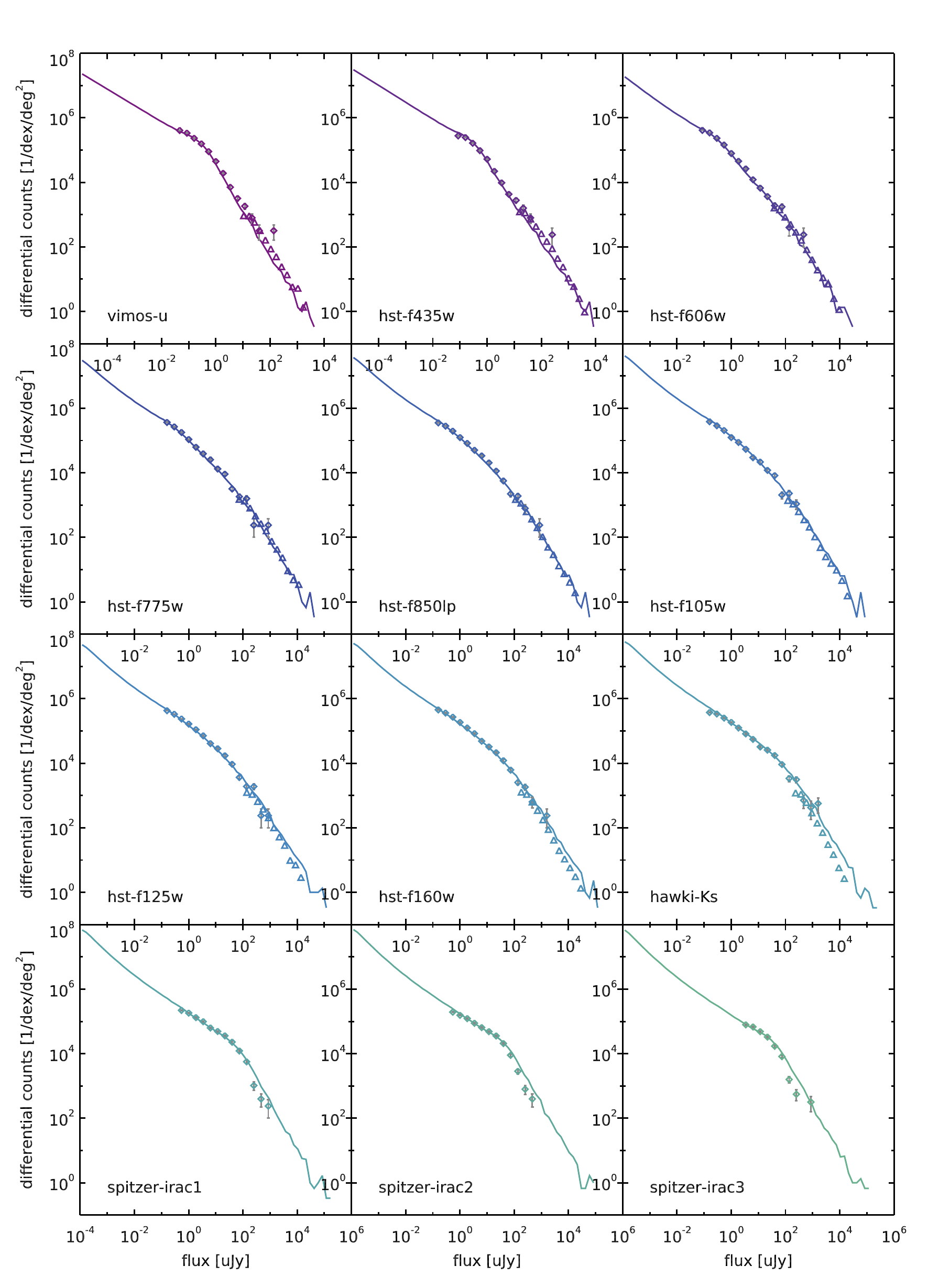}
    \caption{Differential counts produced by \egg (solid lines), compared to observations (open symbols). See text for details and references.}
    \label{FIG:full_counts1}
\end{figure*}

\begin{figure*}
    \centering
    \includegraphics[width=0.95\textwidth]{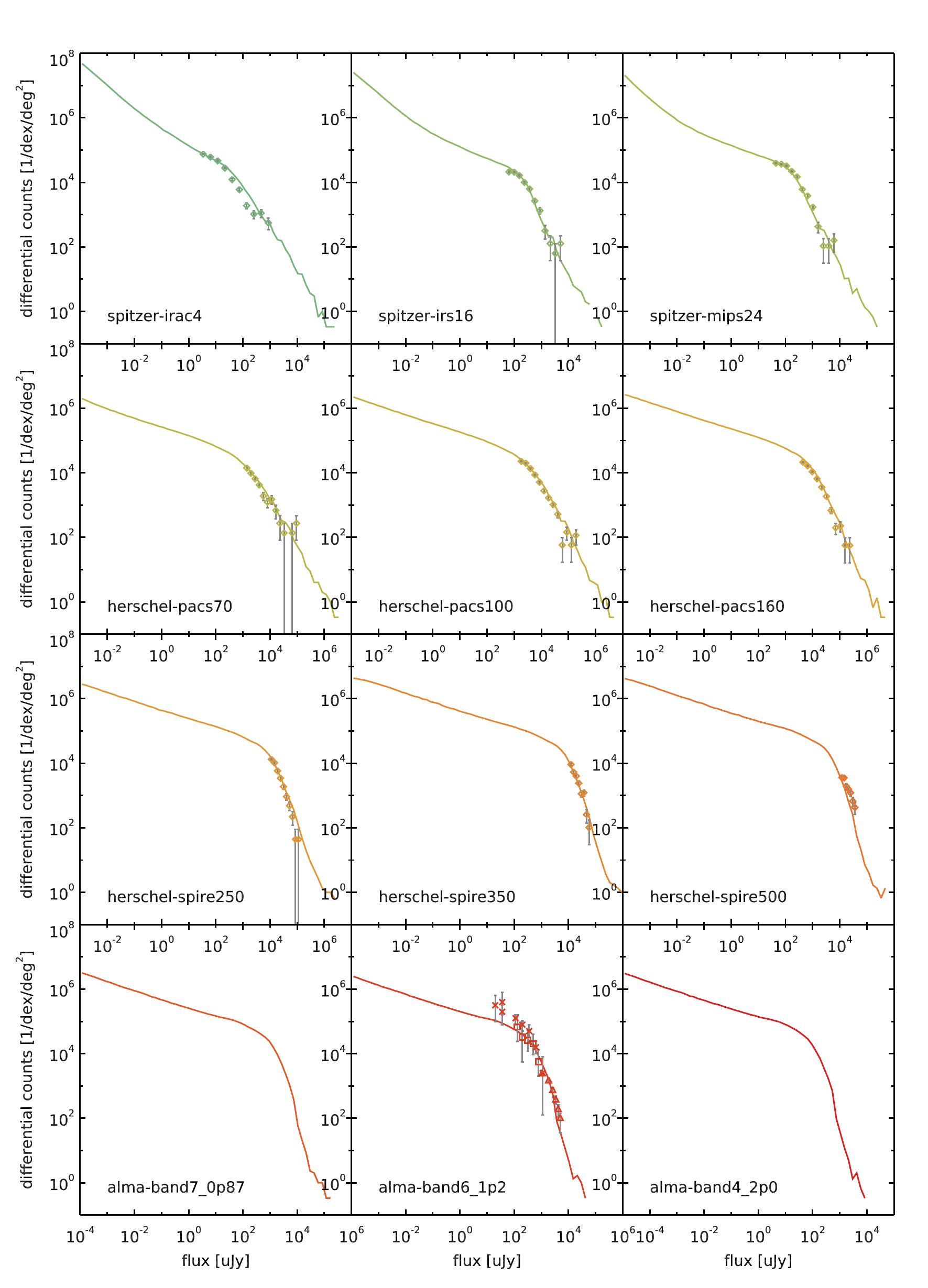}
    \caption{Same as \rfig{FIG:full_counts1} for the remaning bands.}
    \label{FIG:full_counts2}
\end{figure*}

We provide in \rfig{FIG:full_counts1} and \ref{FIG:full_counts2} the differential number counts in an exhaustive list of bands from the $u$ band to ALMA $2\,{\rm mm}$. To obtain a wide dynamic range, we generate three catalogs with a ``wedding cake'' strategy. The first catalog covers $20\,{\rm deg}^2$ down to ${\rm [4.5]} = 20$ (where [4.5] is the AB magnitude in the \spitzer IRAC $4.5\,\um$ band); the second catalog covers $1\,{\rm deg}^2$ down to ${\rm [4.5]} = 30$; and the last catalog covers $0.05\,{\rm deg}^2$ down to ${\rm [4.5]} = 36$. For all three catalogs, the minumum redshift is lowered to $z=0.005$. These catalogs contain $\sim400\,000$, $\sim3\,500\,000$ and $\sim3\,500\,000$ galaxies, respectively. The widest catalog is used for counts below $100\,\dex^{-1}{\rm deg}^{-1}$, and the deepest catalog is used for counts above $300\,000\,\dex^{-1}{\rm deg}^{-1}$.

The counts produced by \egg are compared to the observations in GOODS--{\it South} (open diamonds). For UV-NIR bands, we also show the counts from GAMA \citep[open triangles]{hill2011}, corrected for the different filter response curves. To perform the correction, we generate fluxes in both the \hst bands and the corresponding closest band from SDSS or UKIDSS, compute the median ratio of the two, and then multiply the fluxes from GAMA by this factor. This correction is $1.07$, $0.84$, $0.93$, $1.08$, $1.05$, $1.04$, $1.0$, $0.91$ and $0.97$ for the $u$, $g$, $r$, $i$, $z$, $Y$, $J$, $H$ and $K$ bands, respectively. For clarity, we only display one point every five from the Hill et al.~data. Lastly, for ALMA $1.2\,{\rm mm}$, we show the same literature data as in \rfig{FIG:irflux}.

These counts are tabulated and freely accessible on the \egg website\footnote{\url{http://cschreib.github.io/egg/counts.html}}.

% program: code/egg/paper/allcumul/plot.pro
% file: code/egg/paper/allcumul/diff1.eps

% program: code/egg/paper/allcumul/plot.pro
% file: code/egg/paper/allcumul/diff2.eps

\section{UV and IR luminosity functions \label{APP:lfs}}

\begin{sidewaysfigure}
    \centering
    \includegraphics[width=0.9\textheight]{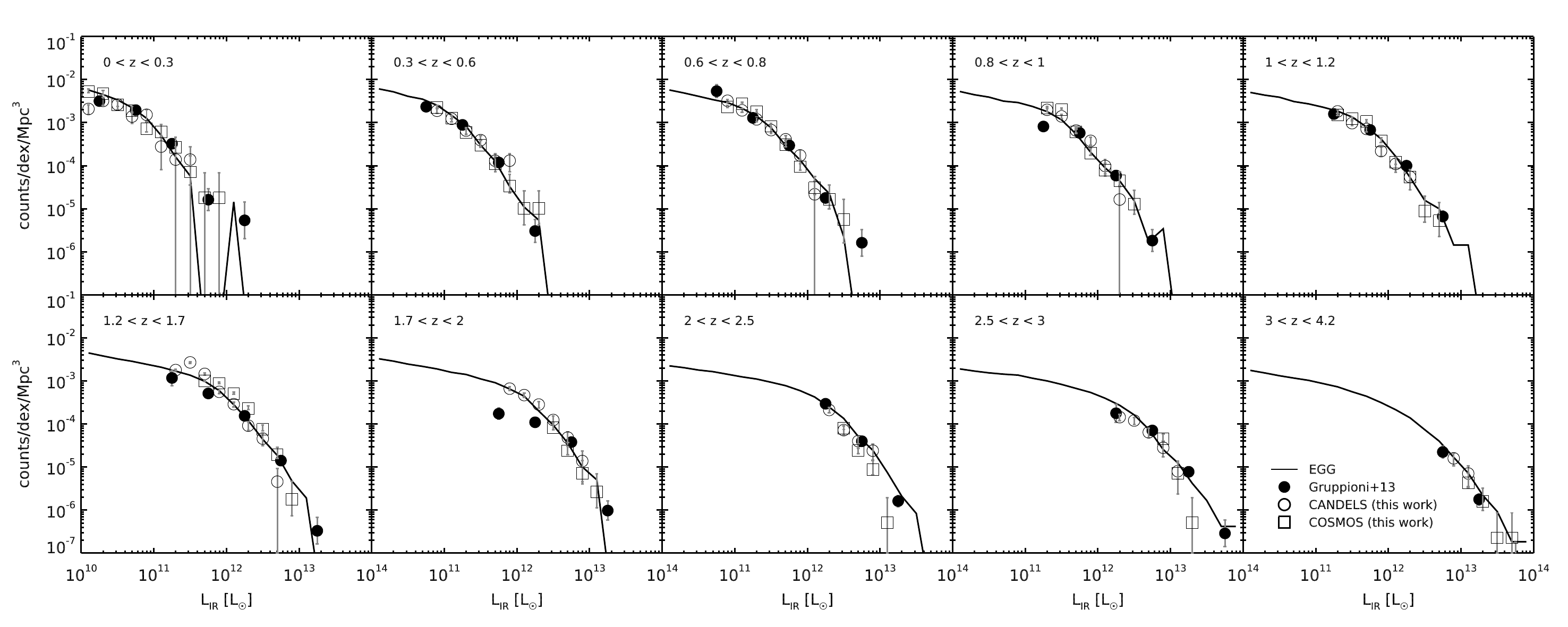}
    \caption{Infrared luminosity ($\lir$) function at various redshifts. The solid line is the output of EGG. It is compared against observations from our own catalogs in CANDELS (open circles) and COSMOS 2$\deg^2$ (open squares), and the results of \cite{gruppioni2013} in the same fields (filled circles). All observations use either \spitzer MIPS or \herschel photometry to derive $\lir$.}
    \label{FIG:lf_ir}
\end{sidewaysfigure}

\begin{figure*}
    \centering
    \includegraphics[width=0.85\textwidth]{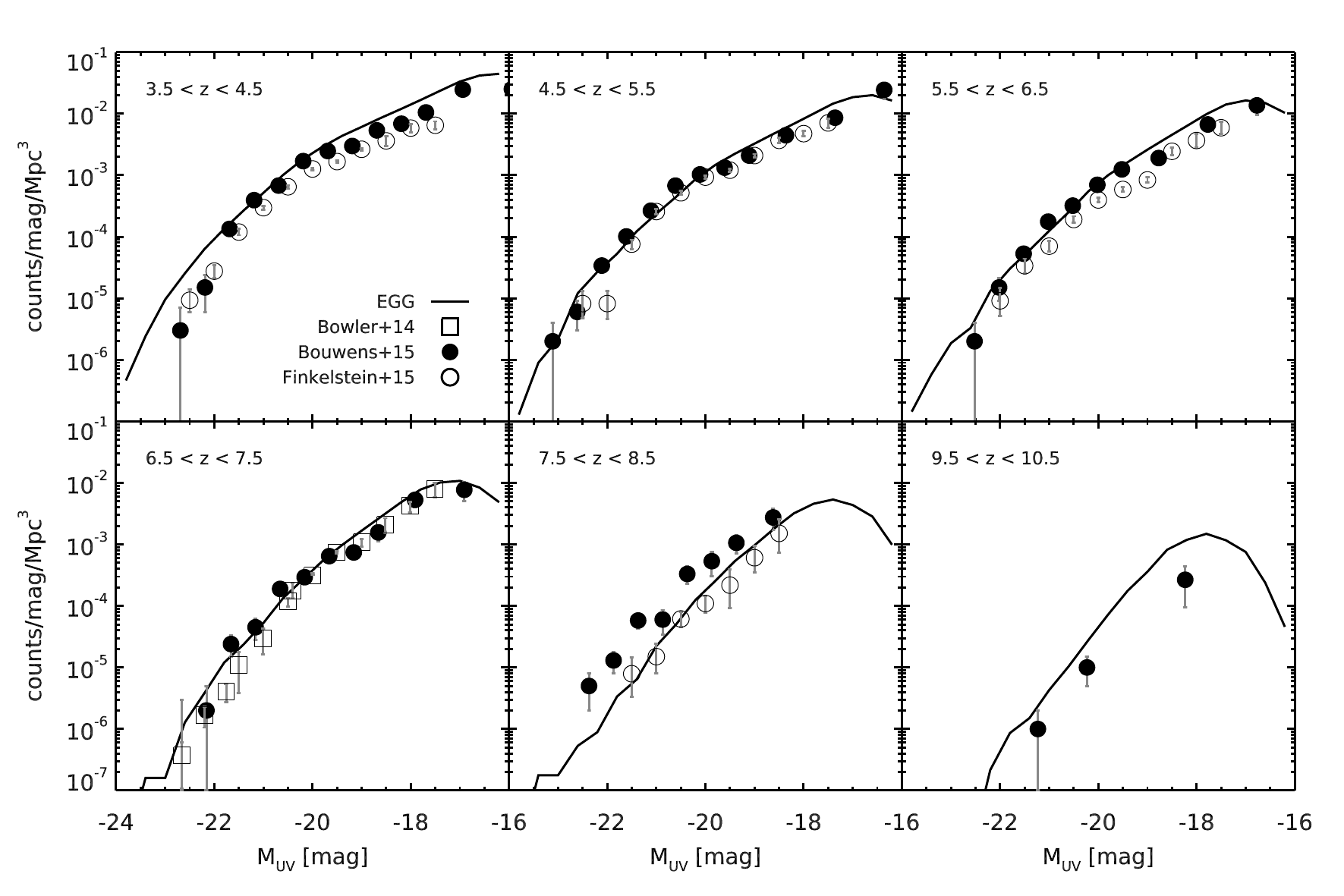}
    \caption{Ultraviolet absolute magnitude ($M_{\rm UV}$, $1600\,\AA$) function at various redshifts. The solid line is the output of EGG. It is compared against observations from \cite{bouwens2015} (filled circles) and \cite{finkelstein2015} (open circles), each compiling multiple fields observed with \hubble WFC3, and \cite{bowler2014} (open squares, only at $z=7$) which uses the UltraVISTA imaging of COSMOS.}
    \label{FIG:lf_uv}
\end{figure*}

\rfigs{FIG:lf_ir} and \ref{FIG:lf_uv} show, respectively, the IR and UV luminosity functions (LFs) produced by EGG in various redshifts bins and compare them against observations from our own catalogs or the literature. The IR LFs are in good agreement with the observations, although this is true almost by construction given that we use the same mass functions and the observed $\lir$--$\mstar$ relation. The UV luminosity is less clearly controlled by our simulation procedure, as it is indirectly determined by the \uvj colors and stellar masses. Still, the EGG LFs reproduce correctly the shape and normalization of the LFs at $3.5<z<7.5$, although our LFs seem to match better the straight power-law shape observed by \cite{bowler2014} at $z=7$, as opposed to an actual Schechter function (especially so at $z>5.5$). At $z=8$, our LF agrees better with the observations of \cite{finkelstein2015} rather than that of \cite{bouwens2015}, which are each based on a different combination of \hubble deep fields. Given the small volumes we are dealing with at these high redshifts, this could be explained by cosmic variance. On the other hand, our stellar mass functions are not observationally constrained at this redshift, hence we could agree with either of these observational data by tweaking the high redshift extrapolation of the mass function.

% program: code/egg/calibrate/luminosity_functions/plot_ir.pro
% file: code/egg/calibrate/luminosity_functions/lf_ir.eps

% program: code/egg/calibrate/luminosity_functions/plot_uv.pro
% file: code/egg/calibrate/luminosity_functions/lf_uv.eps

\end{document}